\documentclass[11pt]{article}
\usepackage[fleqn]{amsmath}
\usepackage[margin=1in]{geometry}
\usepackage{graphicx}
\usepackage{wrapfig}
\usepackage{wrapstuff}
\usepackage{amsfonts, amsmath, amssymb, amsthm}
\usepackage{newtxtext}
\usepackage{enumitem}
\setlist[itemize]{itemsep=0pt, topsep=2pt}
\setlist[enumerate]{itemsep=0pt, topsep=2pt}
\setlist[description]{itemsep=0pt, topsep=2pt}

\usepackage[font={small,it}]{caption}
\usepackage{mathrsfs}  
\usepackage{algorithm} 
\usepackage[noend]{algpseudocode}
\algrenewcommand\algorithmicrequire{\textbf{Input:}}
\algrenewcommand\algorithmicensure{\textbf{Output:}}
\usepackage{color}
\usepackage{array}
\usepackage{multirow}
\usepackage{tabularx}
\usepackage[table,xcdraw,x11names]{xcolor}
\usepackage[inkscapeformat=pdf]{svg}
\usepackage{mathtools}
\usepackage{tabulary} 
\usepackage{arydshln}
\usepackage{pdfpages}
\usepackage{booktabs}
\usepackage{comment}
\usepackage{tikz}
\usetikzlibrary{calc}
\usetikzlibrary{shapes.geometric,arrows.meta}
\usepackage{hyperref}
\usepackage{cleveref}
\definecolor{DarkRed}{rgb}{0.5,0.1,0.1}
\definecolor{DarkBlue}{rgb}{0.1,0.1,0.5}
\definecolor{ForestGreen}{rgb}{0.1333,0.5451,0.1333}
\hypersetup{colorlinks=true, linkcolor=DarkRed,citecolor=ForestGreen}

\theoremstyle{plain}
\newtheorem{theorem}{Theorem}[section]
\newtheorem{lemma}[theorem]{Lemma}

\newtheorem{observation}[theorem]{Observation}
\crefname{observation}{observation}{observations}
\Crefname{observation}{Observation}{Observations}
\newtheorem{claim}[theorem]{Claim}
\crefname{claim}{claim}{claims}
\Crefname{claim}{Claim}{Claims}
\newtheorem{fact}[theorem]{Fact}

\newtheorem{question}{Question}[section]
\theoremstyle{definition}
\newtheorem{definition}[theorem]{Definition}
\newtheorem*{remark}{Remark}
\usepackage{thm-restate}
\usepackage{booktabs}
\usepackage{tabularx}

\DeclareSymbolFont{bbold}{U}{bbold}{m}{n}
\DeclareSymbolFontAlphabet{\mathbbold}{bbold}

\newcommand{\cC}{\mathcal{C}}

\newcommand{\cU}{\mathcal{U}}

\newcommand{\cT}{\mathcal{T}}

\newcommand{\cH}{\mathcal{H}}

\newcommand{\cS}{\mathcal{S}}
\newcommand{\cO}{\mathcal{O}}

\newcommand{\Near}{\mathsf{Near}}

\newcommand{\LCA}{\mathrm{LCA}}

\newcommand{\depth}{\operatorname{depth}}

\newcommand{\before}{\prec_{(x,y)}}
\newcommand{\beforeeq}{\preccurlyeq_{(x,y)}}

\newcommand{\bef}{\prec}
\newcommand{\befeq}{\preccurlyeq}
\newcommand{\aft}{\succ}
\newcommand{\afteq}
{\succcurlyeq}

\title{An Optimal Structure for All-Pairs$^2$ Mincuts and\\ Sensitivity Oracles for All-Pairs Mincuts}

\title{Almost Optimal Oracles for All-Pairs $\times$ All-Pairs Mincuts}

\title{All-Pairs $\times$ All-Pairs Gomory-Hu Structure in $O(n)$ Space}

\title{Beyond Gomory-Hu: All-Pairs $\times$ All-Pairs Mincuts in $O(n)$ Space}

\author{Koustav Bhanja%
\thanks{Weizmann Institute of Science, Israel. Email: \texttt{koustav.bhanja@weizmann.ac.il}. Supported by the Koshland Prize fellowship, and Merav Parter’s European Research Council (ERC) grant under the European Union’s Horizon 2020 research and innovation programme, grant agreement No. 949083.}
\and Merav Parter%
\thanks{Weizmann Institute of Science, Israel. Email: \texttt{merav.parter@weizmann.ac.il}. Supported partially by the European Research Council (ERC) under the European Union’s Horizon 2020 research and innovation programme, grant agreement
No. 949083.}
\and Asaf Petruschka%
\thanks{Weizmann Institute of Science, Israel. Email: \texttt{asaf.petruschka@weizmann.ac.il}. Supported by an Azrieli Foundation fellowship, and by Merav Parter’s European Research Council (ERC) grant under the European Union’s Horizon 2020 research and innovation programme, grant agreement No. 949083.}}

\date{}

\begin{document}

\maketitle
\pagenumbering{roman}

\begin{abstract}
     For over four decades, two classical structures have offered complementary views of minimum cuts. The Gomory--Hu tree gives a remarkably compact representation of minimum cuts in an undirected graph: using only $O(n)$ space, it encodes one $(x,y)$-mincut for every pair of vertices $x,y$. Its limitation is that it retains only one mincut for each pair. In contrast, for a fixed terminal pair $x,y$, the Picard--Queyranne representation admits an $O(n)$ space ordering of vertices that can determine, for any additional pair $u,v$, whether some $(x,y)$-mincut separates $u$ and $v$, and report such a cut whenever one exists. This raises a natural question: can these two complementary guarantees be achieved simultaneously  while retaining the same $O(n)$ space bound?

We show that they can. For every undirected weighted graph, we construct an $O(n)$ space data structure that simultaneously encodes, for every choice of four vertices $x,y,u,v$, an $(x,y)$-mincut separating $u$ and $v$ whenever one exists. More precisely, given any terminal pair $x,y$, the structure recovers in $O(n)$ time the Picard--Queyranne ordering of vertices. After this update, any pair $u,v$ can be tested for separability in $O(1)$ time, and a separating mincut $C$ can be reported in $O(|C|)$ time. Thus, information concerning all $\Theta(n^4)$ choices of $x,y,u,v$ is represented using only $O(n)$ space. This is essentially the strongest information of this form that can fit in $O(n)$ space: if one additionally requires the separating mincut to place $u$ on the side of $x$ and $v$ on the side of $y$, then $\Omega(n^2)$ space is necessary even for a fixed terminal pair $x,y$.

As an immediate consequence, we obtain an $O(n)$ space all-pairs mincuts sensitivity oracle for a single edge failure in undirected multigraphs, with worst-case optimal $O(n)$ time for reporting a new mincut after the failure. This resolves an open question of Baswana and Pandey [SODA 2022], who asked whether subquadratic space and $O(m)$ query time could be achieved simultaneously. Technically, our construction substantially generalizes the recent $O(n)$ space nearest mincut hierarchy [SODA 2027], which reports only the nearest $(x,y)$-mincut. Recovering instead the entire Picard--Queyranne ordering of vertices for an arbitrary terminal pair $x,y$ requires new structural properties of the Dinitz--Vainshtein [STOC 1994] connectivity carcass that extend its use beyond the Steiner mincuts it directly represents.
\end{abstract}

\newpage
{\setlength{\parskip}{0em} \tableofcontents{}}

\newpage
\pagenumbering{arabic}

\section{Introduction}

Let $G = (V,E,w)$ be an undirected graph with $n$ vertices and $m$ edges, with integral positive weights on the edges.
For two vertices $x,y\in V$, a minimum $(x,y)$-cut, or \emph{$(x,y)$-mincut} henceforth, is a set $C \subseteq V$ with $x \in C$, $y \notin C$ such that the total weight of the edges going from $C$ to $\overline{C} = V \setminus C$ is minimized.
The study of \emph{compact representations} for different families of mincuts is an immensely rich field with numerous applications; see e.g.~\cite{dinitz1976structure, DBLP:conf/stoc/DinitzV94, DinitzN95, NagamochiI92, Karger00, DBLP:journals/jacm/KawarabayashiT19, edgeconnectivityaigmentation/naor1997fast, edgeconnaugmentation/cen2022augmenting, DBLP:conf/soda/AbboudKT22, DBLP:journals/talg/BaswanaBP23,  DBLP:conf/stoc/YKKrauthgamer26} and references therein.
Two such classical and \emph{orthogonal} representations are the \emph{Gomory-Hu tree} and the \emph{Picard-Queyranne Representation}.

    \paragraph{Gomory-Hu Tree.}
    Dating back to 1961, the Gomory-Hu (GH) tree~\cite{GomoryH61} is arguably the most well-known mincuts representation.
    It is an $O(n)$ space tree structure that encodes an $(x,y)$-mincut for \emph{each of the possible $\binom{n}{2}$ vertex pairs $x,y$}, using only the $n-1$ cuts induced by its edges.
    However, while GH trees are extremely useful, they have an important limitation:
    a GH tree recovers only \emph{one} of the possibly many $(x,y)$-mincuts. 
    This becomes important, for example, in \emph{sensitivity} settings, where one wishes to determine whether a given edge $e \in E$ belongs to \emph{some} $(x,y)$-mincut, say $C$.
    Viewing $G$ as a multigraph with $w(e)$ the multiplicity of $e$, this is equivalent to asking if whether the deletion (or \emph{failure}) of one copy of $e$ would cause a decrease in connectivity between $x$ and $y$, measured by the $(x,y)$-mincut value.
    If so, the witnessing cut $C$ remains a $(x,y)$-mincut after the failure.
    However, any particular $(x,y)$-mincut encoded by the GH tree may not contain $e$, and therefore does not reveal the affect of its deletion on $(x,y)$-mincuts.
    
    \paragraph{Picard-Queyranne Representation.}
    In 1980, Picard and Queyranne~\cite{DBLP:journals/mp/PicardQ80} showed that, for a \emph{fixed terminal pair} $x,y$, the above sensitivity setting has an $O(n)$ space representation: Given any edge $e \in E$, their structure reports an $(x,y)$-mincut $C$ with $e$ crossing from $C$ to $\overline{C}$, or determines that such does not exist.
    We now briefly explain how the Picard-Querynanne (PQ) representation works.
    For $u,v \in V$, let $u \sim_{x,y} v$ iff every $(x,y)$-mincuts keep $u$ and $v$ on the same side.
    The equivalence classes are called the  \emph{$x,y$-units}.
    The PQ representation gives partial order on the $x,y$-units, whose downward-closed sets bijectively correspond to \emph{all} $(x,y)$-mincuts.
    It is encoded by a DAG with $O(m)$ edges.
    The $O(n)$ space structure is a topological ordering of this DAG.
    Reporting an $(x,y)$-mincut where edge $e = (u,v) \in E$ participates amounts to finding a prefix of this ordering which separates (the $x,y$-units of) $u$ and $v$.
    Note that it is not actually needed that $u,v$ are the endpoints of some edge: for any vertex pair $u,v$ of interest, the $O(n)$ space PQ ordering can report an $(x,y)$-mincut separating $u$ from $v$ (if such exists).

\bigskip
To summarize, the GH tree encodes one $(x,y)$-mincut for all terminal pairs $x,y \in V$, while the PQ representation fixes the terminal pair $x,y$ but encodes $(x,y)$-mincuts separating all possible pairs of vertices $u,v \in V$.
It is thus natural to ask whether the best of both representations can be achieved \emph{simultaneously}:
\begin{question}\label{q:main}
    Is there 
    a compact representation
    that, given any terminal pair $x,y \in V$ and any vertex pair $u,v \in V$, 
    encodes 
    an $(x,y)$-mincut separating $u$ from $v$ whenever such exists?
\end{question}

This idea of ``marrying'' different mincut representations has a long history.
One \emph{exemplar} is 
Dinitz and Vainshtein's \emph{Connectivity Carcass}~\cite{DBLP:conf/stoc/DinitzV94, DBLP:conf/soda/DinitzV95, DBLP:journals/siamcomp/DinitzV00,BaswanaP25}: an ingenious and far-reaching hybrid structure interpolating the PQ representation with the \emph{Cactus Representation} of global mincuts~\cite{dinitz1976structure}. 
The carcass will play a crucial role in this paper.
The specific combination of GH with PQ has also been studied. 
Gusfield and Naor~\cite{DBLP:journals/algorithmica/GusfieldN93} showed that augmenting each edge $(s,t)$ of the GH tree with the PQ DAG of $s,t$ enables the effective recovery of \emph{all} PQ DAGs for all possible $\binom{n}{2}$ terminal pairs.
While this is a natural and elegant strategy, the resulting representation has $O(nm)$ space, so it is not quite compact.

In this paper, we answer Question~\ref{q:main} in the affirmative by providing an \emph{optimally compact} $O(n)$ space representation, presented in the following~\Cref{thm:main}.

\begin{theorem}\label{thm:main}
    Let $G = (V,E,w)$ be an undirected graph with positive integer edge-weights.
    There is an $O(n)$ space data structure that support the following two-stage query process:
    \begin{itemize}
        \item \textbf{(Update)} Given any terminal pair $x,y \in V$, it can report all $x,y$-units, and further determine a topological ordering of them, within $O(n)$ time.

        \item \textbf{(Separation Query)} After updating, given any query of  $u,v \in V$, it can determine whether $u$ and $v$ are separated by some $(x,y)$-mincut in $O(1)$ time.
        If so, such a cut $C$ can be reported in $O(|C|)$ time.
    \end{itemize}
\end{theorem}

\begin{remark}
    The second item follows immediately from the first, as the update here is not merely an $O(n)$ time ``black-box procedure'', but rather explicitly constructs the PQ representation for the terminal pair $x,y$.
\end{remark}

The compactness of our structure is optimal in a very strong sense.
Suppose that instead of encoding an $(x,y)$-mincut separating $u$ and $v$, we wished to encode such an $(x,y)$-mincut with $u$ on the side of $x$ and $v$ on the side of $y$, whenever such exists.
This directed version of separability by mincuts admits an $\Omega(n^2)$ worst-case lower bound \emph{even with the terminal pair $x,y$ is fixed}.\footnote{The lower bound was implicitly given in~\cite{DBLP:journals/talg/BaswanaBP23, DBLP:conf/icalp/Bhanja25}. Consider an arbitrary bipartite graph $H = (L,R,E_H)$ with $n$ vertices on each side. Add two new vertices $x$ and $y$. For each $v$ on the left (resp., $u$ on the right), connect $x$ to $v$ (resp., $u$ to $y$) by an edge of weight $\deg_H (v)$ (resp., $\deg_H (u)$), and call the resulting graph $G$. It can be easily verified that $G$ has an $(x,y)$-mincut keeping $u \in R$ with $x$ and $v \in L$ with $y$ iff $H$ does not have the edge $(u,v)$. But encoding an arbitrary $n$ by $n$ bipartite graph requires $\Omega(n^2)$ space.}

Technically, our structure substantially generalizes the framework laid by the first and last authors with Kenneth-Mordoch in~\cite{BhanjaPP26}, confirming their conjecture that such generalization is achievable.
The structure of~\cite{BhanjaPP26} uses $O(n)$ space and takes $O(n)$ time to report the \emph{nearest} $(x,y)$-mincut $\Near(x,y)$ for any given query of $x,y$.
$\Near(x,y)$ is usually defined as the unique inclusion-minimal $(x,y)$-mincut, but an equivalent definition is simply \emph{the $x,y$-unit containing $x$}.
Thus, our structure gives a strict generalization, as it uses the same space and time bounds to recover \emph{all $x,y$-units}.
From a sensitivity perspective, encoding the nearest mincuts corresponds to \emph{edge insertions}: inserting a new edge $e$ causes the $x,y$-mincut value to \emph{increase} if and only if $e$ goes from $\Near(x,y)$ to $\Near(y,x)$.
As discussed before reporting, determining the $x,y$-units naturally corresponds to the harder sensitivity setting of edge deletions.

\paragraph{Mincut Sensitivity Oracles.}
The study of mincut sensitivity oracles in various settings has recently received considerable attention~\cite{BaswanaGK22,
DBLP:conf/soda/BaswanaP22, DBLP:journals/talg/BaswanaBP23,
DBLP:conf/icalp/BaswanaB24, DBLP:conf/isaac/Bhanja24,
DBLP:conf/esa/BaswanaBR25, DBLP:conf/icalp/Bhanja25,
DBLP:conf/innovations/AhiCPPS26,
BhanjaPP26}.
The discussion below concerns the setting of undirected multigraphs with one edge insertion/failure.
Previous all-pairs oracles for the failure setting were given by Baswana and Pandey~\cite{DBLP:conf/soda/BaswanaP22}.
They showed an $O(n^2)$ space that given any $x,y \in V$ and $e \in E$ can determine in just $O(1)$ time if the failure of (one copy of) $e$ decreases the $(x,y)$-mincut value, and can additionally report a new $(x,y)$-mincut in $O(n)$ time.
They also showed a more compact $O(m)$ space structure for the same task, albeit with larger query times.
Their paper poses the following question: 
\begin{quote}
\itshape
``Does there exist a data structure for all-pairs mincuts sensitivity that occupies $o(n^2)$ space and achieves $O(m)$ query time?''  
\end{quote}
\Cref{thm:main} gives a positive answer, with optimal $O(n)$ space and worst-case optimal $O(n)$ query time.

Focusing on optimally compact $O(n)$ space oracles, \Cref{thm:main} gives some satisfying closure.
The edge insertion setting has followed the trajectory 
\textbf{\emph{Fixed-Pair $\to$ Single-Source $\to$ All-Pairs}},
with~\cite{BaswanaGK22} providing the first arrow, and the aforementioned very recent work of~\cite{BhanjaKP26} the second one.
Our work provides the second arrow of a similar trajectory for the edge failure setting, with the first one given very recently by the authors in~\cite{BhanjaPP26}.
There are also ``vertical'' correspondences. In the single-source setting, the edge failure structure of~\cite{BhanjaPP26} builds on and substantially extends the corresponding edge insertion structure of~\cite{BaswanaGK22}.
Analogously, in the all-pairs setting, our current work gives the edge failure counterpart to the edge insertion oracle of~\cite{BhanjaKP26} by building on and substantially extending it.

\section{Preliminaries}\label{sec:prelim}

Let $G = (V,E,w \colon E \to \mathbb{Z}_+)$ be a weighted undirected connected graph, and denote $n := |V|$, $m := |E|$.
For $A \subseteq V$, its \emph{capacity} is $c(A) = \sum w(e)$ with the sum  over all $e \in E$ going between $A$ and $\overline{A} := V \setminus A$.
Let $S \subseteq V$ be a given \emph{terminal set}.
If $A \cap S = \emptyset$ and $\overline{A} \cap S \neq \emptyset$, then $A$ is an \emph{$S$-cut}.
An $S$-cut of the smallest capacity $\lambda_S$ is an \emph{$S$-mincut}, or \emph{Steiner mincut for $S$}.
When $S = \{x,y\}$, we write $x,y$-(min)cut, $\lambda_{x,y}$ and so on (where we define $\lambda_{x,x} = \infty$).
An $x,y$-(min)cut $A$ s.t.\ $x \in A$, $y \in \overline{A}$ is called an $(x,y)$-(min)cut.

\begin{description}
    \item[$S$-units.] 
    $u,v \in V$ are $S$-equivalent if no $S$-mincut separates them 
    (i.e., no $S$-minuct $A$ has $u \in A$, $v \in \overline{A}$).
    The equivalence classes are called \emph{$S$-units}.
    The \emph{terminal $S$-units} are those intersecting $S$, and their number is $d(S)$
    (for $|S| \leq 1$, $d(S) := 0$).
    Let $\phi_S(\cdot)$ map vertices to identifiers of their $S$-units.

    \item[Valid partitions.]
    A partition $S_1,S_2$ of $S$ is \emph{valid} if it is \emph{induced} by some $S$-mincut $A$, meaning $A \cap S = S_i$ for some $i \in \{1,2\}$.
    %
    Vertex $v$ is \emph{distinguished} by $S_1,S_2$ if there are two different $S$-mincuts $A,B$ inducing $S_1,S_2$, where $A$ keeps $v$ with $S_1$, but $B$ keeps $v$ with $S_2$.
    A vertex is called \emph{stretched} (w.r.t.\ $S$) if it is distinguished by at least one valid partition.
\end{description}

\paragraph{The Connectivity Carcass.}\label{sec:carcass-prelim}

We succinctly describe Dinitz and Vainshtein's \emph{Connectivity Carcass} \cite{DBLP:conf/stoc/DinitzV94, DBLP:conf/soda/DinitzV95, DBLP:journals/siamcomp/DinitzV00}, closely following~\cite{BhanjaPP26}.
The carcass is a highly complex, and we recommend its detailed presentation in~\cite{BaswanaP25}.
See~\Cref{fig:carcass} for an illustration.
Let $S \subseteq V$ be a terminal set.

\definecolor{outline}{RGB}{0,45,62}
\definecolor{guideblue}{RGB}{0,105,155}
\definecolor{guidered}{RGB}{220,65,10}
\definecolor{mapblue}{RGB}{170,210,232}
\definecolor{mapgreen}{RGB}{162,216,145}
\definecolor{maporange}{RGB}{245,180,128}
\definecolor{mappurple}{RGB}{212,106,207}

\begin{figure}[t]
\centering
\resizebox{\textwidth}{!}{%
\begin{tikzpicture}[
  x=0.018cm,y=-0.018cm,
  every node/.style={font=\normalsize},
  vertex/.style={circle,draw=outline,line width=.65pt,
    minimum size=4.32mm,inner sep=0pt,outer sep=0pt},
  terminal/.style={vertex,fill=black,text=white},
  nonterminal/.style={vertex,fill=gray!20,text=black},
  junction/.style={vertex,fill=white,text=black},
  edge/.style={black,line width=.7pt},
  blueguide/.style={guideblue,line width=.65pt,dash pattern=on 4pt off 3pt},
  redguide/.style={guidered,line width=.65pt,dash pattern=on 4pt off 3pt},
  separator/.style={orange!85!yellow,line width=.65pt,dash pattern=on 4pt off 3pt},
  projection/.style={line width=1.1pt},
  weight/.style={font=\small,inner sep=1pt,fill=white},
  title/.style={font=\rmfamily\mdseries\normalsize}
]
\path[use as bounding box] (15,-8) rectangle (1131,394);
\node[title] at (219,7) {Graph};

\foreach \name/\xx/\yy in {
 f/40/53,e/86/94,d/121/146,g/179/120,z/181/69,
 x/40/153,y/39/231,a/39/318,b/85/290,c/122/225,
 j/180/248,w/203/197,h/235/154,i/235/220,
 k/299/103,l/368/48,m/433/39,u/409/130,v/408/238,
 n/300/269,o/372/316,p/435/339,q/393/376
}{\coordinate (G-\name) at (\xx,\yy);}

\draw[blueguide] (20,105)
  .. controls (55,130) and (96,166) .. (121,166)
  .. controls (145,166) and (165,153) .. (192,127.893)
  .. controls (209,108) and (214,76) .. (213,52);
\draw[blueguide] (20,190)
  .. controls (55,200) and (96,191) .. (121,186)
  .. controls (153,180) and (182,157) .. (207,137)
  .. controls (227,117) and (235,88) .. (235,66);
\draw[blueguide] (20,280)
  .. controls (55,272) and (96,214) .. (121,206)
  .. controls (160,194) and (198,166) .. (222,146.107)
  .. controls (242,126) and (254,101) .. (255,80);
\draw[separator] (272,67)--(272,310);
\draw[redguide] (291,58)--(433,197);
\draw[redguide] (324,41)
  .. controls (343,79) and (363,98) .. (434,96);

\foreach \aa/\bb in {
 f/e,e/d,d/g,g/z,g/h,d/c,c/b,b/a,
 f/x,x/y,y/a,c/j,j/w,w/h,j/i,h/i,h/k,i/n,
 k/l,l/m,l/u,u/v,v/o,n/o,n/u,k/v,o/p,o/q,p/q%
}{\draw[edge] (G-\aa)--(G-\bb);}

\foreach \lab/\xx/\yy in {
 5/70/65,5/116/109,3/143/120,3/204/119,
 2/131/202,5/113/266,5/69/324,
 3/142/245,2/213/244,7/243/187,3/259/119,
 3/260/255,4/326/69,6/395/32,2/405/85,
 4/323/300,2/402/287,3/407/323,3/376/351,7/423/368
}{\node[weight] at (\xx,\yy) {$\lab$};}

\foreach \name in {a,b,c,d,e,f,g,h,i,j,k,l,m,n,o,p,q}
  \node[terminal] at (G-\name) {$\name$};
\foreach \name in {x,y,z,w,u,v}
  \node[nonterminal] at (G-\name) {$\name$};

\begin{scope}[xshift=1.2cm]
\node[title] at (779,7) {Connectivity Carcass: Skeleton and Projections};
\foreach \name/\xx/\yy in {
 f/498/34,e/545/56,d/591/95,
 A/627/150,B/627/226,C/684/120,D/684/249,E/755/179,F/871/179,
 a/497/333,b/544/320,c/591/292,j/684/302,g/683/70,
 k/901/118,l/950/63,m/1023/46,n/901/244,o/972/291,
 hi/810/180,pq/1022/350,
 x/583/175,y/584/216,z/683/36,w/751/242,u/955/112,v/955/235
}{\coordinate (S-\name) at (\xx,\yy);}

\draw[projection,guideblue] (498,62)
  .. controls (537,85) and (578,112) .. (599,140)
  .. controls (610,152) and (611,169) .. (610,191)
  .. controls (609,214) and (610,237) .. (598,250)
  .. controls (571,275) and (535,292) .. (505,311);
\draw[blueguide] (521,105)
  .. controls (556,130) and (602,166) .. (627,166)
  .. controls (651,166) and (673,157) .. (700.482,133.696)
  .. controls (719,115) and (731,85) .. (730,62);
\draw[blueguide] (521,190)
  .. controls (556,200) and (602,191) .. (627,186)
  .. controls (659,180) and (694,168) .. (719.5,149.5)
  .. controls (740,130) and (754,102) .. (754,80);
\draw[blueguide] (521,280)
  .. controls (556,272) and (602,214) .. (627,206)
  .. controls (666,194) and (714,181) .. (738.518,165.304)
  .. controls (762,145) and (781,119) .. (782,98);
\draw[separator] (848,81)--(848,324);
\draw[redguide] (872,57)--(994,199);
\draw[redguide] (904,45)
  .. controls (926,82) and (950,102) .. (1014,99);

\draw[projection,green!45!black] (703,308)
  .. controls (703,290) and (702,277) .. (705,266)
  .. controls (709,253) and (743,211) .. (758,200)
  .. controls (768,193) and (777,195) .. (786,195);
\draw[projection,guidered] (962,78)
  .. controls (948,92) and (927,116) .. (914,138)
  .. controls (901,158) and (890,169) .. (890,180)
  .. controls (890,195) and (901,218) .. (912,229);
\draw[projection,violet] (926,138)
  .. controls (913,158) and (902,169) .. (902,180)
  .. controls (902,195) and (913,218) .. (924,229)
  .. controls (935,247) and (967,263) .. (990,273);

\foreach \aa/\bb in {
 f/e,e/d,d/A,A/B,A/C,C/g,C/E,
 B/c,c/b,b/a,B/D,D/j,D/E,E/hi,hi/F,
 F/k,k/l,l/m,F/n,n/o,o/pq%
}{\draw[edge] (S-\aa)--(S-\bb);}

\foreach \name in {A,B,C,D,E,F}
  \node[junction] at (S-\name) {};
\foreach \name in {a,b,c,d,e,f,g,j,k,l,m,n,o}
  \node[terminal] at (S-\name) {$\name$};

\node[terminal,minimum size=7.5mm]
  at (S-hi) {$h,i$};
\node[terminal,minimum size=7.5mm]
  at (S-pq) {$p,q$};

\node[vertex,fill=mapblue] at (S-x) {$x$};
\node[vertex,fill=mapblue] at (S-y) {$y$};
\node[vertex,fill=red!70] at (S-z) {$z$};
\node[vertex,fill=mapgreen] at (S-w) {$w$};
\node[vertex,fill=maporange] at (S-u) {$u$};
\node[vertex,fill=mappurple] at (S-v) {$v$};
\draw[red,line width=.7pt] (S-g) circle[radius=3.7mm];
\end{scope}
\end{tikzpicture}%
}
\caption{The connectivity carcass. Graph edges without weight-labels have weight $1$; terminals are black and non-terminals are gray; dashed lines are $S$-mincuts, color-coded by their representing canonical cuts.}
\label{fig:carcass}
\end{figure}
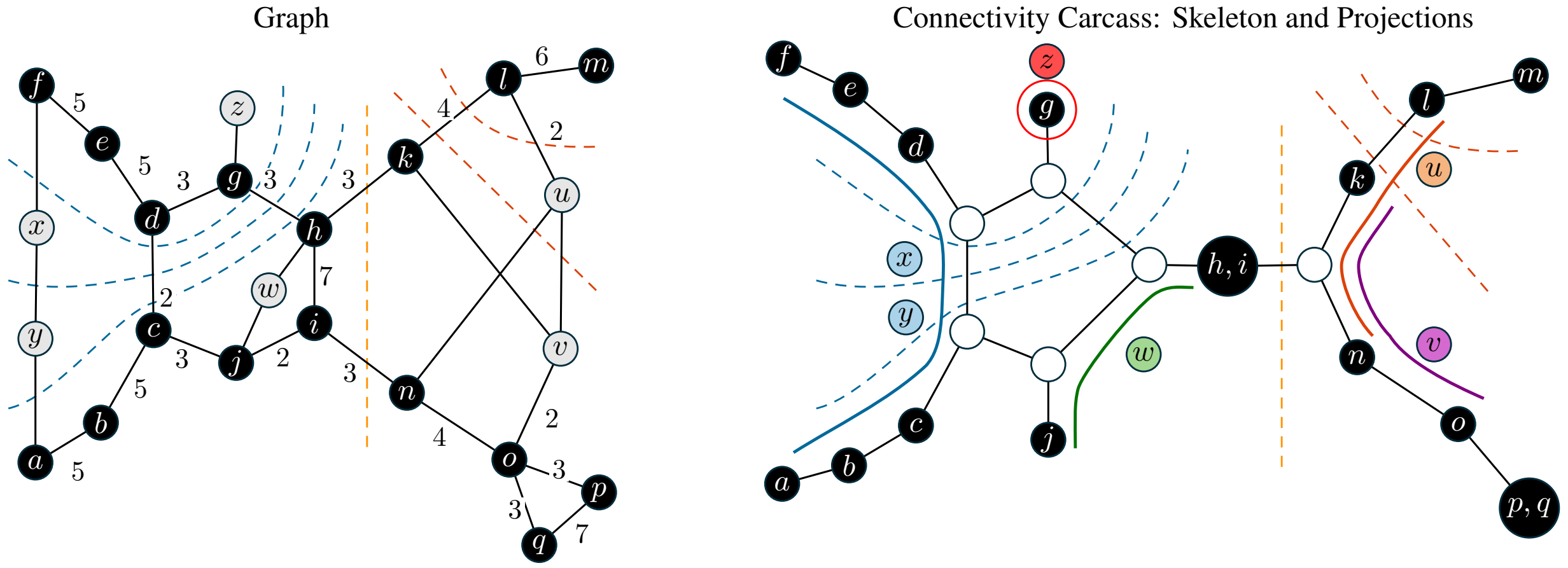

\begin{definition}[Cactus]
    In this paper, a \emph{cactus} $\cH$ is an undirected connected graph where no two cycles share a node.
    Edges/nodes in cycles are called \emph{cycle-edges/nodes}; the rest are~\emph{tree-edges/nodes}.
    A \emph{canonical cut} $C$ is either one tree-edge or a pair of non-adjacent cycle-edges from the same cycle;
    removing $C$ from $\cH$ leaves two connected components called its \emph{sides}.
    A proper path is a simple path with at most one edge from each cycle; it can be specified by its endpoints as the unique proper path connecting them.
\end{definition}

\begin{theorem}[\protect{Skeleton}]
\label{thm:skeleton}
    There exists a cactus $\cH_S$ called the \emph{skeleton} of $S$, with the following properties.
    \begin{itemize}
        \item $\cH_S$ has $d(S)$ 
        \emph{nonempty} (or \emph{terminal}) tree-nodes, representing the sets $\{S \cap U \mid \text{$U$ terminal $S$-unit}\}$.
        
        \item $\cH_S$ may further have $O(d(S))$ \emph{empty} nodes, each of degree at least $3$, and exactly $3$ for cycle-nodes.

        \item Canonical cuts $C$ of $\cH_S$ bijectively correspond to valid partitions $S_1,S_2$ of $S$: 
        $S_1$ are the terminals from one side of $C$, and $S_2$ are those from the other side.
        An $S$-mincut is said to be represented by $C$ if it induces $S_1,S_2$.
        A vertex is said to be distinguished by $C$ if it is distinguished by $S_1,S_2$.
        
    \end{itemize}
\end{theorem}


\begin{theorem}[Projections]
\label{thm:proj-vertex}
    Each vertex $v \in V $ has a \emph{projection} $\pi_S (v)$ in $\cH_S$ such that:
    \begin{itemize}
        \item If $v$ is stretched, $\pi_S (v)$ is a proper path that starts and ends by tree-edges.
        A canonical cut $C$ distinguishes $v$ iff $\pi_S (v)$ has endpoints in both sides of $C$ (i.e., iff $C$ has some edge of $\pi_S (v)$).
    
        \item If $v$ is non-stretched, $\pi_S (v)$ is a tree-node. 
        If $v \in S$, $\pi_S (v)$ is the terminal node containing $v$.

        \item The following holds:
        Let $C$ be a canonical cut corresponding to valid partition $S_1,S_2$.
        If $C$ keeps $\pi_S (v)$ entirely on the side of $S_1$,
        then every $S$-mincut inducing $S_1,S_2$ keeps $v$ on the side of $S_1$.
    \end{itemize}
\end{theorem}

\begin{theorem} [``Distinctness'' property]
\label{lem: Distinctness}
     Let $u,v\in V$, and let $C$ be a canonical cut of $\cH_S$ that distinguishes $u$ or $v$ (or both).
     If there is no $S$-mincut represented by $C$ that separates $u$ from $v$, then $\phi_S (u) = \phi_S (v)$.
\end{theorem}

\paragraph{Submodularity.}
It is well-known that $c(\cdot)$ is \emph{sub/posimodular}:
$c(A) + c(B) \geq c(A \cap B) + c(A \cup B)$ and $c(A) + c(B) \geq c(A \setminus B) + c(B \setminus A)$
for every $A,B \subseteq V$.
Another inequality of similar flavor~\cite[Problem 6.48(iii)]{lovasz1979combinatorial} is that, for every $A,B,C \subseteq V$, 
\begin{equation}\label{eq:four-components}
    c(A) + c(B) + c(C) \geq c(A \setminus (B \cup C)) + c(B \setminus (A \cup C)) + c(C \setminus (A \cup B)) + c(A \cap B \cap C).
\end{equation}
The next lemmas summarize implications of these facts on mincuts, whose variants are widely observed (specifically,~\Cref{lem:sub-posi-general} slightly extends~\cite[Lemmas 2.1 and 4.2]{BhanjaKP26}).
The proofs appear in Appendix~\ref{sec:missing-proofs}.

\begin{lemma}\label{lem:sub-posi-general}
    Let $S,T \subseteq V$, and suppose $A$ is an $S$-mincut and $B$ is a $T$-mincut.
    \begin{itemize}
        \item (``Submodularity'') If one of $A \cap B$ and $A \cup B$ is an $S$-cut and the other is a $T$-cut, then the former is an $S$-mincut, and the latter is a $T$-mincut.
        
        \item (``Posimodularity'') If one of $A \setminus B$ and $B \setminus A$ is an $S$-cut and the other is a $T$-cut, then the former is an $S$-mincut, and the latter is a $T$-mincut.
        \item (``Hybrid'') If, among the four cuts $A \cap B, A \cup B, A \setminus B, B \setminus A$, two are $S$-cuts and the other two are $T$-cuts, then the former two are $S$-mincuts and the latter two are $T$-mincuts.
    \end{itemize}
    Furthermore, one of the above cases must hold.
\end{lemma}

\def\SUBANDPOSI{
\begin{proof}[Proof of~\Cref{lem:sub-posi-general}]
    We first prove the correctness of each case:
    
    \begin{itemize}
        \item 
        (Submodularity) The $S$-cut must have capacity at least $c(A)$, and the $T$-cut must have capacity at least $c(B)$.
        Summing both these inequalities, we obtain that $c(A) + c(B) \leq c(A \cap B) + c(A \cup B)$.
        But, by submodularity, the latter must hold with equality, and hence also the two original inequalities.
    
        \item
        (Posimodularity) Symmetric to submodularity.
    
        \item
        (Hybrid)
        In a similar manner to the sub/posimodularity cases, now we have four inequalities, and summing them yields $2c(A) + 2c(B) \leq c(A \cap B) + c(A \cup B) + c(A \setminus B) + c(B \setminus A)$.
        But now, by applying submodularity and posimodularity together, the latter must hold with equality, and hence also the four original inequalities.
    \end{itemize}
    
    We now prove that one of the cases must hold.
    Choose $I = \{s_1,s_2,t_1,t_2\}$ s.t.\
    $s_1 \in S \cap A$, $s_2 \in S \cap \overline{A}$, $t_1 \in T \cap B$, $t_2 \in T \cap \overline{B}$.
    
        If each of $A \cap B, \overline{A \cup B}, A \setminus B, B\setminus A$ intersects $I$: Then, as these four sets are disjoint, each contains \emph{exactly} one vertex of $I$.
        Those with $s_1$ and with $s_2$ give two different $S$-cuts, and those with $t_1$ and $t_2$ give the two $T$-cuts.

        Otherwise, one of the four sets has empty intersections with $I$.
        If $(A \cap B) \cap I = \emptyset$, then $s_1 \in A \setminus B$ and $t_1 \in B \setminus A$, so the former is an $S$-cut and the latter is a $T$-cut, hence we get the posimodularity case.
        If $(\overline{A \cup B}) \cap I = \emptyset$, then $s_2 \in B \setminus A$ and $t_2 \in A \setminus B$, and again we get the posimodularity case.
        If $(A \setminus B) \cap I = \emptyset$, then $s_1 \in A \cap B$ and $t_2 \in \overline{A \cup B}$, so we get the submodularity case.
        Finally, if $(B \setminus A) \cap I = \emptyset$, then $t_1 \in A \cap B$ and $s_2 \in \overline{A \cup B}$, so again we get the submodularity case.
\end{proof}
}

\begin{lemma}[Four Components Lemma]\label{lem: simple-four-comps}
    Let $S,T,W \subseteq V$.
    Suppose $A$ is an $S$-mincut, $B$ is a $T$-mincut, and $C$ is a $W$-mincut.
    Consider the four sets $A \setminus (B \cup C)$, $B \setminus (A \cup C)$, $C \setminus (A \cup B)$ and $A \cap B \cap C$.
    If among them, three distinct sets form an $S$-cut, a $T$-cut and a $W$-cut, then the fourth set must be empty. 
\end{lemma}
\def\FOURCOMPS{
\begin{proof}[Proof of~\Cref{lem: simple-four-comps}.]
    The $S$-cut must have capacity at least $\lambda_S = c(A)$.
    Similarly, the $T$-cut and $W$-cut has capacities at least $c(B)$ and $c(C)$.
    Subtracting $c(A) + c(B) + c(C)$ from both sides of~\Cref{eq:four-components} and applying these observations, we see that the fourth set has non-positive capacity.
    In a connected graph, there are only two such sets: $V$ and $\emptyset$.
    But $V$ is ruled out, as the other three sets are disjoint from the fourth and non-empty.
\end{proof}
} 

\section{Technical Overview}\label{sec:overview}

This section presents the central ideas behind the data structure.
Our exposition aims to give not only high-level concepts but also sufficient technical detail to grasp them.
It assumes only the preliminary tools from~\Cref{sec:prelim} as black boxes.
We include several (short) proofs, which we regard as most insightful for demonstrating the techniques; nevertheless, they may be skipped if one wishes to focus on the big picture.

Our approach builds on the framework recently introduced by the first and last authors with Kenneth-Mordoch~\cite{BhanjaKP26}, who showed an $O(n)$ space structure for all-pairs \emph{nearest} mincuts.
Our structure strictly generalizes it in a highly white-box manner.
Almost every aspect requires non-trivial generalizations, and several new key insights and ingredients are introduced.
We do not rely on  familiarity with~\cite{BhanjaKP26}, and explain everything from scratch.
A detailed comparison with~\cite{BhanjaKP26} is provided in Appendix~\ref{sec:comparison-BKP}.

We focus the more detailed discussion on showing how to design a data structure with just $O(n)$ space that, given a query of $x,y\in V$, recovers the partition of $V$ into $x,y$-units.
This constitutes the major part of the overview, in~\Cref{sec:skeleton-hierarchy,sec:strategy,sec:defining-alpha-comps,sec:projections-overview}.
We then outline the broad strokes of how this structure is extended further to also provide a topological ordering of the $x,y$-units, in \Cref{sec:ordering-overview}.

\subsection{The skeleton hierarchy and LCAs} \label{sec:skeleton-hierarchy}
Our starting point is the \emph{skeleton hierarchy tree} $\cT$.
    Its nodes are subsets of $V$.
    The root is $V$.
    There are $n$ leaves, one for (the singleton of) each vertex.
    Each non-leaf $\mu$ has $d(\mu)$ children $\{\mu \cap U \mid \text{$U$ terminal $\mu$-unit}\}$, and  
    stores its skeleton $\cH_\mu$ (so $\mu$'s children are $\cH_\mu$'s nonempty nodes).
Thus, $\cT$ has at most $2n-1$ nodes and $\sum_{\mu \in \cT} d(\mu) \leq 2n-2$ edges.
As $|\cH_\mu| = O(d(\mu))$, the skeleton hierarchy only requires $O(n)$ space.

For a given query $x,y \in V$, the hierarchy alone already provides non-trivial information on the $x,y$-units.
Consider the lowest common ancestor $\mu = \LCA_\cT (x,y)$,
and its children $\alpha,\beta$ with $x \in \alpha$ and $y \in \beta$.

\begin{definition}[$\alpha\beta$-components]\label{def:alphabeta-comps}
    Two nodes in $\cH_\mu$ are $\alpha\beta$-equivalent if they are on the same side of every canonical cut separating $\alpha$ from $\beta$.
    The \emph{$\alpha\beta$-components} are the equivalence classes with a nonempty node.
\end{definition}

Without causing ambiguity, we  identify an $\alpha\beta$-component with the set of terminals within it, so the $\alpha\beta$-components partition $\mu$.
Their structure is illustrated in~\Cref{fig:inside-lca}.
Let $P_{\alpha\beta}$ be the \emph{cactus-path} of tree-edges and cycles between $\alpha$ and $\beta$ in $\cH_\mu$.
Each $\alpha\beta$-component $T$ intersects $P_{\alpha\beta}$ at exactly one node, and the only nodes of $P_{\alpha\beta}$ not included in any $\alpha\beta$-component are the entry and exit point of cycles.%
\footnote{To be accurate, there is a corner case where a cycle $\cO$ on $P_{\alpha\beta}$ has its entry and exit nodes connected directly by one cycle-edge. In this case, the stated~\Cref{def:alphabeta-comps} includes the entry and exit nodes in $\alpha\beta$-components that intersect $P_{\alpha\beta}$ in their other neighbors of entry and exit nodes from the cycle. We fix this case $\alpha\beta$-components by explicitly excluding the empty entry and exit nodes of the cycle from participating in $\alpha\beta$-components.}

\input{inside-lca}

Since $\mu$ is the LCA of $x,y$ in $\cT$, we see that $x,y$-mincuts are $\mu$-mincuts. 
Further, they are precisely those $\mu$-mincuts represented by canonical cuts of $\cH_\mu$ which separate $\alpha$ from $\beta$.
We therefore have: 
\begin{observation}\label{lem:atomicity-LCA}
    Every $\alpha\beta$-component is contained in one $x,y$-unit without any other $\alpha\beta$-components.
\end{observation}

\paragraph{Projections in the LCA.}
However, our goal is to recover the \emph{entire} partition of $V$ into $x,y$-units, rather than just the LCA.
This will require utilizing the carcass beyond just skeletons.
To warm up, we show a simple $O(n^2)$ space solution:
For each $v \in V$ and non-leaf node $\mu \in \cT$, store its projection %
$\pi_\mu (v)$ into $\cH_\mu$, augmented with the $\mu$-unit identifier $\phi_\mu (v)$.
Recall that a projection is specified by its two endpoints, so storing it takes $O(1)$ space.
Using \Cref{thm:proj-vertex}, we find a \emph{dichotomy} relating projections to $\alpha\beta$-components:

\begin{lemma}[Dichotomy in LCA]\label{lem:LCA-dichotomy}
    Let $\mu = \LCA_{\cT} (x,y)$, with children $\alpha$ and $\beta$ containing $x$ and $y$ respectively.
    Then, for 
    every vertex $v$, the following holds:
    \begin{itemize}
        \item {\bf (``Merged'')} If $\pi_\mu (v)$ is contained within one $\alpha\beta$-component $T$, then $v$ is with $T$ in the same $x,y$-unit.
        \item {\bf (``Split'')} Else, the $x,y$-unit of $v$ does not intersect $\mu$.
    \end{itemize}
\end{lemma}
\begin{proof}
    Let $T$ be some $\alpha\beta$-component.
    If $\pi_\mu (v)$ is contained in $T$, then each canonical cut separating $\alpha$ from $\beta$ keeps $T$ and $\pi_\mu (v)$ on the same side.
    As $x,y$-mincuts are represented by such canonical cuts, they cannot separate $T$ from $v$ by~\Cref{thm:proj-vertex}.
    Conversely, if $\pi_\mu (v)$ is not contained in $T$, then by definition of $\alpha\beta$-components, some canonical cut $C$ which separates $\alpha$ from $\beta$ keeps $T$ on one side and an endpoint of $\pi_\mu (v)$ on the other.
    So, by~\Cref{thm:proj-vertex}, $C$ represents some $x,y$-mincut $X$ which separates $T$ from $v$.
\end{proof}

Applying the dichotomy to every vertex, we recover all of the $x,y$-units that contain $\alpha\beta$-components, i.e., that intersect $\mu$.
To recover the rest, containing the ``split'' vertices, we utilize distinctness (\Cref{lem: Distinctness}):

\begin{lemma}[Distinctness in LCA] \label{lem:distinctness of xy units}
    Let $\mu = \LCA_{\cT} (x,y)$, with children $\alpha$ and $\beta$ containing $x$ and $y$ respectively.
    Let $u, v \in V$ be vertices whose $x,y$-units do not intersect $\mu$.
    Then 
    $\phi_{x,y}(u) = \phi_{x,y} (v)$ iff $\phi_\mu (u) = \phi_\mu (v)$.
    That is, $u$ and $v$ are in the same $x,y$-unit iff they are in the same $\mu$-unit.
\end{lemma}

\begin{proof}
    (`If')
    Follows since $x$ and $y$ belong to different $\mu$-units, so every $x,y$-mincut is also a $\mu$-mincut.

    (`Only if')
    Let $C$ be a canonical cut separating $\alpha$ from $\beta$ that has an edge of $\pi_\mu (u)$.
    Such must exists: otherwise, $\pi_\mu (u)$ is contained in one $\alpha\beta$-component, so $u$'s $x,y$-unit would intersect $\mu$ by~\Cref{lem:LCA-dichotomy}.
    By~\Cref{thm:proj-vertex}, $u$ is distinguished by $C$.
    Every $\mu$-mincut represented by $C$ is an $x,y$-mincut, so it does not separate $u$ from $v$ by $\phi_{x,y} (u) = \phi_{x,y} (v)$.
    Thus, \Cref{lem: Distinctness} yields that $\phi_\mu (u) = \phi_\mu (v)$.
\end{proof}

\subsection{The strategy: building up with representative projections}\label{sec:strategy}

\begin{wrapstuff}[type=figure,r,width=0.24\textwidth,top=0]
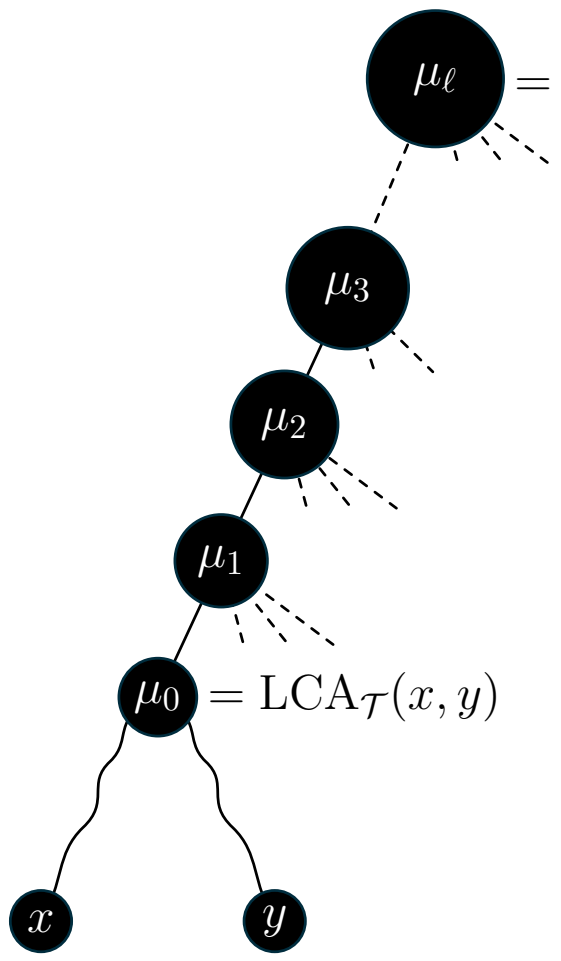
\caption{The query path.}
\label{fig:hierarchy-chain}
\centering
\begingroup
\definecolor{hierarchyoutline}{RGB}{0,45,62}
\resizebox{\linewidth}{!}{%
\begin{tikzpicture}[
  vertex/.style={circle,draw=hierarchyoutline,line width=.8pt,
    fill=black,text=white,minimum size=7.4mm,inner sep=1pt,
    outer sep=0pt,font=\Large},
  edge/.style={black,line width=.8pt,line cap=round},
  omitted/.style={edge,dash pattern=on 2.5pt off 3pt}
]
\coordinate (m0) at (1.2,2.3);
\coordinate (m1) at (1.85,3.7);
\coordinate (m2) at (2.5,5.1);
\coordinate (m3) at (3.15,6.5);
\coordinate (ml) at (4.05,8.65);
\coordinate (x) at (0,0);
\coordinate (y) at (2.4,0);

\draw[edge] (m0)--(m1)--(m2)--(m3);
\draw[edge] (m3)--(3.36,7.0);
\draw[omitted] (3.36,7.0)--(3.84,8.15);
\draw[edge] (3.84,8.15)--(ml);

\foreach \parent in {m1,m2,ml}{%
  \begin{scope}[shift={(\parent)}]
    \foreach \dx in {.25,.75,1.25}{%
      \draw[omitted] (0,0)--({.92*\dx},-.86);
    }
  \end{scope}
}
\begin{scope}[shift={(m3)}]
  \foreach \dx in {.3,.95}{%
    \draw[omitted] (0,0)--({.92*\dx},-.86);
  }
\end{scope}

\draw[edge] (m0)
 .. controls (.7,2.08) and (.95,1.87) .. (.71,1.64)
 .. controls (.46,1.41) and (.72,1.23) .. (.43,.96)
 .. controls (.15,.69) and (.24,.40) .. (x);
\draw[edge] (m0)
 .. controls (1.7,2.08) and (1.45,1.87) .. (1.69,1.64)
 .. controls (1.94,1.41) and (1.68,1.23) .. (1.97,.96)
 .. controls (2.25,.69) and (2.16,.40) .. (y);

\foreach \p/\s/\diameter in {m0/0/8,m1/1/9.5,m2/2/11,m3/3/12.5,ml/\ell/14}
  \node[vertex,minimum size=\diameter mm] at (\p) {$\mu_{\s}$};
\node[vertex,minimum size=6.3mm] at (x) {$x$};
\node[vertex,minimum size=6.3mm] at (y) {$y$};
\node[anchor=west,font=\Large,inner sep=1pt] at (1.67,2.3)
  {$=\mathrm{LCA}_{\mathcal T}(x,y)$};
\node[anchor=west,font=\Large,inner sep=1pt] at (4.82,8.65)
  {$=V$};
\end{tikzpicture}%
}%
\endgroup
\end{wrapstuff}

We now discuss the high-level strategy for recovering the $x,y$-units using only $O(n)$ space.

Consider the path from $\LCA_\cT (x,y)$ to the root $V$ (\Cref{fig:hierarchy-chain}):
$
\LCA_{\cT} (x,y) = \mu_0 \subset \mu_1 \subset \mu_2 \subset \cdots \subset \mu_\ell = V. 
$
We call it the \emph{query path} for $x,y$.
The idea is to construct the $x,y$-units in a bottom-up fashion. Initially, we find the partition of the LCA $\mu_0$ into $x,y$-units.
Then, we work our way up: Step $i$ gets the partition of $\mu_{i-1}$, and extends it to  $\mu_i$.
The main challenge of course is implementing the steps.
This relies mainly on the notion of \emph{representative projections}.

For every child-parent pair $\alpha \subseteq \mu$ in $\cT$, we carefully define (at preprocessing) a small-on-average
subset $R(\alpha,\mu)$ of $\mu \setminus \alpha$, called the \emph{$\alpha$-representatives of $\mu$}.
For each representative $v \in R(\alpha,\mu)$, we store its projection $\pi_\alpha (v)$ into the child's skeleton $\cH_{\alpha}$. 
These projections are \emph{augmented} with additional $O(1)$ information.
For example, we augment $\pi_\alpha (v)$ with the $\alpha$-unit identifier $\phi_\alpha (v)$, and more sophisticated augmentations are discussed later.

The representative projections have three key properties: 

\par
\wrapstuffclear

\begin{itemize}
    \item \textbf{\emph{Informative:}} The augmented projections of $R(\mu_{i-1}, \mu_i)$ within $\cH_{\mu_{i-1}}$, along with the already-known partition of $\mu_{i-1}$ into $x,y$-units, allow us to also understand how $\mu_{i-1} \cup R(\mu_{i-1}, \mu_i)$ is partitioned.

    \item \textbf{\emph{Representing:}}
    It is enough to understand how $\mu_{i-1} \cup R(\mu_{i-1}, \mu_i)$ is
    partitioned into $x,y$-units in order to fully determine the partition of the parent $\mu_i$.

    \item \textbf{\emph{Succinct:}} 
    The total number of representatives, summed over all child-parent pairs in $\cT$, is only $O(n)$.
\end{itemize}
Thus, to implement a step and extend the known partition by $x,y$-units from $\mu_{i-1}$ to $\mu_i$, we first exploit the \emph{Informative} property to classify the representatives $R(\mu_{i-1},\mu_i)$ into their $x,y$-units, and then apply the \emph{Representing} property to further determine all of $\mu_i$.
The succinct property ensures that storing $O(1)$ information per representative uses $O(n)$ total space.

In~\Cref{sec:defining-alpha-comps}, we discuss how to \emph{define} the representatives so that they are indeed \emph{Representing} but also \emph{Succinct}.
Then, \Cref{sec:projections-overview} deals with the \emph{Informative} property: how can we augment (at preprocessing) and use (at query time) the stored projections in a given node on the query path, so that we can use them to classify the projected vertices to their $x,y$-units.


\subsection{Defining the representatives: the $\alpha$-components of the skeleton $\cH_\mu$}\label{sec:defining-alpha-comps}

\paragraph{Finding the atoms for $x,y$-units above the LCA.}
Let $\alpha \subseteq \mu$ be a child-parent pair in $\cT$. 
Recall that the purpose of defining $R(\alpha,\mu)$, the $\alpha$-representatives in $\mu$, is to support any (yet-unknown) query pair $x,y \in \alpha$.
We thus focus our attention at the tree-node $\alpha$ as a ``focal point'' within $\cH_{\mu}$.

Consider one tree-edge $g = (\alpha, \beta)$ adjacent to $\alpha$.
Let $H$ be the connected component of $\cH_\mu \setminus g$ containing $\beta$.
We call $H$ the \emph{sub-skeleton rooted by $g$ in direction $\beta$}, and without causing ambiguity we may identify $H$ with its terminals.
In many cases, we can prove that $H$ is an \emph{atom}: it lies completely within one $x,y$-unit.
To demonstrate why this is useful, let us imagine that sub-skeletons are \emph{always} atoms.
With this simplifying assumption, we can take $R(\alpha,\mu)$ as one terminal from each sub-skeleton.

\begin{itemize}
    \item \emph{Representing:} If we know how the $x,y$-units partition $\alpha \cup R(\alpha, \mu)$, we can extend it to all of $\mu$ by including each sub-skeleton $H$ in the $x,y$-unit containing its representative.

    \item \emph{Succinct:} We took $O(\deg_{\cH_\mu} (\alpha))$ representatives in $R(\alpha, \mu)$.
    So, summing over all child-parent pairs,
    \begin{equation}\label{eq:rep-calculation}
    \sum_{\mu \in \cT} \sum_{\alpha \text{ child of } \mu} |R(\alpha,\mu)|
    \leq O\big( 
    \sum_{\mu \in \cT} \sum_{\alpha \text{ child of } \mu} \deg_{\cH_\mu} (\alpha) \big) \leq O\big( \sum_{\mu \in \cT} |\cH_\mu|\big) \leq O(n).
    \end{equation}
\end{itemize}

We now present the \emph{actual} atoms, called \emph{$\alpha$-components}.
They play an analogous role to the $\alpha\beta$-components in the LCA from~\Cref{def:alphabeta-comps}.
However, there is an important distinction: different $\alpha$-components \emph{can} get merged into the same $x,y$-unit.
The formal definition of $\alpha$-components (\Cref{def:alpha-comps} below) is rather technical, essentially tailored to ensure atomicity: for $x,y \in \alpha$, no $x,y$-mincut can split an $\alpha$-component.
This is proved in~\Cref{lem: atomicity}, and we will use atomicity repeatedly (and implicitly) from now on.
See~\Cref{fig:alpha-components} for an illustration.

\definecolor{outline}{RGB}{0,45,62}
\definecolor{mapblue}{RGB}{170,210,232}
\definecolor{mapgreen}{RGB}{162,216,145}

\definecolor{projblue}{RGB}{0,114,178}
\definecolor{projorange}{RGB}{213,125,0}
\definecolor{projpurple}{RGB}{119,70,172}
\definecolor{projmagenta}{RGB}{194,51,116}

\newcommand{\graycomponent}{%
 \path[branch] (-23,125)--(23,125)--(23,72)
  .. controls (23,53) and (13,40) .. (0,40)
  .. controls (-13,40) and (-23,53) .. (-23,72)--cycle;
}
\newcommand{\pendant}[2]{%
 \begin{scope}[shift={(#1)},x=.022cm,y=.022cm,rotate=#2]
  \path[region] (-30,140)--(30,140)--(30,75)
   .. controls (30,51) and (14,30) .. (0,30)
   .. controls (-14,30) and (-30,51) .. (-30,75)--cycle;
  \draw[edge] (0,0)--(0,40);
  \graycomponent
 \end{scope}
}

\begin{wrapstuff}[type=figure,r,width=0.45\textwidth,top=0]
\centering
\resizebox{!}{.27\textheight}{%
\begin{tikzpicture}[
 x=.022cm,y=-.022cm,
 every node/.style={font=\Large},
 vertex/.style={circle,draw=outline,line width=.65pt,
   minimum size=4.32mm,inner sep=0pt,outer sep=0pt},
 terminal/.style={vertex,fill=black,text=white},
 junction/.style={vertex,fill=white},
 region/.style={draw=outline,fill=mapblue!40,line width=.65pt},
 branch/.style={draw=black,fill=gray!20,line width=.7pt},
 edge/.style={black,line width=.7pt},
 projection/.style={green!50!black,line width=2.4pt,line cap=round,line join=round},
 example projection/.style={line width=2.4pt,line cap=round,line join=round},
 type/.style={font=\rmfamily\mdseries\Large,inner sep=0pt}
]

\foreach \name/\xx/\yy in {
 alpha/500/325,L/500/465,T/500/185,
 A/340.508/265.249,B/330.259/227,
 C/368.508/216.751,D/378.756/255,
 R/378.756/395,P/621.244/255,
 E/621.244/395,F/662.053/376.830,G/691.945/410.028,
 H/669.609/448.715,I/625.913/439.427
}{\coordinate (\name) at (\xx,\yy);}

\begin{scope}[shift={(L)},x=.022cm,y=.022cm,rotate=180]
 \path[region] (-95,145)--(95,145)--(95,35)
  .. controls (95,3) and (47,-18) .. (0,-18)
  .. controls (-47,-18) and (-95,3) .. (-95,35)--cycle;
 \foreach \xx in {-58,0,58}{
  \draw[edge] (0,0)--(\xx,50);
  \begin{scope}[shift={(\xx,10)}]\graycomponent\end{scope}
 }
\end{scope}

\begin{scope}[shift={(T)},x=.022cm,y=.022cm,rotate=0]
 \path[region] (-74,145)--(74,145)--(74,45)
  .. controls (74,9) and (37,-14) .. (0,-14)
  .. controls (-37,-14) and (-74,9) .. (-74,45)--cycle;
 \foreach \xx in {-35,35}{
  \draw[edge] (0,0)--(\xx,50);
  \begin{scope}[shift={(\xx,10)}]\graycomponent\end{scope}
 }
\end{scope}

\begin{scope}[shift={(354.508,241)},x=.022cm,y=.022cm,rotate=60]
 \path[region] (-95,145)--(95,145)--(95,35)
  .. controls (95,-13) and (47,-55) .. (0,-55)
  .. controls (-47,-55) and (-95,-13) .. (-95,35)--cycle;
 \draw[edge] (-28,0)--(-58,50);
 \draw[edge] (0,28)--(0,50);
 \draw[edge] (28,0)--(58,50);
 \foreach \xx in {-58,0,58}{
  \begin{scope}[shift={(\xx,10)}]\graycomponent\end{scope}
 }
\end{scope}

\begin{scope}[shift={(R)},x=.022cm,y=.022cm,rotate=120]
 \path[region] (-36,140)--(36,140)--(36,50)
  .. controls (36,8) and (19,-18) .. (0,-18)
  .. controls (-19,-18) and (-36,8) .. (-36,50)--cycle;
 \draw[edge] (0,0)--(0,40);
 \graycomponent
\end{scope}

\begin{scope}[shift={(P)},x=.022cm,y=.022cm,rotate=-60]
\foreach \xx in {-35,35}{
 \begin{scope}[shift={(\xx,50)}]
  \path[region] (-30,140)--(30,140)--(30,75)
   .. controls (30,51) and (14,30) .. (0,30)
   .. controls (-14,30) and (-30,51) .. (-30,75)--cycle;
  \graycomponent
 \end{scope}
 \draw[edge] (0,0)--(\xx,90);
 \node[type,rotate=-60] at (\xx,207) {Type-$2$};
}
\end{scope}
\begin{scope}[shift={(654.153,414.000)},x=.022cm,y=.022cm,rotate=-120]
\foreach \name/\xx in {F/-105,G/-35,H/35,I/105}{
 \begin{scope}[shift={(\xx,12)}]
  \path[region] (-30,140)--(30,140)--(30,75)
   .. controls (30,51) and (14,30) .. (0,30)
   .. controls (-14,30) and (-30,51) .. (-30,75)--cycle;
  \graycomponent
 \end{scope}
 \coordinate (tip-\name) at (\xx,52);
 \draw[edge] (\name)--(tip-\name);
 \node[type,rotate=60] at (\xx,169) {Type-$3$};
}
\end{scope}

\foreach \aa/\bb in {
 alpha/L,alpha/T,alpha/D,alpha/R,alpha/P,alpha/E,
 A/B,B/C,C/D,D/A,E/F,F/G,G/H,H/I,I/E%
}{\draw[edge] (\aa)--(\bb);}

\draw[projection] (488,185)--(488,274)
 .. controls (488,285) and (478,298) .. (469,293.246)
 --(384.756,244.608);
\node[vertex,fill=mapgreen,minimum size=5.4mm] at (459,264) {$z$};

\begin{scope}[shift={(L)},x=.022cm,y=.022cm,rotate=180]
 \draw[example projection,projblue] (-66,77)
  .. controls (-77,94) and (-78,111) .. (-51,120);
 \node[vertex,fill=projblue!30,minimum size=5.4mm] at (-52,93) {$u$};
\end{scope}

\begin{scope}[shift={(P)},x=.022cm,y=.022cm,rotate=-60]
 \draw[example projection,projorange,rounded corners=2pt]
  (-25,145)--(-25,92)--(0,28)--(25,92)--(25,145);
 \node[vertex,fill=projorange!30,minimum size=5.4mm] at (0,68) {$v$};
\end{scope}

\draw[example projection,projpurple] (494,314.608)--(294.814,429.608);
\node[vertex,fill=projpurple!30,minimum size=5.4mm] at (388,355) {$w$};

\draw[example projection,projmagenta,rounded corners=2pt]
 (658.55,365.33)--(748.18,337.55)--(800.14,367.55);
\node[vertex,fill=projmagenta!30,minimum size=5.4mm] at (713,320) {$q$};

\foreach \name in {L,T,A,B,C,D,P,E,F,G,H,I}
 \node[junction] at (\name) {};
\node[terminal] at (R) {};
\node[terminal,minimum size=16mm,font=\LARGE] at (alpha) {$x\;y$};
\node[font=\Huge,anchor=north,inner sep=0pt] at (457,367) {$\alpha$};

\node[type] at (500,23) {Type-$1(i)$};
\node[type] at (500,627) {Type-$1(ii)$};
\node[type,rotate=60] at (214.212,160) {Type-$1(i)$};
\node[type,rotate=-60] at (242.790,473.5) {Type-$1(i)$};
\end{tikzpicture}%
}
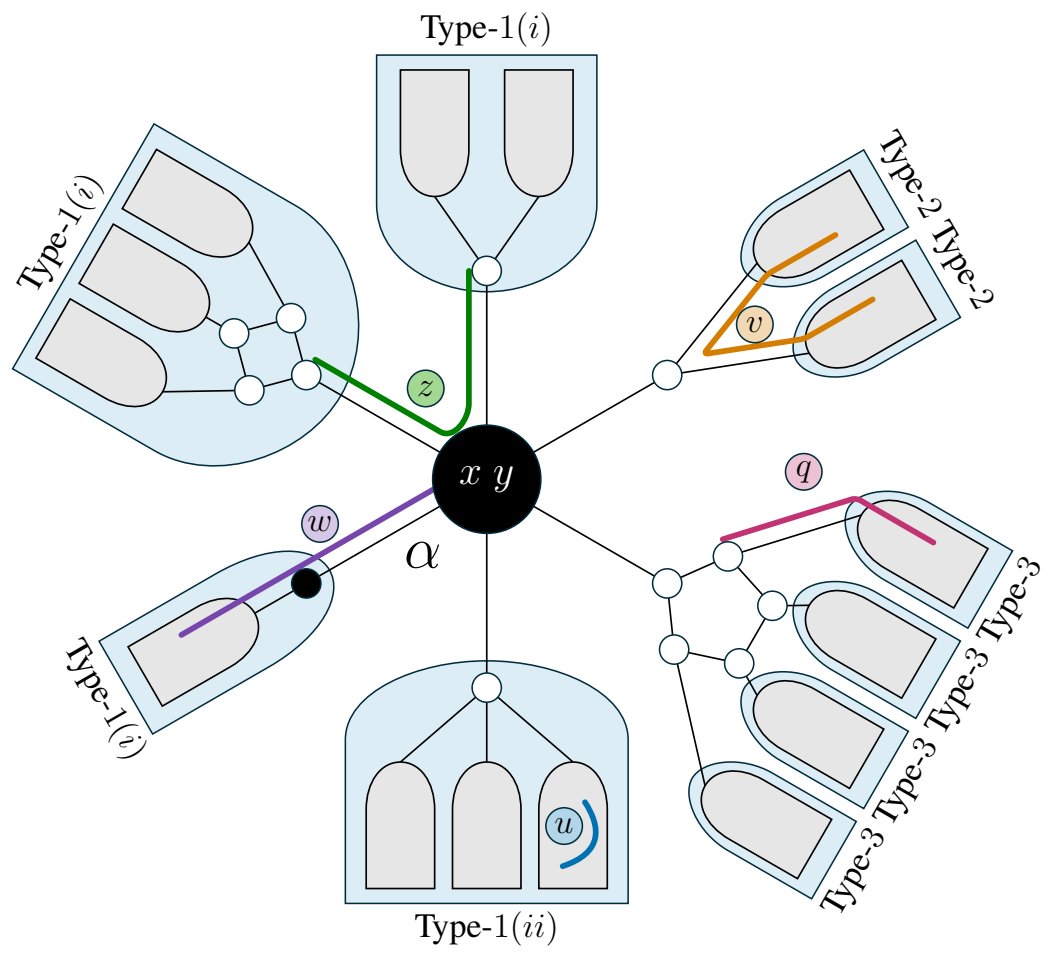
\caption{The $\alpha$-components in $\cH_\mu$ (Def.~\ref{def:alpha-comps}). Some projections in $\cH_\mu$ are illustrated. $\pi_\mu (z)$ causes the sub-skeletons it touches to become Type-1(i).}
\label{fig:alpha-components}

\end{wrapstuff}

\begin{definition}[$\alpha$-components]\label{def:alpha-comps}
Let $g = (\alpha, \beta)$ be a tree-edge of $\cH_\mu$ emanating from $\alpha$, and $H$ be the sub-skeleton of $g$ in direction $\beta$.
Then $H$ gives rise to the following $\alpha$-components:

\begin{description}[leftmargin=*]
    \item[Type-1.] 
    If (i) for some $z \in V$, $\pi_\mu (z)$ intersects $H$ only at $\beta$, or (ii) $\deg_{\cH_\mu} (\beta) \geq 4$, then $H$ is an $\alpha$-component.

    \item[Type-2.] Else, if $\beta$ is a tree-node s.t.\ $\deg_{\cH_\mu}(\beta)=3$, let  $g_1=(\beta,\gamma_1)$ and $g_2=(\beta,\gamma_2)$ be its other two adjacent edges.
    For $i \in \{1,2\}$, the sub-skeleton $T_i$ rooted by $g_i$ in direction $\gamma_i$ is an $\alpha$-component.

    \item[Type-3.] 
    Else, there is a cycle $\cO = (\beta_0, \beta_1, \dots, \beta_k, \beta_0)$ with $\beta_0 = \beta$.
    For every $i \in [1,k]$, let $g_i = (\beta_i, \gamma_i)$ be the unique tree-edge of $\beta_i$.
    The sub-skeleton $T_i$ rooted at $g_i$ in direction $\gamma_i$ is an $\alpha$-component.
\end{description}
\end{definition}

$\,$
$\,$
$\,$
$\,$
$\,$
$\,$
$\,$
$\,$
$\,$
$\,$
$\,$
$\,$
$\,$
$\,$
$\,$
$\,$

\paragraph{Overcoming the space challenge from skeleton cycles.}
By atomicity, taking one representative from each $\alpha$-component ensures the \emph{Representing} property: Once we classify these representatives, we understand which $\alpha$-components get merged to the same $x,y$-unit.
However, this definition does not ensure the \emph{Succinct} property required to keep the space at only $O(n)$.
Note that there are at most $2 \deg_{\cH_\mu} (\alpha)$ Type-1/2 $\alpha$-components.
So, the problem is with Type-3: just one cycle $\cO = (\beta_0, \beta_1, \dots, \beta_k, \beta_0)$ may give rise to many different $\alpha$-components $T_1, \dots, T_k$.

We tackle this challenge by a structural insight, showing that
$T_1, \dots, T_k$ are \emph{strongly correlated}:
Either $T_1, \dots, T_k$ are all in the same $x,y$-unit, or each $T_i$ is in a unique $x,y$-unit without other vertices from $\mu$.
We refer to this property as \emph{cyclic consistency}, formally proved in~\Cref{lem:cyc-dichotomy}.
Thus, it suffices for us to take representatives only from $T_1$ and $T_k$, but not from the $\alpha$-components $T_i$ at the middle part of the cycle (i.e. with $1 < i < k$).
With this modification, we still ensure the \emph{Representing} property, since determining the $x,y$-units of $T_1$ and $T_k$ lets us further determine the $x,y$-units of $T_1, \dots, T_k$ through the cyclic dichotomy.
But now, we now have that $|R(\alpha,\mu)| \leq 2 \deg_{\cH_\mu} (\alpha)$,
so the \emph{Succinct} property (\Cref{eq:rep-calculation}) goes through.

\paragraph{Conceptual insight: utilizing the carcass beyond Steiner mincuts.}
Let us briefly take a step back and discuss the bigger picture.
Intuitively, the LCA was quite straightforward to deal with because $x,y$-mincuts were exactly those Steiner mincuts by which the $\alpha\beta$-components of its skeleton were defined.
Above the LCA, however, it is much less clear that $\alpha$-components have anything to do with $x,y$-mincuts.
The $\alpha$-components correspond to sides of valid partitions induced by $\mu$-mincuts, but now $x$ and $y$ are two terminals from $\alpha$, so they are \emph{not} separated by any Steiner mincut.
The key \emph{conceptual} insight from the above discussion is that the ``centered view'' at $\alpha$ of the skeleton $\cH_\mu$ in~\Cref{fig:alpha-components}, which contracts large chunks as $\alpha$-components, does in fact relate to $x,y$-mincuts in intricate ways, like the atomicity and cyclic consistency properties.
In the following section, we extend this insight further beyond just the skeleton $\cH_\mu$.
We show that, by delicate structural analysis and some clever augmentations,  projections in $\cH_\mu$ can be utilized to shed light not only on $\mu$-mincuts, but also on $x,y$-mincuts for terminal pairs $x,y \in \mu$ not separated by any Steiner mincut.

\subsection{Determining the $x,y$-units of representatives with augmented projections}\label{sec:projections-overview}

We turn to discuss the \emph{Informative} property described in~\Cref{sec:strategy}, which is the most challenging and intricate one.
The setting is the following:
\begin{itemize}
\item
We already know the partition of some node $\mu$ on the query path by the $x,y$-units.
\item
We have a set $R$ of vertices whose augmented projections in $\cH_\mu$ are known to us.
\item
The goal is to extend the partition by $x,y$-units from $\mu$ to $\mu \cup R$.
\end{itemize}
The idea is that $R$ contains the representatives from the parent of $\mu$, and their assignment into $x,y$-units will then be used to extend the partition to all of $\mu$'s parent (as just explained in~\Cref{sec:defining-alpha-comps}).
These aspects are now abstracted away: we think of $R$ as having arbitrary vertices with known augmented projections in $\cH_\mu$.

For $\mu = \LCA_{\cT} (x,y)$ we already saw in~\Cref{sec:skeleton-hierarchy} how the \emph{LCA dichotomy} and \emph{distinctness} (\Cref{lem:LCA-dichotomy,lem:distinctness of xy units}) achieve our goal (there we considered $R = V$, but the same arguments work for any $R \subseteq V$).
We now focus on a node $\mu$ above $\LCA_{\cT} (x,y)$.
As before, we denote by $\alpha$ the child of $\mu$ such that $x,y \in \alpha$.

\paragraph{Generalizing the dichotomy above the LCA.}
At the LCA, the dichotomy hinged on the interaction of a given projection with the $\alpha\beta$-components.
We now show that an analogous but more involved dichotomy also holds above the LCA, based on the interaction of a projection $\pi_\mu (v)$ with the $\alpha$-components in $\cH_\mu$.
This interaction is formalized by the simple notion of \emph{$\alpha$-components tied to $v$} in~\Cref{def:tied-component} below, expressed mainly through~\Cref{obs:tied-comp} following it (which is immediate by the projection theorem, \Cref{thm:proj-vertex}).

\begin{definition}[Tied $\alpha$-component]\label{def:tied-component}
    An $\alpha$-component $T$ is said to be \emph{tied} to vertex $v$ if $\pi_\mu (v)$  intersects $T$.
\end{definition}

\begin{observation}\label{obs:tied-comp}
    Let $\alpha$ be a nonempty node of $\cH_\mu$ with $x,y\in \alpha$.
    If an $\alpha$-component $T$ is tied to vertex $v$, then there exists a $\mu$-mincut $A$ with $A \cap \mu = T$ and $v \in A$.
    All vertices which are not tied to $T$, and in particular, the terminals $x,y \in \alpha$, are in the complement $\overline{A}$.
\end{observation}

The \emph{tied-component dichotomy} in $\mu$ is the following:

\begin{lemma}[Tied component dichotomy]\label{lem:tied-comp-dichotomy}
    Let $\alpha$ be a nonempty node of $\cH_\mu$ with $x,y\in \alpha$.
    If $T$ is an $\alpha$-component tied to vertex $v$, then (exactly) one of the following holds:
    \begin{itemize}
        \item {\bf (``Merged'')} The same $x,y$-unit contains both $v$ and $T$.
        \item {\bf (``Split'')} The $x,y$-unit of $v$ does not contain any $u \in \mu$, 
        nor any $u$ who is not tied to $T$.
    \end{itemize}
\end{lemma}

\begin{wrapstuff}[type=figure,r,width=0.14\textwidth,top=1]
\centering
\begingroup
\definecolor{vennoutline}{RGB}{0,45,62}
\definecolor{vennblue}{RGB}{170,210,232}
\resizebox{\linewidth}{!}{%
\begin{tikzpicture}[every node/.style={font=\LARGE},
  set/.style={draw=vennoutline,line width=.8pt}]
  \fill[vennblue!30] (-.9,0) circle[radius=1.6cm];
  \fill[vennblue!30] (.9,0) circle[radius=1.6cm];
  \begin{scope}
    \clip (-.9,0) circle[radius=1.6cm];
    \fill[vennblue!55] (.9,0) circle[radius=1.6cm];
  \end{scope}
  \draw[set] (-.9,0) circle[radius=1.6cm];
  \draw[set] (.9,0) circle[radius=1.6cm];
  \node at (-1.5,1.85) {$X$};
  \node at (1.5,1.85) {$A$};
  \node at (-1.45,0) {$x$};
  \node at (1.45,0) {$T$};
  \node at (0,.45) {$v$};
  \node at (0,-.45) {$u$};
  \node at (0,2.0) {$y$};
\end{tikzpicture}%
}%
\endgroup
\label{fig:tied-comp-dichotomy}
\end{wrapstuff}
\begin{proof}
    Say $v$ and $T$ are not in the same $x,y$-unit, so there is an $x,y$-mincut $X$ with $v \in X$ and $T \subseteq \overline{X}$. (The illustration shows w.l.o.g.\ $x \in X$, $y \in \overline{X}$.)
    Let $A$ be the $\mu$-mincut of~\Cref{obs:tied-comp}: $A \cap \mu = T$, $v \in A$, $x,y \notin A$.
    Then $A \setminus X$ is a $\mu$-cut (separates $T$ from $x$), and $X \setminus A$ is an $x,y$-cut.
    By posimodularity (\Cref{lem:sub-posi-general}), $X \setminus A$ is an $x,y$-mincut, and note that $v \notin X \setminus A$.
    Now, let $u$ be from the $x,y$-unit of $v$, i.e., $u$ and $v$ are on the same side of every $x,y$-mincut.
    Thus, 
    $u \in X \cap A$.
    As $X \cap A \cap \mu = X \cap T = \emptyset$, we have $u \notin \mu$.
    Also, as $u \in A$, it must be tied to $T$ by~\Cref{obs:tied-comp}.
\end{proof}


Observe that, by the definition of $\alpha$-components (\Cref{def:alpha-comps}) and the structural properties of projections in $\cH_\mu$ from \Cref{thm:proj-vertex}, every vertex $v$ has at most two tied $\alpha$-components.
Furthermore, $v$ has no tied $\alpha$-components iff $\pi_\mu (v) = \alpha$, i.e., $v$ is projected to just the node $\alpha$ with $x,y$.
In this overview, we will ignore this last corner-case and assume that every given $v \in R$ has some tied component.
In a nutshell, to handle the corner case, we will store the projection of such $v$ at some node further below $\mu$, and ``solve it there''.
This is discussed in depth and formally in~\Cref{sec:ds-construction-units,sec:algorithm-units}.


\paragraph{Deciding on the dichotomy.}
Note that there is a major difference between the LCA dichotomy in~\Cref{lem:LCA-dichotomy} and the one of~\Cref{lem:tied-comp-dichotomy}.
In the LCA, the projection of $v$ allowed us to also \emph{decide} between the ``merged'' and ``split'' cases.
In contrast, \Cref{lem:tied-comp-dichotomy} merely states the two possible options, but does not provide a way to distinguish between them.
We now discuss this challenge.

To gain intuition, let us first consider the simplest case: $\pi_\mu (v)$ is completely contained in the $\alpha$-component $T$.
Intuitively, this is analogous to having the projection completely inside an $\alpha\beta$-component of the LCA, and thus $v$ and $T$ should be merged into the same $x,y$-unit.
This is indeed true: 
\begin{lemma}\label{lem:projection-inside}
    Let $\alpha$ be a nonempty node of $\cH_\mu$ with $x,y\in \alpha$.
    If $\pi_\mu (v)$ is contained in an $\alpha$-component $T$, then $v$ and $T$ are in the same $x,y$-unit.
\end{lemma}    
\begin{wrapstuff}[type=figure,r,width=0.14\textwidth,top=0]
\centering
\begingroup
\definecolor{vennoutline}{RGB}{0,45,62}
\definecolor{vennblue}{RGB}{170,210,232}
\resizebox{\linewidth}{!}{%
\begin{tikzpicture}[every node/.style={font=\LARGE},
  set/.style={draw=vennoutline,line width=.8pt}]
  \fill[vennblue!30] (-.9,0) circle[radius=1.6cm];
  \fill[vennblue!30] (.9,0) circle[radius=1.6cm];
  \begin{scope}
    \clip (-.9,0) circle[radius=1.6cm];
    \fill[vennblue!55] (.9,0) circle[radius=1.6cm];
  \end{scope}
  \draw[set] (-.9,0) circle[radius=1.6cm];
  \draw[set] (.9,0) circle[radius=1.6cm];
  \node at (-1.5,1.85) {$X$};
  \node at (1.5,1.85) {$A$};
  \node at (-1.45,0) {$x$};
  \node at (1.45,0) {$T$};
  \node at (0,0) {$v$};
  \node at (0,2.0) {$y$};
\end{tikzpicture}%
}%
\endgroup
\end{wrapstuff}

\begin{proof}
    Seeking contradiction, let $X$ be an $x,y$-mincut with $v \in X$ and $T \subseteq \overline{X}$.
    Let $A$ be the $\mu$-mincut of~\Cref{obs:tied-comp}: $A \cap \mu = T$, $v \in A$, $x,y \notin A$.
    Then $X \setminus A$ is an $x,y$-cut, and $A \setminus X$ is a $\mu$-cut.
    By posimodularity (\Cref{lem:sub-posi-general}), $A \setminus X$ is a $\mu$-mincut separating $v$ from $T$.
    But $\pi_\mu (v)$ is contained in $T$, so this contradicts~\Cref{thm:proj-vertex}.
\end{proof}

Next, consider the case where $v$ has two tied $\alpha$-components $T_1$ and $T_2$.
Carrying over intuition from the LCA, if $T_1$ and $T_2$ are not in the same $x,y$-unit, then $\pi_\mu (v)$ ``goes between different $x,y$-units'' in $\cH_\mu$, so $v$ should be split from $\mu$.
Conversely, if $T_1$ and $T_2$ are merged, then $\pi_\mu (v)$ ``stays in one $x,y$-unit'' in $\cH_\mu$ and thus $v$ should be merged with them.
Again, it can indeed be formalized by utilizing~\Cref{obs:tied-comp}.

\begin{lemma}\label{lem:two-tied-comps}
    Let $\alpha$ be a nonempty node of $\cH_\mu$ with $x,y\in \alpha$.
    If vertex $v$ has two tied $\alpha$-components $T_1$ and $T_2$, then $v$ is merged to the same $x,y$-unit as its tied $\alpha$-components iff $T_1$ and $T_2$ are in the same $x,y$-unit.
\end{lemma}
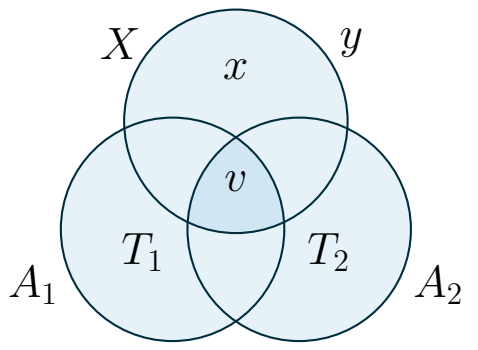
\begin{proof}
\begin{wrapstuff}[type=figure,r,width=0.19\textwidth,top=0]
\centering
\begingroup
\definecolor{vennoutline}{RGB}{0,45,62}
\definecolor{vennblue}{RGB}{170,210,232}
\resizebox{\linewidth}{!}{%
\begin{tikzpicture}[every node/.style={font=\LARGE},
  set/.style={draw=vennoutline,line width=.8pt}]
  \fill[vennblue!30] (-.85,-.5) circle[radius=1.5cm];
  \fill[vennblue!30] (.85,-.5) circle[radius=1.5cm];
  \fill[vennblue!30] (0,.95) circle[radius=1.5cm];
  \begin{scope}
    \clip (-.85,-.5) circle[radius=1.5cm];
    \clip (.85,-.5) circle[radius=1.5cm];
    \fill[vennblue!55] (0,.95) circle[radius=1.5cm];
  \end{scope}
  \draw[set] (-.85,-.5) circle[radius=1.5cm];
  \draw[set] (.85,-.5) circle[radius=1.5cm];
  \draw[set] (0,.95) circle[radius=1.5cm];
  \node[anchor=east] at (-2.25,-1.25) {$A_1$};
  \node[anchor=west] at (2.25,-1.25) {$A_2$};
  \node at (-1.55,2) {$X$};
  \node at (-1.25,-.8) {$T_1$};
  \node at (1.25,-.8) {$T_2$};
  \node at (0,1.65) {$x$};
  \node at (0,.15) {$v$};
  \node at (1.55,2) {$y$};
\end{tikzpicture}%
}%
\endgroup
\label{fig:two-tied-components}
\end{wrapstuff}
    The `only if' is by~\Cref{lem:tied-comp-dichotomy}, as $v$ is either merged to all of its tied $\alpha$-components, or split from all of them.
    For the `if', suppose towards contradiction that $T_1$ and $T_2$ are in the same $x,y$-unit, but $v$ is not with them.
    Let $X$ be an $x,y$-mincut with $v \in X$, so $T_1, T_2 \subseteq \overline{X}$.
    By~\Cref{obs:tied-comp}, there are $\mu$-mincuts $A_i$ with $A_i \cap \mu = T_i$, $v \in A_i$ and $x,y \notin A_i$, for $i \in \{1,2\}$.
    Then $A_1 \setminus (X \cup A_2)$ and $A_2 \setminus (X \cup A_1)$ are $\mu$-cuts (separating $T_1$ and $T_2$) and $X \setminus (A_1 \cup A_2)$ is an $x,y$-cut.
    So, by~\Cref{lem: simple-four-comps}, $\emptyset = A_1 \cap A_2 \cap X \ni v$, a contradiction.
\end{proof}

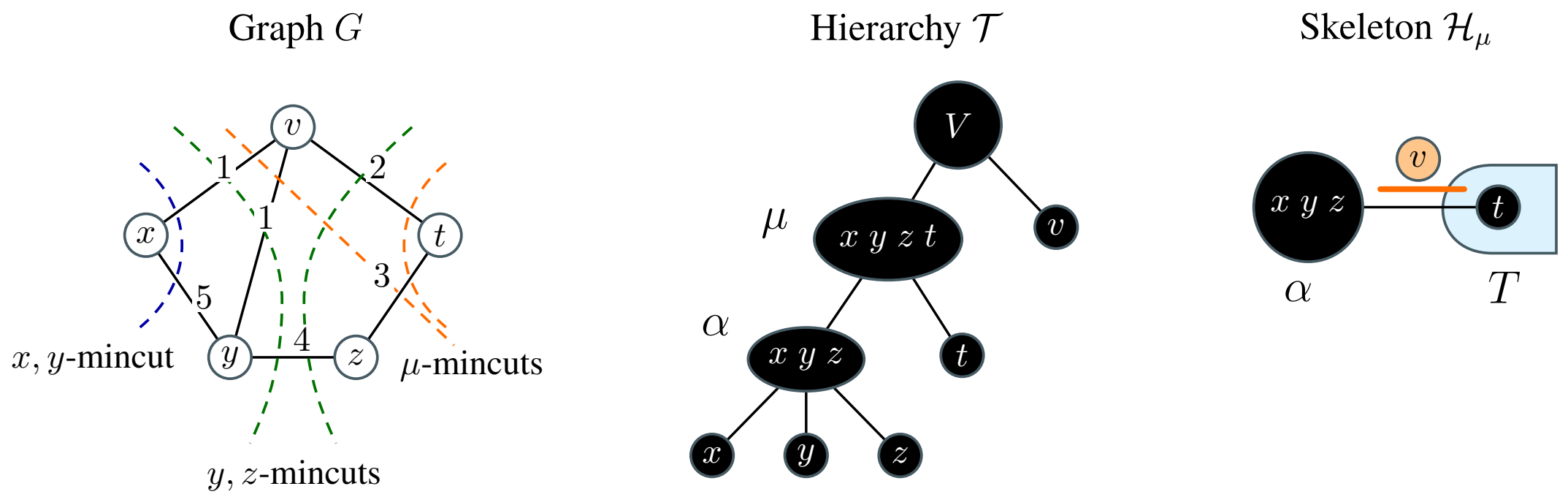
\begin{figure}[t]
\centering
\resizebox{0.85\linewidth}{!}{%
\begin{tikzpicture}[
 x=.022cm,y=-.022cm,
 every node/.style={font=\normalsize},
 edge/.style={black,line width=.8pt},
 vertex/.style={circle,draw=black!75!cyan,line width=.85pt,
   fill=white,minimum size=4.7mm,inner sep=0pt},
 weight/.style={fill=white,inner sep=1pt,font=\normalsize},
 cut/.style={line width=.85pt,dash pattern=on 4pt off 3pt},
 title/.style={font=\large,inner sep=0pt},
 region/.style={draw=black!75!cyan,line width=.85pt,fill=cyan!12}
]
\node[title] at (170,10) {Graph $G$};
\coordinate (gx) at (95,112);
\coordinate (gv) at (168.5,58);
\coordinate (gt) at (242,112);
\coordinate (gy) at (137,173);
\coordinate (gz) at (200,173);
\draw[edge] (gx)--(gv);
\draw[edge] (gv)--(gt);
\draw[edge] (gv)--(gy);
\draw[edge] (gx)--(gy);
\draw[edge] (gy)--(gz);
\draw[edge] (gz)--(gt);

\draw[cut,blue!65!black] (92,76)
 .. controls (120,99) and (120,135) .. (92,158);
\draw[cut,green!45!black] (109,58)
 .. controls (145,90) and (163,115) .. (163,145)
 .. controls (163,174) and (157,195) .. (147,216);
\draw[cut,green!45!black] (228,58)
 .. controls (192,90) and (174,115) .. (174,145)
 .. controls (174,174) and (180,195) .. (190,216);
\draw[cut,orange!85!red] (135,59)--(249,167);
\draw[cut,orange!85!red] (245,76)
 .. controls (217,99) and (217,135) .. (245,158);
\node[weight] at (134,78) {$1$};
\node[weight] at (211,78) {$2$};
\node[weight] at (155,103) {$1$};
\node[weight] at (124,143) {$5$};
\node[weight] at (173,164) {$4$};
\node[weight] at (213,132) {$3$};
\foreach \p/\s in {gx/x,gv/v,gt/t,gy/y,gz/z}
 \node[vertex] at (\p) {$\s$};
\node[anchor=east,inner sep=1pt] at (111,174) {$x,y$-mincut};
\node[anchor=west,inner sep=1pt] at (220,177) {$\mu$-mincuts};
\node[inner sep=1pt] at (169,232) {$y,z$-mincuts};

\node[title] at (475,10) {Hierarchy $\mathcal{T}$};
\coordinate (root) at (501,57);
\coordinate (mu) at (466,114);
\coordinate (alpha) at (425,174);
\coordinate (hv) at (550,108);
\coordinate (ht) at (503,172);
\coordinate (hx) at (378,222);
\coordinate (hy) at (425,222);
\coordinate (hz) at (472,222);
\foreach \a/\b in {root/mu,root/hv,mu/alpha,mu/ht,alpha/hx,alpha/hy,alpha/hz}
 \draw[edge] (\a)--(\b);
\node[vertex,fill=black,text=white,minimum size=9.5mm] at (root) {$V$};
\node[draw=black!75!cyan,line width=.85pt,fill=black,text=white,
 ellipse,minimum width=14mm,minimum height=9mm,inner sep=1pt]
 at (mu) {$x\;y\;z\;t$};
\node[draw=black!75!cyan,line width=.85pt,fill=black,text=white,
 ellipse,minimum width=12mm,minimum height=7mm,inner sep=1pt]
 at (alpha) {$x\;y\;z$};
\foreach \p/\s in {hv/v,ht/t,hx/x,hy/y,hz/z}
 \node[vertex,fill=black,text=white] at (\p) {$\s$};
\node[anchor=east,inner sep=1pt,font=\Large] at (418,106) {$\mu$};
\node[anchor=east,inner sep=1pt,font=\Large] at (389,157) {$\alpha$};

\node[title] at (720,10) {Skeleton $\mathcal{H}_\mu$};
\path[region] (768,77)--(800,77)--(800,121)--(768,121)
 .. controls (735,121) and (735,77) .. (768,77)--cycle;
\draw[edge] (676,98)--(771,98);
\draw[orange!85!red,line width=1.8pt,line cap=round]
 (712,89)--(754,89);
\node[vertex,fill=black,text=white,minimum size=12mm]
 at (676,98) {$x\;y\;z$};
\node[vertex,fill=black,text=white] at (771,98) {$t$};
\node[vertex,fill=orange!45] at (731,74) {$v$};
\node[font=\Large] at (671,140) {$\alpha$};
\node[font=\Large] at (774,138) {$T$};
\end{tikzpicture}%
}
\caption{Vertex $v$ has one tied $\alpha$-component $T$ and $\alpha \in \pi_\mu (v)$.
For the query pair $x,y \in \alpha$, $T$ and $v$ are merged to the same $x,y$-unit.
But for the query pair $y,z \in \alpha$, they are split into different $y,z$-units.}
\label{fig:maximizer-example}
\end{figure}

We now turn to the more complicated case, where $v$ has just one tied $\alpha$-component $T$, but $\pi_\mu (v)$ is not contained in $T$.
It turns out that if $\pi_\mu (v)$ is contained in one sub-skeleton $H$ emanating out of $\alpha$ (which contains $T$), then we can generalize~\Cref{lem:projection-inside} and still determine that $v$ should be merged with $T$; this is proved in~\Cref{lem:one-tied-simple-case}.
Thus, the remaining and most challenging case is when $\pi_\mu (v)$ \emph{has $\alpha$ as an endpoint}.
In fact, in this case both the ``merged'' and ``split'' options are possible: see such an example in~\Cref{fig:maximizer-example}.
So, to decide between them, we \emph{must} further augment $\pi_\mu (v)$.
Since $\alpha$ is one of the two endpoints of $\pi_\mu (v)$, at preprocessing, we may augment $\pi_\mu (v)$ with $O(1)$ information depending on $\alpha$ and on $T$
(which is determined as the $\alpha$-component with the other endpoint of $\pi_\mu (v)$, and does not depend on $x,y$).
We will call this augmentation the \emph{maximizer augmentation} for $\alpha$, presented next.

\paragraph{Augmenting with maximizers.}
Consider the cut family
\[
\cC_{T,v} := \{C \mid \text{$C$ is an $a,b$-mincut for some $a,b \in \alpha$, such that $T \subseteq C$ and $v \notin C$}\} ~.
\]
The reason for focusing on $\cC_{T,v}$ is simple: $v$ and $T$ are split by the $x,y$-units iff $\cC_{T,v}$ has an $x,y$-mincut.
However, we cannot afford to store all pairs $a,b \in \alpha$ with an $a,b$-mincut in $\cC_{T,v}$. 
One key insight to ``compress'' $\cC_{T,v}$ is the following: while the cuts in $\cC_{T,v}$ may cross in arbitrary ways, this family is \emph{closed under unions},
and thus admits a \emph{maximizer}, i.e., a unique $M \in \cC_{T,v}$ that contains every $C \in \cC_{T,v}$ (or $\cC_{T,v}$ is empty).
This is proved in~\Cref{lem:maximizer}.
Let $w,z \in \alpha$ be such that $M$ is a $(w,z)$-mincut ($w \in M$, $z \notin M$).
The maximizer augmentation stores $(w,z)$, and the canonical cut $C_M$ representing $M$ in $\cH_{\LCA_{\cT} (w,z)}$ (or a flag indicating $\cC_{T,v}$ is empty).
The second key insight is that this $O(1)$ information is enough for us to determine at query time if $\cC_{T,v}$ has an $x,y$-mincut.

To get a feeling for how this is done, let us ask: are $T$ and $z$ in the same $x,y$-unit? 
Note that $T \cup \{z\} \subseteq \mu$, so we can indeed check this. If the answer is \emph{yes}, then we argue that $v$ must be merged to the same $x,y$-unit with them: otherwise, there would be some $x,y$-mincut $X$ with $T \cup \{z\} \subseteq X$ and $v \notin X$, but then $X \in \cC_{T,v}$ and $X \not\subseteq M$, contradicting $M$'s maximality.
When the answer is \emph{no}, we prove (\Cref{lem:using-maximizer}) that $T$ and $v$ are merged iff $M \cap \{x,y\} = \emptyset$.
If $x,y \in \LCA_{\cT} (w,z)$, then we can find $M \cap \{x,y\}$ by using $C_M$ in the skeleton of $\LCA_{\cT}(w,z)$.
Otherwise, we can in fact prove that $M \cap \{x,y\} \neq \emptyset$ (\Cref{lem:wz mincut is larger than xymincut}).

\begin{figure}[t]
\centering
\resizebox{\linewidth}{!}{%
\begin{tikzpicture}[
  font=\large,
  question/.style={draw=black,line width=.85pt,rounded corners=3mm,
    fill=white,align=center,text width=2.7cm,minimum height=1.45cm,
    inner sep=2mm},
  result/.style={draw=black,line width=.85pt,rounded corners=2mm,
    fill=gray!45,align=center,minimum width=1.9cm,
    minimum height=.9cm,inner sep=2mm,font=\large\sffamily},
  arrow/.style={-{Stealth[length=2.1mm,width=1.5mm]},
    line width=.9pt},
  answer/.style={fill=white,inner sep=1.5pt,font=\large}
]
\node[question,text width=3.5cm] (count) at (0,0)
  {How many tied\\$\alpha$-components?};
\node[question] (two) at (4.7,1.4)
  {$T_1$ and $T_2$ in\\one $x,y$-unit?};
\node[question,text width=2.1cm] (one) at (4.7,-1.4)
  {$\alpha\in\pi_\mu(v)$?};
\node[result] (two-merged) at (8.3,2.2) {MERGED};
\node[result] (two-split) at (8.3,.6) {SPLIT};
\node[question,text width=2.5cm] (unit) at (10.7,-1.4)
  {$z$ and $T$ in\\one $x,y$-unit?};
\node[result] (one-merged) at (8.7,-3.25) {MERGED};
\node[result] (unit-merged) at (14.3,.15) {MERGED};
\node[question] (lca) at (14.3,-2.25)
  {$x,y\in$\\$\mathrm{LCA}_{\mathcal T}(w,z)$?};
\node[question,text width=2.8cm] (side) at (18.3,-1.35)
  {$C_M$ has one of\\$x,y$ on the\\side of $w$?};
\node[result] (lca-split) at (18.3,-3.3) {SPLIT};
\node[result] (final-split) at (22,-.45) {SPLIT};
\node[result] (final-merged) at (22,-2.25) {MERGED};

\draw[arrow] (-5.15,0) -- node[above,align=center,inner sep=2mm]
  {Augmented\\projection $\pi_\mu(v)$} (count.west);
\draw[arrow] (count.east) -- (two.west);
\draw[arrow] (count.east) -- (one.west);
\node[answer] at (1.9,1.4) {Two: $T_1,T_2$};
\node[answer] at (2.7,-1.05) {One: $T$};
\draw[arrow] (two.east) -- node[answer,above] {yes} (two-merged.west);
\draw[arrow] (two.east) -- node[answer,below] {no} (two-split.west);
\draw[arrow] (one.east) -- node[answer,above,align=center]
  {yes: use\\maximizer} (unit.west);
\draw[arrow] (one.east) -- node[answer,below] {no} (one-merged.west);
\draw[arrow] (unit.east) -- node[answer,above] {yes} (unit-merged.west);
\draw[arrow] (unit.east) -- node[answer,below] {no} (lca.west);
\draw[arrow] (lca.east) -- node[answer,above] {yes} (side.west);
\draw[arrow] (lca.east) -- node[answer,below] {no} (lca-split.west);
\draw[arrow] (side.east) -- node[answer,above] {yes} (final-split.west);
\draw[arrow] (side.east) -- node[answer,below] {no} (final-merged.west);
\end{tikzpicture}%
}
\caption{Deciding on the dichotomy case from~\Cref{lem:tied-comp-dichotomy}. (If we reach the "use maximizer" arrow but no maximizer exists, we return MERGED.)}
\label{fig:merge-split-flowchart}
\end{figure}
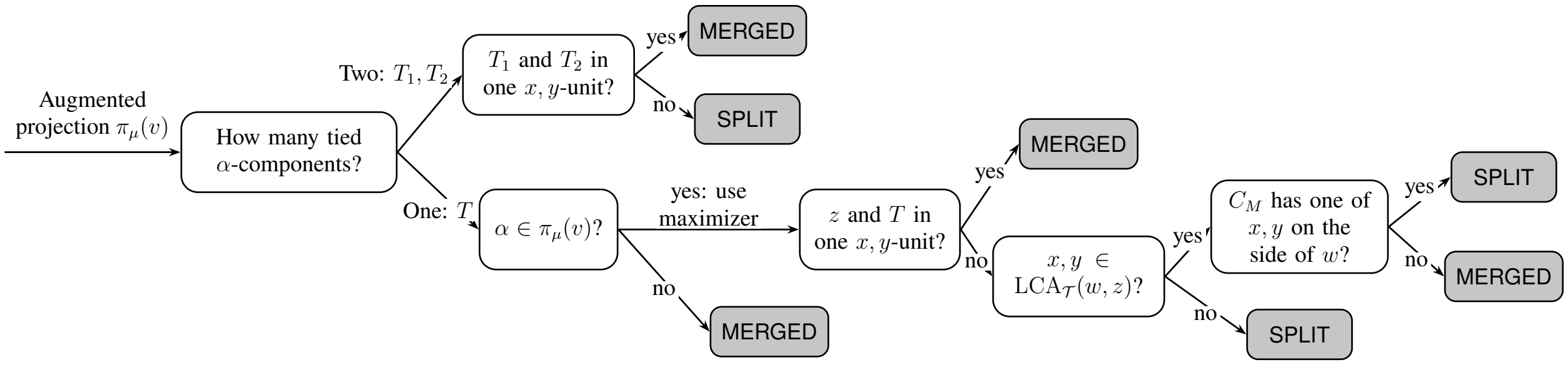

This concludes our discussion on deciding the dichotomy: For each $v \in R$, given $\pi_\mu (v)$ with the maximizer augmentation, we can  decide between the ``merged'' and ``split'' cases for $v$ and its tied $\alpha$-component(s) from~\Cref{lem:tied-comp-dichotomy}.
\Cref{fig:merge-split-flowchart} shows a schematic flowchart of the decision process.

\paragraph{Deciding the $x,y$-units of ``split'' vertices via a new distinctness lemma.}
As was the case in the LCA, after merging all those vertices from $R$ that should be merged with their tied components into the $x,y$-units that intersect $\mu$, we are now left with the classification of the remaining ``split'' vertices from $R$.
By~\Cref{lem:tied-comp-dichotomy}, these cannot be merged with any existing $x,y$-unit that intersects $\mu$.
However, two ``split'' vertices might still belong to the same $x,y$-unit.
In the LCA, deciding this was done by simply comparing the $\mu$-unit identifiers $\phi_\mu (\cdot)$, exploiting the distinctness property of the carcass from~\Cref{lem: Distinctness}.
While the argument using~\Cref{lem: Distinctness} no longer applies (again, since $x,y$-mincuts are not Steiner mincuts), we prove that the same strategy remains valid, which fits into the general theme.
Thus, the following~\Cref{lem: extended-distinctness} can be seen as a new distinctness lemma which goes beyond classical carcass distinctness to handle $x,y$-mincuts when $x,y$ are two terminals that are \emph{not} separated by a Steiner mincut.

\begin{lemma}[Extended Distinctness] \label{lem: extended-distinctness}
    Let $\alpha$ be a nonempty node of $\cH_\mu$ with $x,y\in \alpha$.
    Let $u,v \in V$ be vertices whose $x,y$-units do not intersect $\mu$, such that at least one of $u,v$ has some tied $\alpha$-component.
    Then
    $\phi_{x,y} (u) = \phi_{x,y} (v)$ iff $\phi_\mu (u) = \phi_\mu (v)$.
    That is, $u,v$ are in the same $x,y$-unit iff they are in the same $\mu$-unit.
\end{lemma}
\begin{proof}
\begin{wrapstuff}[type=figure,r,width=0.14\textwidth,top=1]
\centering
\begingroup
\definecolor{vennoutline}{RGB}{0,45,62}
\definecolor{vennblue}{RGB}{170,210,232}
\resizebox{\linewidth}{!}{%
\begin{tikzpicture}[every node/.style={font=\LARGE},
  set/.style={draw=vennoutline,line width=.8pt}]
  \fill[vennblue!30] (-.9,0) circle[radius=1.6cm];
  \fill[vennblue!30] (.9,0) circle[radius=1.6cm];
  \begin{scope}
    \clip (-.9,0) circle[radius=1.6cm];
    \fill[vennblue!55] (.9,0) circle[radius=1.6cm];
  \end{scope}
  \draw[set] (-.9,0) circle[radius=1.6cm];
  \draw[set] (.9,0) circle[radius=1.6cm];
  \node at (-1.5,1.85) {$X$};
  \node at (1.5,1.85) {$B$};
  \node at (-1.45,0) {$v$};
  \node at (1.45,0) {$y$};
  \node at (0,.45) {$x$};
  \node at (0,-.45) {$u$};
  \node at (0,2.0) {$t$};
\end{tikzpicture}%
}%
\endgroup
\label{fig:distinctness-only-if}
\end{wrapstuff}

    (`Only if') We prove the contrapositive.
    Suppose a $\mu$-mincut $B$ separates $u$ from $v$.
    As $x,y$ are terminals from the same node $\alpha$ of $\cH_\mu$, they are not separated by $B$.
    Choose the sides such that $x,y \in B$, and without loss of generality say $u \in B$ and $v \notin B$ (otherwise swap $u,v$ in the following).
    Fix some $t \in \mu \cap \overline{B}$.
    Because $u$'s $x,y$-unit does not intersect $\mu$, there is an $x,y$-mincut $X$ such that $u \in X$ and $t \notin X$.
    If $v \notin X$ we are done.
    Otherwise, note that $X \cap B$ is an $x,y$-cut, and that $X \cup B$ is a $\mu$-cut (separating $x$ from $t$).
    Thus, by submodularity (\Cref{lem:sub-posi-general}), $X \cap B$ is an $x,y$-mincut, separating $u$ from $v$ as needed.

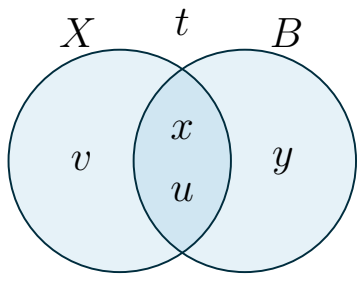
\begin{wrapstuff}[type=figure,r,width=0.14\textwidth,top=1]
\centering
\begingroup
\definecolor{vennoutline}{RGB}{0,45,62}
\definecolor{vennblue}{RGB}{170,210,232}
\resizebox{\linewidth}{!}{%
\begin{tikzpicture}[every node/.style={font=\LARGE},
  set/.style={draw=vennoutline,line width=.8pt}]
  \fill[vennblue!30] (-.9,0) circle[radius=1.6cm];
  \fill[vennblue!30] (.9,0) circle[radius=1.6cm];
  \begin{scope}
    \clip (-.9,0) circle[radius=1.6cm];
    \fill[vennblue!55] (.9,0) circle[radius=1.6cm];
  \end{scope}
  \draw[set] (-.9,0) circle[radius=1.6cm];
  \draw[set] (.9,0) circle[radius=1.6cm];
  \node at (-1.5,1.85) {$X$};
  \node at (1.5,1.85) {$A$};
  \node at (-1.45,0) {$x$};
  \node at (1.45,.45) {$v$};
  \node at (1.45,-.45) {$t_2$};
  \node at (0,.45) {$u$};
  \node at (0,-.45) {$t_1$};
  \node at (0,2.0) {$y$};
\end{tikzpicture}%
}%
\endgroup
\label{fig:venn-x-a-terminals}
\end{wrapstuff}
    (`If') 
    By symmetry, suppose $v$ has a tied $\alpha$-component $T$.
    By~\Cref{obs:tied-comp}, there is a $\mu$-mincut $A$ with $A \cap \mu = T$, $v \in A$, $x,y \notin A$.
    As $\phi_\mu (u) = \phi_\mu (v)$, also $u \in A$.
    Seeking contradiction, let $X$ be an $x,y$-mincut with $u \in X$ and $v \notin X$.
    Then $X \cup A$ and $X \setminus A$ are both $x,y$-cuts.
        If there exists $t_1 \in T \cap X$, then $X \cap A$ is a $\mu$-cut
        (separating $t_1$ from one of $x,y$), so by submodularity (\Cref{lem:sub-posi-general}) it is a $\mu$-mincut, and it has $u$ inside and $v$ outside.
        Else, there must be some $t_2 \in T \setminus X$, so $A \setminus X$ is a $\mu$-cut (separating $t_2$ from one of $x,y$), thus by posimodularity (\Cref{lem:sub-posi-general}) it is a $\mu$-mincut, and has $v$ inside and $u$ outside.
    So, in both cases, we got a contradiction to $\phi_\mu (u) = \phi_\mu (v)$. This concludes the proof.
\end{proof}

Thus, we finally achieved our goal of extending the known partition by $x,y$-units from $\mu$ to $\mu \cup R$, and showed how to obtain the \emph{Informative} property from~\Cref{sec:strategy}.
This gives the final brick for implementing the bottom-up strategy presented there, thus yielding the desired $O(n)$ space structure for recovering the $x,y$-units of any given  query pair $x,y \in V$.

All missing structural lemmas mentioned in this overview (e.g., atomicity, cyclic consistency, maximizer existence) appear in~\Cref{sec:structural-analysis-for-units}.
\Cref{sec:reporting-units} then gives the details for an $O(n)$ space structure that reports the $x,y$-units within $O(n)$ time, treating also the corner case of representative projections equal to $\alpha$, which we have ignored here for the sake of clarity.

\subsection{Topologically ordering the $x,y$-units}\label{sec:ordering-overview}

We now give a high-level outline of how we further recover a topological ordering of the $x,y$-units, so that the union of every nonempty proper prefix is an $(x,y)$-mincut. We follow the same bottom-up construction along the query path, maintaining an ordering as the known partition grows. ``Merged'' vertices inherit the position of their existing unit; our task is to position the newly created units.

At the LCA, we first order the $\alpha\beta$-components along the cactus path $P_{\alpha\beta}$ from $\alpha$ to $\beta$.
The projections then guide the placement of the “split” representatives: each is assigned to a \emph{bucket} immediately before its rightmost $\alpha\beta$-component, as determined by its projection along $P_{\alpha\beta}$. 
Choosing one representative from each new unit, this gives a placement of the new units between the already ordered components, while leaving the internal order of each bucket undetermined.

Above the LCA, newly introduced cycle components can be placed in their cycle order between the two boundary components. 
Placing split representatives requires a further structural insight: a split vertex $v$ is comparable to each of its tied $\alpha$-components in the $(x,y)$-mincuts partial order. 
If there are two tied components, \(v\) must appear between them. 
If there is just one tied component \(T\), the maximizer augmentation from~\Cref{sec:projections-overview} also determines on which side of $T$ to place $v$.
We can therefore again assign each new unit to a bucket immediately beside the unit containing an appropriate component. 
In both cases, we prove that the placement creates no conflicts between different buckets or with the existing ordering.

It remains to order the new units within each bucket. 
Here, the geometry of projections again provides most of the information we need. After grouping vertices by their associated component, vertices whose projections reach deeper into the same component $T$ can be placed closer to $T$ in the ordering. When the relevant endpoints coincide, we use a precomputed ordering associated with the Steiner mincuts represented by their common last tree-edge. This requires only $O(1)$ additional information for each stored projection, preserving the total $O(n)$ space. 
The full construction and its analysis appear in~\Cref{sec:reporting-ordering}.

\subsection*{AI methodology} The scientific content, including the ideas, results, statements, and proofs, was developed entirely by the authors without use of AI.
The authors used ChatGPT-5.6 \& 6 Astra and Claude Opus 5.5 as an assistive tools during the preparation of this manuscript, restricted to refining the exposition and organization, polishing language, detecting minor gaps and typos, and assisting with the development of figures and visual illustrations.

\section{Completing the Structural Analysis for $x,y$-Units}\label{sec:structural-analysis-for-units}

This section complements \Cref{sec:defining-alpha-comps,sec:projections-overview} from the overview, and provides the more involved missing proofs that were only mentioned there.

\subsection{Atomicity and cyclic consitency for $\alpha$-components}

\begin{lemma}[Atomicity]\label{lem: atomicity}
    If $x,y \in \alpha$, then no $x,y$-mincut can split an $\alpha$-component.
    Namely, each $\alpha$-component is fully contained in one $x,y$-unit.
\end{lemma}
\begin{proof}
    Let $T$ be an $\alpha$-component within a sub-skeleton $H$, where $H$ is rooted at tree-edge $g=(\alpha,\beta)$, and let $u,v \in T$.
    We prove that no $x,y$-mincut can separate $u,v$.
    
    For the canonical cut defined by $g$, by~\Cref{thm:skeleton}, there is a $\mu$-mincut $A$ with $H \subseteq A$ and $\alpha \subseteq \overline{A}$.
    We now fork into cases: In each, we choose a vertex $z \in A$ such that every $\mu$-mincut that separates $u$ from $v$ must keep $z$ and $\alpha$ on the same side.
    \begin{itemize}
        \item Suppose $T = H$ is of Type 1(i).
        Then we select $z$ as the promised vertex whose projection $\pi_\mu (z)$ intersects $H$ only in the node $\beta$. Thus, by~\Cref{thm:proj-vertex}, we may choose the $\mu$-cut $A$ so that $z \in A$.
        As $u,v \in H$, any $\mu$-mincut separating them is represented by a canonical cut within $H$.
        In each such canonical cut, $\beta$ is on the side of $\alpha$, and hence $z$ must also be on the side of $\alpha$ by~\Cref{thm:proj-vertex}.
        
        \item Suppose $T = H$ is of Type 1(ii).
        Then $\beta$ is a tree-node with at least three incident tree-edges inside $H$.
        Thus, we can find one tree-edge $g' = (\beta, \beta')$ among them, such that $g'$ does not lead to $u$ nor to $v$.
        Let $H'$ be the connected component of $H \setminus g'$ with $u,v, \beta \in H'$.
        We select any terminal from $H \setminus H'$ as $z$.
        Now, as $u,v \in H'$, any $\mu$-mincut separating them is represented by a canonical cut within $H'$, and such a canonical cut keeps all of $H \setminus H'$ on the side of $\alpha$.

        \item Suppose $T$ is of Type-2 or Type-3.
        Then there exists at least one other $\alpha$-component $T'$ within $H$.
        We select any terminal from $T'$ as $z$.
        Since $u,v \in T$, every $\mu$-mincut separating them must be represented by a canonical cut within $T$, and such a canonical cut keeps all of $T'$ on the side of $\alpha$.
    \end{itemize}

    Now, seeking contradiction, suppose $X$ is an $x,y$-mincut which separates $u,v$.
    Without loss of generality, say $x,u \in X$ and $y,v \notin X$.
    As $x,y \notin A$, we have that $X \setminus A$ and $X \cup A$ are $x,y$-cuts.
    Also, $A \setminus X$ and $A \cap X$ are $\mu$-cuts, as they both separate $u$ from $v$.
    By sub/posimodularity (\Cref{lem:sub-posi-general}), the latter are $\mu$-mincuts.
    Both of them keep $\alpha$ outside, so by the property of $z$, they must also keep $z$ outside.
    Thus $z \notin (A \setminus X) \cup (A \cap X) = A$, a contradiction.
\end{proof}

\begin{lemma}[Cyclic consistency]\label{lem:cyc-dichotomy}
    Let $T_1, \dots, T_k$ be the Type-3 $\alpha$-components in $\cH_\mu$, arising from a cycle $\cO$ within sub-skeleton $H$ (recall~\Cref{def:alpha-comps}).
    Let $x,y \in \alpha$.
    Then, (exactly) one of the following holds:
    \begin{itemize}
        \item \textbf{(``Merged'')} All of $T_1, \dots, T_k$ are in one $x,y$-unit.
        \item \textbf{(``Split'')} For every $i \in [1,k]$, the $x,y$-unit of $T_i$ does not contain any vertex $v$ who is not tied to $T_i$.
    \end{itemize}
\end{lemma}
\begin{proof}
    The heavy lifting of the proof is in showing the following:
    \begin{claim}\label{lem:cyc-consistency}
        The following conditions are equivalent:
        \begin{enumerate}
            \item[(1)] \emph{There exist} $i \neq j$ such that $T_i$ and $T_j$ are in different $x,y$-units.
            \item[(2)] \emph{For every} $i \neq j$ it holds that $T_i$ and $T_j$ are in different $x,y$-units.
        \end{enumerate}
    \end{claim}

    Before proving~\Cref{lem:cyc-consistency}, let us use it to obtain the lemma. 
    Suppose $T_1, \dots, T_k$ are not all in one $x,y$-unit, so by~\Cref{lem:cyc-consistency}, $T_1, \dots, T_k$ are in distinct $x,y$-units.
    Let $v$ be a vertex not tied to $T_i$.
    If $v$ is tied to some $T_j$ with $j \neq i$, then by~\Cref{lem:tied-comp-dichotomy} it either be in the $x,y$-unit of $T_j$, or in an $x,y$-unit that does not intersect $\mu$ at all; in any case, $v$ cannot be in the $x,y$-unit of $T_i$.
    
    Otherwise, since a projection cannot be contained within the cycle $\cO$ (see \Cref{thm:proj-vertex}), $\pi_\mu (v)$ cannot be contained in $H$. 
    Thus, by~\Cref{thm:proj-vertex}, there is a $\mu$-mincut $A$ with $A \cap \mu = H$ and $v, x, y \notin A$.
    Now, take some $j \in [1,k]$ s.t.\ $j \neq i$, and let $X$ be an $x,y$-mincut with $T_j \subseteq X$ and $T_i \subseteq \overline{X}$.
    Then $X \cap A$ is a $\mu$-cut (separating $T_j$ from $T_i$), and $X \cup A$ is an $x,y$-cut.
    By submodularity (\Cref{lem:sub-posi-general}), $X \cup A$ is an $x,y$-mincut.
    One of $X$ or $X \cup A$ must separate $v$ and $T_i$ (if $v \in X$ it is $X$, and if $v \notin X$ it is $X \cup A$), as needed.
\end{proof}

\begin{proof}[Proof of~\Cref{lem:cyc-consistency}.]
    Let us define a third condition: (3) $T_1$ and $T_k$ are separated by some $x,y$-mincut.
    We prove that $(1) \Rightarrow (3) \Rightarrow (2)$, and $(2) \Rightarrow (1)$ is trivial.

    $(1) \Rightarrow (3)$:
    Let $X$ be an $x,y$-mincut separating $T_i$ from $T_j$, $i \neq j$.
    Suppose towards a contradiction that $T_1$ and $T_k$ are on the same side of $X$. Without loss of generality say $T_1, T_i,  T_k \subseteq X$ and $T_j \subseteq \overline{X}$.
    Take $A$ as a $\mu$-mincut represented by the canonical cut $g$ of $\cH_\mu$, with sides chosen so that $T_1, \dots, T_k \subseteq A$ and $\alpha \subseteq \overline{A}$.
    As $x,y \notin A$, we see that $X \cup A$ is an $x,y$-cut.
    As $T_i \subseteq X \cap A$ and $T_j \subseteq \overline{X}$, we see that $X \cap A$ is a $\mu$-cut.
    Thus, by submodularity (\Cref{lem:sub-posi-general}), $X \cap A$ is a $\mu$-mincut.
    Now, observe that in the skeleton $\cH_\mu$, any canonical cut which keeps $T_1$ and $T_k$ on the same side and $T_j$ on the other side must keep $\alpha$ with $T_1$ and $T_k$.
    However, $X \cap A$ keeps $T_1,T_k$ inside and $\alpha$ outside.
    Thus, this $\mu$-mincut cannot be represented by any canonical cut of the skeleton $\cH_\mu$, a contradiction to~\Cref{thm:skeleton}.

    $(3) \Rightarrow (2)$:
    Let $X$ be an $x,y$-mincut separating $T_1$ from $T_k$.
    Fix $T_i$ and $T_j$ for $i \neq j$; we should show an $x,y$-mincut separating them.
    If $X$ is such, we are done. Otherwise, without loss of generality, suppose that $T_1, T_i, T_j \subseteq X$ and $T_k \subseteq \overline{X}$, and that $i < j$.
    Consider the canonical cut of $\cH_\mu$ defined by the two cycle-edges $(\beta_0, \beta_k)$ and $(\beta_{j-1}, \beta_j)$ on $\cO$ (these two edges are non-adjacent since $j \notin \{1,k\}$).
    Let $C$ be a $\mu$-mincut represented by this canonical cut.
    Note that $C$ keeps $T_j, \dots, T_k$ on the one side, and keeps $\alpha$ with $T_1, \dots T_{j-1}$ on the other side;
    we may choose the sides so that $T_j, \dots, T_k \subseteq C$.
    Then $C \setminus X$ is a $\mu$-cut: it keeps $T_k$ inside and $T_1$ outside.
    Also, $X \setminus C$ is an $x,y$-cut, as $x,y \in \alpha$ and $\alpha \subseteq \overline{C}$.
    Thus, by posimodularity (\Cref{lem:sub-posi-general}), $X \setminus C$ is an $x,y$-mincut, with $T_i$ inside and $T_j$ outside, as required.   
\end{proof}

\subsection{Determining the dichotomy with one tied $\alpha$-component}

\begin{lemma}\label{lem:one-tied-simple-case}
    Let $x,y \in \alpha$.
    If vertex $v$ has one tied $\alpha$-component $T$ and $\alpha \notin \pi_\mu (v)$, then $v$ is merged with $T$ to the same $x,y$-unit.
\end{lemma}
\begin{proof}
    Let $H$ be the sub-skeleton containing $T$ rooted at the tree-edge $g = (\alpha,\beta)$.
    Since $\pi_\mu (v)$ can either be a tree-node or a proper path starting and ending in tree-edges (\Cref{thm:proj-vertex}), there are two possibilities: (i) $\pi_\mu (v)$ is contained in $T$, or (ii) $T$ is of Type-2 or Type-3.
    In both, $\pi_\mu (v)$ is contained in $H$.
    Let $A$ be a $\mu$-mincut represented by $g$ with $A \cap \mu = H$, so $v \in A$ by~\Cref{thm:proj-vertex}.
    
    Seeking contradiction, suppose $X$ is an $x,y$-mincut with $T \subseteq X$ and $v \notin X$.
    As $x,y \in \alpha \subseteq \overline{A}$, we have: $x \in X \setminus A$, $v \in A \setminus X$, $T \subseteq X \cap A$, $y \in \overline{X \cup A}$.
    By submodularity (\Cref{lem:sub-posi-general}), we now see that $X \cap A$ is a $\mu$-mincut, keeping $T$ inside and $v$ outside.
    In case (i), this is already a contradiction to~\Cref{thm:proj-vertex}.
    In case (ii), \Cref{thm:proj-vertex} implies that $X \cap A$ must be represented by the rooting edge of $T$, so $X \cap A \cap \mu = T$.
    Then considering some other $\alpha$-component $T'$ within $H$, we have $T' \subseteq A \setminus X$.
    Now, applying posimodularity (\Cref{lem:sub-posi-general}) on $X$ and $A$, we obtain that $A \setminus X$ is a $\mu$-mincut, keeping $v$ inside and $T$ outside.
    But now, we again invoke \Cref{thm:proj-vertex} which implies that $\overline{A \setminus X} \cap \mu = T$.
    This is a contradiction since $x,y \in \overline{A \setminus X}$ are terminals, but they are not in $T$.
\end{proof}

\begin{lemma}[Maximizer]\label{lem:maximizer}
    Let $T$ be an $\alpha$-component tied to vertex $v$, and
    consider the family of cuts
    \[
    \cC_{T,v} := \{C \mid \text{$C$ is an $(a,b)$-mincut with $a,b \in \alpha$ s.t. $T \subseteq C$ and $v \notin C$}\}
    \]
    If the family $\cC_{T,v}$ is nonempty, then it has unique maximal element  with respect to containment: i.e., there exists a unique $M \in \cC_{T,v}$ such $C \subseteq M$ for every $C \in \cC_{T,v}$.
\end{lemma}
\begin{proof}
    Seeking contradiction, suppose $M_1$ and $M_2$ are two different maximal elements in $\cC_{T,v}$, where each $M_i$ is an $(a_i,b_i)$-cut for $a_i,b_i\in \alpha$.
    By~\Cref{lem:sub-posi-general}, one of its three cases holds for $M_1$ and $M_2$. 
    It cannot be ``submodularity'' or ``hybrid'': Indeed, then $M_1 \cup M_2$ would be an $(a_i,b_i)$-mincut for some $i \in \{1,2\}$, with $T \subseteq M_1 \cup M_2$ and $v \notin M_1 \cup M_2$, so $M_1 \cup M_2 \in \cC_{T,v}$, contradicting the maximality of $M_1$ and $M_2$.
    Thus, the ``posimodularity'' case holds: one of $M_1 \setminus M_2, M_2 \setminus M_1$ is an $a_1,b_1$-mincut and the other is an $a_2,b_2$-mincut.
    Let $A$ be the $\mu$-mincut of~\Cref{obs:tied-comp}: $A \cap \mu = T$ and $v \in A$.
    As $a_1, b_1, a_2, b_2$ are terminals within the node $\alpha$, which is not in $T$, we obtain that $M_1 \setminus (M_2 \cup A)$ and $M_2 \setminus (M_1 \cup A)$ form an $a_1,b_1$-cut and a $a_2,b_2$-cut.
    Also, $M_1 \cap M_2 \cap A$ forms a $\mu$-cut (separating $T$ from $\alpha$).
    Thus, by~\Cref{lem: simple-four-comps} we obtain $\emptyset = A \setminus (M_1 \cup M_2) \ni v$, a contradiction.
\end{proof}

\begin{lemma}\label{lem:using-maximizer}
    Let $T$ be an $\alpha$-component tied to vertex $v$.
    Let $M$ be the maximizer of $\cC_{T,v}$ from~\Cref{lem:maximizer}, and let $w,z \in \alpha$ be such that $M$ is a $(w,z)$-mincut (so $w \in M$, $z \notin M$).
    Then, for every $x,y \in \alpha$,
    \[
    \text{$v$ and $T$ are in different $x,y$-units $\iff$ $z$ and $T$ are in different $x,y$-units, and $M \cap \{x,y\} \neq \emptyset$}.
    \]
\end{lemma}
\begin{proof}
    ($\Rightarrow$) Let $X$ be an $x,y$-mincut with $T \subseteq X$ and $v \notin X$.
    Then $X \in \cC_{T,v}$, hence $X \subseteq M$.
    As $z \in \overline{M}$, this means that $X$ separates $T$ from $z$.
    Also, $X$ contains one of $x,y$, so $M \cap \{x,y\} \neq \emptyset$.

    ($\Leftarrow)$
    Let $X$ be an $x,y$-mincut with $T \subseteq X$ and $z \notin X$.
    If $v \notin X$ we are done, so suppose $v \in X$.
    Note that $\overline{X \setminus M}$ keeps $T$ inside, $v$ outside, and contains $M \cup \{z\}$.
    Thus, it cannot be an $a,b$-mincut for any $a,b \in \alpha$, as this would contradict the maximality of $M$ in $\cC_{T,v}$.
    This eliminates the posimodularity and hybrid cases of~\Cref{lem:sub-posi-general} for the cuts $X$ and $M$, hence the submodularity case must hold.
    Without loss of generality, say $x \in X$ and $y \notin X$.
    If $x \in X \cap M$, then submodularity implies that $X \cap M$ is an $x,y$-mincut, which keeps $T$ inside and $v$ outside, so we are done.
    We show that this must indeed occur:
    Otherwise, for submodularity to hold, $X \cup M$ must be a $x,y$-mincut, hence $y \notin M$, so $M \cap \{x,y\} = \emptyset$, a contradiction.
\end{proof}

\begin{claim}\label{claim:LCAs-without-hierarchy}
    For $x,y,w,z \in V$,
    if $\min\{\lambda_{x,z}, \lambda_{y,z}\} < \lambda_{w,z}$ then $w,z$ are not separated by any $x,y$-mincut.
\end{claim}
\begin{proof}
    Say $\lambda_{x,z} < \lambda_{w,z}$ (otherwise swap $x$ and $y$).
    Then any $x,z$-mincut must keep $w$ on the opposite side to $x$, and thus $\lambda_{x,w} \leq \lambda_{x,z} < \lambda_{w,z}$.
    Thus, we have that $\lambda_{w,z} > \max\{\lambda_{x,w},\lambda_{x,z}\}$.

    Seeking contradiction, suppose an $(x,y)$-mincut $X$ separates $w,z$, say $w \in X$, $z \notin X$ (otherwise, swap $w$ and $z$ in the following argument).
    Take an $(x,z)$-mincut $Z$.
    Then $X \cap Z$ is an $x,y$-cut, and $X \cup Z$ is an $x,z$-cut. 
    By submodularity (\Cref{lem:sub-posi-general}), $X \cup Z$ is an $x,z$-mincut.
    But $w \in X \cup Z$, so it is also an $w,z$-cut.
    Hence, $\lambda_{w,z} \leq \lambda_{x,z}$, a contradiction.
\end{proof}

\begin{lemma} \label{lem:wz mincut is larger than xymincut}
    In the conditions of~\Cref{lem:using-maximizer},
    if $z$ and $T$ are in different $x,y$-units and $\lambda_{w,z} > \min\{\lambda_{x,z},\lambda_{y,z}\}$, then $v$ and $T$ are in different $x,y$-units. 
\end{lemma}
\begin{proof}
    Repeat the proof of direction ($\Leftarrow)$ in~\Cref{lem:using-maximizer} with the last sentence (``Otherwise, ...'') changed as follows: Otherwise, for submodularity to hold, $X \cap M$ must be a $w,z$-mincut, so $w \in X \cap M$.
    Thus $X$ separates $w$ from $z$, so \Cref{claim:LCAs-without-hierarchy} gives $\lambda_{w,z} \leq \min \{\lambda_{x,z},\lambda_{y,z}\}$, a contradiction.
\end{proof}

\section{Data Structure for Reporting $x,y$-Units}\label{sec:reporting-units}

In this section we give the formal details for the $O(n)$ space data structure that, given query $x,y \in V$, reports all of the $x,y$-units within $O(n)$ time, which was outlined in \Cref{sec:skeleton-hierarchy,sec:strategy,sec:defining-alpha-comps,sec:projections-overview} of the technical overview.

\subsection{Construction}\label{sec:ds-construction-units}

\paragraph{How we store projections.}
As mentioned in the overview, there is one corner case which creates some complications: for a given child-parent pair $\alpha \subseteq \mu$, a representative vertex $v$ which is projected into $\cH_\mu$ might have no tied $\alpha$-components.
This occurs precisely when $\pi_\mu (v) = \alpha$, i.e., $v$'s projection in $\cH_\mu$ is just the nonempty node $\alpha$.
From now on, call a projection \emph{trivial} if it is just one nonempty node.
Our problem thus stems from the possibility of trivial projections.

The key observation is that trivial projections can be stored \emph{implicitly} by using the hierarchy $\cT$.
This frees up space for us to store \emph{more information} for $v$  if it happens to be trivially projected as $\pi_\mu (v) = \alpha$.
Specifically, we will be able to find the projection of $v$ ``further down'' towards $\LCA_{\cT} (x,y)$ of the given query $x,y \in \alpha$, and classify its $x,y$-unit already there, instead of at the higher node $\mu$.
This is formally shown in~\Cref{sec:algorithm-units} on the query algorithm for recovering the $x,y$-units.

We now explain the details of how we store projections. 
Suppose that, during preprocessing, we wish to store the projection of some vertex $v$ in the skeleton $\cH_\mu$ of node $\mu \in \cT$.
This request triggers~\Cref{alg:store-projection}: starting at node $\mu$, check if the projection of $v$ is trivial in the skeleton of the current node, and if so continue to its child which is this projection.
The process halts at some descendant $\eta$ of $\mu$ such that $v$ is non-trivially projected in $\cH_\eta$ (or $\eta$ is a leaf of $\cT$).
Finally, we store the tuple $(v, \mu, \eta, \pi_\eta (v))$, with $\pi_\eta (v)$ interpreted as a sentinel value $\perp$ when $\eta$ is a leaf.

\begin{algorithm}[H]
\caption{\textsc{Store-Projection}$(v,\mu)$}
\label{alg:store-projection}
\begin{algorithmic}[1]
\State $\eta \gets \mu$
\While{$\eta$ is not a leaf of $\cT$ and $\pi_\eta (v)$ is trivial (i.e., a nonempty node in $\cH_\eta$)}
    \State $\eta \gets \pi_\eta (v)$
\EndWhile
\State Store $(v,\mu,\eta,\pi_\eta (v))$
\end{algorithmic}
\end{algorithm}

If $\eta \neq \mu$, the trivial projection of $v$ in $\mu$ is \emph{implicitly} stored, as the child of $\mu$ which is the ancestor of $\eta$.
In a similar fashion, we can in fact determine the projection of $v$ in all of the skeletons on the $\cT$-path between $\mu$ and $\eta$.
We call $\pi_\eta (v)$ an explicitly stored projection.

\paragraph{Augmenting the explicit projections.}
As described in the technical overview  (\Cref{sec:projections-overview}), we not only store projections, but also augment them with some additional information. 
Formally, each \emph{explicitly stored} projection $\pi_\mu (v)$ is augmented by storing (next to it) the following information:
\begin{itemize}
    \item Store the $\mu$-unit identifier $\phi_\mu (v)$.
    \item If $\pi_\mu (v)$ is not a single node, then for each endpoint $\alpha$ of $\pi_\mu (v)$ which is nonempty:
    \begin{itemize}
        \item Let $T$ be the $\alpha$-component with the other endpoint of $\pi_\mu (v)$.
        \item Let $M$ be the $(w,z)$-mincut with $w,z \in \alpha$ which is the maximizer of $\cC_{T,v}$ from~\Cref{lem:maximizer}.
        If the maximizer $M$ does not exist, i.e., $\cC_{T,v} = \emptyset$, just store a flag indicating this situation.
        When $M$ exists, store the ordered pair $(w,z)$ and the canonical cut $C_M$ representing $M$ in $\cH_{\LCA_{\cT} (w,z)}$.    
    \end{itemize}
\end{itemize}

\paragraph{The main structure.}
We are now ready to formally describe our main structure, built on the notion of representatives described in~\Cref{sec:defining-alpha-comps}.
Recall that for every child-parent pair $\alpha \subseteq \mu$, we defined the $\alpha$-components in~\Cref{def:alpha-comps}.
For each $\alpha$-component $T$, we choose an arbitrary terminal $v_T \in T$.
The set of $\alpha$-representatives in $\mu$, denoted $R(\alpha,\mu)$, contains the following vertices:
\begin{itemize}
    \item For each $\alpha$-component $T$ of Type-1 or Type-2, we include $v_T$ in $R(\alpha,\mu)$.
    \item For each cycle $(\beta_0, \beta_1, \dots, \beta_k, \beta_0)$ not within a Type-1/2 $\alpha$-component,  and is adjacent to $\alpha$ through tree-edge $g = (\alpha,\beta_0)$, there are $k$ corresponding $\alpha$-components $T_1,\dots,T_k$ of Type-3 (see~\Cref{def:alpha-comps}).
    We take $v_{T_1}$ and $v_{T_k}$ into $R(\alpha,\mu)$.
\end{itemize}

To construct our main structure, we construct the skeleton hierarchy $\cT$ using $O(n)$ space.
Then, we invoke the procedure $\textsc{Store-Projection}(v,\alpha)$ (\Cref{alg:store-projection}) for every child-parent pair $\alpha \subseteq \mu$ and every $v \in R(\alpha,\mu)$.
Each such invocation allocates $O(1)$ more space (a tuple with an augmented explicit projection).
As $|R(\alpha,\mu)| \leq 2 \deg_{\cH_\mu}(\alpha)$, the short calculation in~\Cref{eq:rep-calculation} shows  this sums to $O(n)$ space overall.

\paragraph{Auxiliary data structures.}
For query efficiency, we further augment the hierarchy tree $\cT$ and every skeleton $\cH_\mu$ with standard auxiliary data structures. These do not add any new conceptual component. They are used only to support the operations required by our query algorithm in constant time.

These auxiliary structures are essentially the standard optimal structures for LCA and level-ancestor queries on trees:

\begin{fact}[\cite{HT84,BenderF04}]
\label{fact:tree-data-structures}
For any rooted tree $T$ on $N$ nodes, there is an $O(N)$-space data structure supporting the following operations in $O(1)$ time:
\begin{itemize}
    \item LCA: Given two nodes $u,v\in T$, report $\LCA_T(u,v)$;
    \item Level-ancestor: Given a node $u\in T$ and integer $0 \leq d \leq \depth_T (u)$, report the ancestor of $u$ at depth $d$.
\end{itemize}
\end{fact}

We apply Fact~\ref{fact:tree-data-structures} to the hierarchy tree $\cT$ itself.
Also, for every skeleton $\cH_\mu$ stored by the hierarchy, we apply Fact~\ref{fact:tree-data-structures} to the tree obtained by contracting each cycle of $\cH_\mu$ into a node (storing also the cyclic order in which its incident edges appeared as the edges emanating out of the cycle).
This is the essential data structure for working with cactus graphs, appearing e.g. in~\cite{DBLP:conf/soda/BaswanaP22,BaswanaP25,BhanjaPP26}.
For each oriented tree-edge $g=(\alpha,\beta)$ with $\alpha$ nonempty, we also store which case of~\Cref{def:alpha-comps} applies to the sub-skeleton rooted by $g$ in direction $\beta$.
Using this basic structure, it is straightforward to implement an $O(|\cH_\mu|)$ space data structure that supports all ``skeleton operations'' we need in constant time.
For example:
\begin{itemize}
    \item Given (the endpoints of) $\pi_\mu (v)$ and a nonempty node $\alpha$, identify the tied $\alpha$-components of $v$.
    \item Given canonical cut $C$ in $\cH_\mu$ and $a,b \in \mu$, determine whether $a$ and $b$ are on the same side of $C$.
    \item Given a Type-3 $\alpha$-component $T$, identify the cycle adjacent to $\alpha$ that gave rise to $T$, and the location of $T$ along this cycle.
\end{itemize}
As $\sum_{\mu \in \cT} |\cH_\mu| = O(n)$, these auxiliary structures only require additional $O(n)$ space.

\subsection{Query algorithm}\label{sec:algorithm-units}

Given a query $x,y \in V$, we define the \emph{query path} as the path $\LCA_{\cT} (x,y) = \mu_0 \subset \mu_1 \subset \cdots \subset \mu_\ell = V$ from $\LCA_{\cT}(x,y)$ to the root of $\cT$.

\paragraph{Assigning  representatives on the path.}
As a preliminary step, we \emph{assign} every representative vertex $v \in R(\mu_{i-1}, \mu_i)$, $i \in [1,\ell]$ to a node $\mu_j$ on the query path such that:
\begin{enumerate}
    \item $\mu_j$ is between $\mu_{i-1}$ and $\mu_0$ (i.e., $0 \leq j \leq i-1$).
    \item The augmented projection $\pi_{\mu_j} (v)$ of $v$ in $\cH_{\mu_j}$ is known.
    \item If $\mu_j$ is above $\mu_0 = \LCA_{\cT} (x,y)$ (i.e., if $j > 0$), then $v$ \emph{has at least one tied $\mu_{j-1}$-component} in $\cH_{\mu_j}$.
\end{enumerate}
We denote the set of representatives assigned to $\mu_j$ by $R_j$.
The point of this assignment process is precisely to ensure the last condition, allowing us to classify $v$ into $x,y$-units when we process $\mu_j$, since, when $j > 0$, it must have some tied $\mu_{j-1}$-component there. 

Let us explain how $v \in R(\mu_{i-1}, \mu_i)$ is assigned.
Recall that by invoking $\textsc{Store-Projection}(v,\mu_{i-1})$, we have stored the resulting tuple $(v,\mu_{i-1},\eta,\pi_\eta (v))$.
We take $\mu_j = \LCA_{\cT} (\mu_0, \eta)$.
\begin{itemize}
    \item As $\mu_{i-1}$ is a common ancestor of $\mu_0$ and $\eta$, and $\mu_j$ is their \emph{lowest} common ancestor, we have item (1).

    \item If $\mu_j = \eta$, then we have explicitly stored the augmented projection $\pi_{\mu_j} (v)$, which is non-trivial. Thus, when $j > 0$, $v$ must have some tied $\mu_{j-1}$-component.
    So, we have items (2) and (3) in this case.

    \item Otherwise, $\mu_j$ is a strict ancestor of $\eta$.
    Then $v$ has trivial projection in $\mu_j$ determined as $\pi_{\mu_j} (v) = \beta$, where $\beta$ is the child of $\mu_j$ on the way to $\eta$.
    By~\Cref{thm:proj-vertex}, every $\mu_j$-mincut keeps $v$ with $\beta$, so we can determine $\phi_{\mu_j} (v)$ as equal to $\phi_{\mu_j} (t)$ of every $t \in \beta$.
    (At preprocessing, for each non-root node $\beta \in \cT$, we also store $\phi_{parent_{\cT}(\beta)} (t)$ for some arbitrary $t \in \beta$.)
    As $\pi_{\mu_j} (v)$ is a node, $\phi_{\mu_j} (v)$ is the only augmentation required for this projection.
    Crucially, when $j > 0$, we have $\pi_{\mu_j} (v) = \beta \neq \mu_{j-1}$ (as otherwise $\mu_{j-1}$ would be a lower common ancestor of $\mu_0$ and $\eta$ than $\mu_j$).
    So, $v$ has a tied $\mu_{j-1}$-component (the one with $\beta$).  Thus, we have items (2) and (3) also in this case.   
\end{itemize}
Note that assigning $v$ and finding its augmented projection in the assigned node takes only $O(1)$ time (invoking one LCA and possibly one level-ancestor query on $\cT$).

\paragraph{Constructing the $x,y$-units.}
We now describe the main query procedure which recovers the $x,y$-units.
The procedure maintains a family $\cU$ of disjoint subsets of $V$, which gradually grows to become the full partition of $V$ into $x,y$-units.
It works by processing the nodes on the query path in their order $\mu_0, \mu_1, \dots, \mu_\ell$ from $\mu_0 = \LCA_{\cT} (x,y)$ towards $\mu_\ell = V$.
The following two invariants are maintained:
\begin{itemize}
    \item[{\bf (I1)}] At all times, $\cU$ forms the partition of $\bigcup \cU$ by the $x,y$-units, where $\bigcup \cU$ is the union of all $U \in \cU$.
    \item[{\bf (I2)}] Immediately after $\mu_i$ is processed, it holds that $\bigcup \cU = \mu_i \cup R_{\leq i}$, where $R_{\leq i} = \bigcup_{j=0}^i R_j$.
\end{itemize}

The algorithm for processing the LCA $\mu_0$ is~\Cref{alg:process-LCA}, and the one for processing a higher node $\mu_i$ is given in~\Cref{alg:process-ancestor}, with the helper procedure $\Call{IsMerged}{}$ given in~\Cref{alg:is-merged}.
\begin{algorithm}[H] 
\caption{Processing node $\mu_0 = \LCA_{\cT} (x,y)$}\label{alg:process-LCA}
\begin{algorithmic}[1]
\State Let $\alpha$ and $\beta$ be the nodes of $\cH_{\mu_0}$ with $x \in \alpha$, $y \in \beta$
\State Initialize $\cU$ as the family of $\alpha\beta$-components in $\cH_{\mu_0}$ \Comment{\Cref{lem:atomicity-LCA}}
\For{each $v \in R_0$} 
    \If{$\pi_{\mu_0} (v)$ is contained in  $\alpha\beta$-component $T$} 
        insert $v$ to the $U \in \cU$ s.t.\ $T \subseteq U$ \Comment{\Cref{lem:LCA-dichotomy}}
    \Else
        ~mark $v$ as \emph{split}
    \EndIf
\EndFor
\State Let $\cS$ be the partition of the \emph{split} vertices from $R_0$ by their $\phi_{\mu_0} (\cdot)$ values
\State $\cU \gets \cU \cup \cS$ \Comment{\Cref{lem:LCA-dichotomy,lem:distinctness of xy units}}
\end{algorithmic}
\end{algorithm}

\begin{algorithm}[H] 
\caption{Processing node $\mu_i$ s.t.\ $1 \leq i \leq \ell$}\label{alg:process-ancestor}
\begin{algorithmic}[1]
\State Denote $\mu := \mu_i$ and $\alpha := \mu_{i-1}$.
\For{each $\alpha$-component $T$ s.t. $v_T \in R(\alpha,\mu)$}
    \State Insert all vertices in $T$ to the $U \in \cU$ s.t.\ $v_T \in U$ \Comment{\Cref{lem: atomicity}}
\EndFor
\For{each $\alpha$-component $T$ s.t. $v_T \notin R(\alpha,\mu)$} \Comment{$T$ is Type-3 from the middle of the cycle}
    \State Let $(\beta_0, \beta_1, \dots, \beta_k, \beta_0)$ be the cycle such that $T = T_j$ for some $1 < j < k$.
    \If{$v_{T_1}$ and $v_{T_k}$ are in the same $U \in \cU$}
        insert all vertices of $T$ to this $U$ \Comment{\Cref{lem:cyc-dichotomy}}
    \Else
    ~$\cU \gets \cU \cup \{T\}$ \Comment{\Cref{lem:cyc-dichotomy}}
    \EndIf
\EndFor
\For{each $v \in R_i$} \Comment{$v$ has some tied $\alpha=\mu_{i-1}$-component as it is assigned to $\mu_i$}
    \If{\Call{IsMerged}{$v,\alpha,\mu,\cU$}}
        \State Let $T$ be an $\alpha$-component tied to $v$
        \State Insert $v$ to $U \in \cU$ s.t. $T \subseteq U$ \Comment{\Cref{lem:tied-comp-dichotomy}}
    \Else
        ~mark $v$ as \emph{split}
    \EndIf
\EndFor
\State Let $\cS$ be the partition of the \emph{split} vertices from $R_i$ by their $\phi_\mu (\cdot)$ values
\State $\cU \gets \cU \cup \cS$ \Comment{\Cref{lem:tied-comp-dichotomy,lem: extended-distinctness}}
\end{algorithmic}
\end{algorithm}

\begin{algorithm}[H] 
\caption{Determining the dichotomy of~\Cref{lem:tied-comp-dichotomy} (see the flowchart in~\Cref{fig:merge-split-flowchart})}\label{alg:is-merged}
\begin{algorithmic}[1]
\Procedure{IsMerged}{$v,\alpha,\mu,\cU$}
\If{$v$ has two tied $\alpha$-components $T_1,T_2$ contained in same $U \in \cU$}
    \Return \textsc{TRUE} \Comment{\Cref{lem:two-tied-comps}}
\EndIf
\If{$v$ has one tied $\alpha$-component $T$ and $\alpha \notin \pi_\mu (v)$}
    \Return \textsc{TRUE} \Comment{\Cref{lem:one-tied-simple-case}}
\EndIf
\If{$v$ has one tied $\alpha$-component $T$ and $\alpha \in \pi_\mu (v)$} \Comment{$\alpha$ is an endpoint of $\pi_\mu (v)$ by~\Cref{def:alpha-comps}}
    \If{there is no maximizer for $v$ w.r.t.\ $\alpha$}
        \Return \textsc{TRUE} \Comment{$T,v$ in same $x,y$-unit as $\cC_{T,v} = \emptyset$}
    \Else
        ~let $(w,z)$ and $C_M$ be the maximizer augmentation of $v$ w.r.t.\ $\alpha$
        \If{$z$ and $T$ are contained in same $U \in \cU$}
            \Return \textsc{TRUE} \Comment{\Cref{lem:using-maximizer}}
        \ElsIf{at least one of $x,y$ is not in $\LCA_{\cT} (w,z)$}
            \Return \textsc{FALSE} \Comment{\Cref{lem:wz mincut is larger than xymincut}}
        \ElsIf{both $x,y$ are on side with $z$ of $C_M$ in $\cH_{\LCA_\cT (w,z)}$}
            \Return \textsc{TRUE} \Comment{\Cref{lem:using-maximizer}}
        \EndIf
    \EndIf
\EndIf
\State \Return \textsc{FALSE}
\EndProcedure
\end{algorithmic}
\end{algorithm}

\paragraph{Correctness.}
For processing the LCA $\mu_0$, the maintenance of (I1) is justified by the lemmas in the comments on each line of pseudocode updating $\cU$. After processing $\mu_0$, (I2) clearly holds.

Consider some higher $\mu_i$ with $1 \leq i \leq \ell$.
In the following discussion, we use $\mu := \mu_i$ and $\alpha := \mu_{i-1}$ to match the given pseudocodes.
When we start processing $\mu_i$, it holds that $\bigcup \cU = \alpha \cup R_{\leq i-1}$ by (I2), and $R(\alpha,\mu) \subseteq R_{\leq i-1}$.
This means that we can indeed execute lines 1-7, which insert into $\bigcup \cU$ the different $\alpha$-components, i.e., all  vertices in $\mu \setminus \alpha$.
Thus, after lines 2–7 have been completed, we have $\bigcup \cU = \mu \cup R_{\leq i-1}$.
This shows that we can indeed execute the following lines 8-14, including the calls to $\textsc{IsMerged}()$, as every 
$\alpha$-component is found within $\cU$.
Lines 8-14 insert all vertices in $R_i$ into $\cU$, so indeed (I2) holds when we finish processing $\mu$.

We now discuss why (I1) is preserved, assuming the correctness of the $\textsc{IsMerged}()$ procedure for determining the tied component dichotomy of~\Cref{lem:tied-comp-dichotomy} (which we discuss afterwards).
The justifying lemmas for every update of $\cU$ appear as comments in the corresponding lines of the pseudocode.
There are two types of updates: adding vertices to existing sets $U \in \cU$ such that $U$ intersects $\mu$ (lines 3, 6 and 11), or creating entirely new sets in $\cU$ (lines 7 and 14).
While the correctness of the first-type updates (namely, the ``merged'' cases) is immediate from the justifying lemmas, the second-type updates (the ``split'' cases) might be trickier to see.
The delicate point to notice is that $R_{\leq i-1} \setminus \mu_i$ contains only vertices $u$ such that $\pi_\mu (u) = \alpha$, and hence no vertex in $R_{\leq i-1} \setminus \mu_i$ is tied to any $\alpha$-component. This is proved as follows.

\begin{claim}\label{claim:projection-is-alpha}
    If $u \in R_{\leq i-1} \setminus \mu_i$, then $\pi_{\mu_i} (u) = \mu_{i-1}$
\end{claim}
\begin{proof}
    Since $u \notin \mu_i$, it must belong to $R(\mu_{r-1},\mu_r)$ for some $r > i$.
    Let $\eta$ be the node where the explicit projection of $u$ is stored via $\textsc{Store-Projection}(u,\mu_{r-1})$.
    As $u \in R_j$ for some $j \leq i-1$, we have $\mu_j = \LCA_{\cT} (\mu_0, \eta)$, so $\mu_j$ is an ancestor of $\eta$.
    As $j < i < r$, this means that the path from $\mu_{r-1}$ to $\eta$ goes through $\mu_i$ and $\mu_{i-1}$.
    So this path implicitly determines that $\pi_{\mu_i} (u) = \mu_{i-1}$ (as explained after $\Call{Store-Projection}{}$ (\Cref{alg:store-projection}) is presented).
\end{proof}


With this observation, we justify (I1) for the second-type updates (the ``split'' cases) as follows.
\begin{itemize}
    \item Line 7: 
    This line corresponds to the ``split'' case of~\Cref{lem:cyc-dichotomy}, asserting that the $x,y$-unit of $T = T_j$ does not contain any $u$ who is not tied to $T$.
    Consider the state of $\cU$ immediately before executing Line 7, so $T \cap \bigcup \cU = \emptyset$ (since $T$ is a Type-3 $\alpha$-component from the middle of the cycle and has no stored representative).
    Fix $u \in \bigcup \cU$.
    If $u \in \mu$, then it is not in $T$, so $\pi_\mu (u)$ is a nonempty node in $\cH_\mu$ outside of $T$.
    Else, $u \in R_{\leq i-1} \setminus \mu$, so $\pi_\mu (u) = \alpha$ by~\Cref{claim:projection-is-alpha}.
    In any case, $u$ is not tied to $T$.
    So, by~\Cref{lem:cyc-dichotomy}, adding $T$ as a new set in $\cU$ indeed preserves (I1).

    \item Line 14: When this line is reached, $\cU$ consists only of sets $U$ such that $U \cap \mu \neq \emptyset$ or $U \subseteq R_{\leq i-1} \setminus \mu$.
    This is because the only vertices from $R_i$ previously inserted to sets in $\cU$ were the ``merged'' ones, and they went into sets containing their tied $\alpha$-components (and hence intersecting $\mu$).
    By~\Cref{lem:tied-comp-dichotomy},
    each ``split'' vertex $v \in R_i$, which is tied to some $\alpha$-component $T$, lies in an $x,y$-unit that neither intersects $\mu$ nor contains any vertex $u$ that is not tied to $T$.
    Thus, no vertex from the current $\bigcup \cU$ can be in $v$'s $x,y$-unit.
    We have that $\cS$ forms the partition of the ``split'' vertices in $R_i$ by the $x,y$-units, according to~\Cref{lem: extended-distinctness}.
    Thus, the update $\cU \gets \cU \cup \cS$ indeed maintains (I1).
\end{itemize}

Finally, we explain the correctness of the $\Call{IsMerged}{v,\alpha,\mu,\cU}$ procedure.
Note that when this procedure is called, we already have that $\mu \subseteq \bigcup \cU$, hence $\cU$ correctly determines the partition of $\mu$ by the $x,y$-units from (I1).
Thus, the correctness of most lines is immediately justified by the lemmas in the comments.
The only line that requires some attention is line 8.
For it, we have the following:
\begin{observation}
    $\LCA_{\cT} (w,z) =\{v \in V \mid \lambda_{v,z} \geq \lambda_{w,z}\}$
\end{observation}
\begin{proof}
    Denote $\nu := \LCA_{\cT} (w,z)$, so $\lambda_{w,z} = \lambda_\nu$.
    If $v \in \nu$, then any cut separating $v$ from $z$ is a $\nu$-cut, and hence $\lambda_{v,z} \geq \lambda_\nu$.
    If $v \notin \nu$, then the leaf of $v$ in $\cT$ is not in the subtree rooted by $\nu$, but $z$ does belong to that subtree. Thus, $\omega := \LCA_{\cT} (v,z)$ is a strict ancestor of $\nu$, so $\lambda_{v,z} = \lambda_{\omega} < \lambda_{\nu}$.
\end{proof}
Thus, $\lambda_{w,z} > \min\{\lambda_{x,z},\lambda_{y,z}\}$ iff one of $x,y$ does not belong to $\LCA_{\cT} (w,z)$.
This shows that~\Cref{lem:wz mincut is larger than xymincut} indeed applies when the condition of line 8 is met.
This concludes the correctness discussion.

\paragraph{Query time.}
Assigning the representatives to the sets $R_i$ takes $O(1)$ time per representative as previously explained, hence $O(n)$ time overall.

Processing $\mu_0$ takes $O(|\mu_0| + |R_0|)$ time, and processing a higher $\mu_i$ takes $O(|\mu_i \setminus \mu_{i-1}| + |R_i|)$ time, by using the standard auxiliary skeleton data structures for finding tied $\alpha$-components, locating $\alpha$-components along cycles, determining sides of canonical cuts, and so forth.
(The $|\mu_i \setminus \mu_{i-1}|$ term is for lines 2-7 in~\Cref{alg:process-ancestor}, and $|R_i|$ is for lines 8-14. Similarly, the term $|\mu_0|$ is for line 2 in~\Cref{alg:process-LCA}, and $|R_0|$ is for lines 3-7.)
As $\mu_{i-1} \subseteq \mu_i$ for each $i$, and the sets $\{R_i\}_{i=0,\dots,\ell}$ are disjoint, we get total query time of the order of
\begin{equation}\label{eq:query-time-calculation}
    |\mu_0| + \sum_{i=1}^\ell(|\mu_i|-|\mu_{i-1}|) + \sum_{i=0}^\ell |R_i| = |\mu_{\ell}| + |R_{\leq \ell}| \leq |V| + |V| = O(n).
\end{equation}

\section{Reporting a Topological Ordering of $x,y$-Units}\label{sec:reporting-ordering}

In this section, we extend the data structure of~\Cref{sec:reporting-units} to also yield a topological order of the $x,y$-units.
We start by presenting some general definitions and terminology regarding such orderings.

\begin{definition}
The notations $\beforeeq$ and $\before$ are
defined as follows.
\begin{itemize}
    \item Write $u\preceq_{(x,y)}v$ if every $x,y$-mincut with $v$ on the side of $x$ also has $u$ on the side of $x$.
    
    \item Denote $u \before v$ if $u \beforeeq v$ but $v \not\beforeeq u$.
    
    \item We call $u$ and $v$ \emph{comparable} w.r.t.\ $x,y$ if $u \beforeeq v$ or $v \beforeeq u$, \emph{incomparable} otherwise.

    \item     When $U,U' \subseteq V$ are ``atomic'' sets, namely, each is contained within one $x,y$-unit, we extend the notations $U \beforeeq U'$  and $U \before U'$ in the obvious way.

    \item  When $\cU$ is a collection of sets forming the partition of $\bigcup \cU$ into $x,y$-units, a \emph{topological ordering} of $\cU$ is an ordered list $U_1, U_2, \dots$ of all sets in $\cU$, such that for every $i < j$ it holds that $U_j \not\beforeeq U_i$.
\end{itemize}
\end{definition}

Whenever $x$ and $y$ are clear from context, we omit the subscript from the notation, and the ``w.r.t.\ $x,y$'', for ease of readability.
Note that we (arbitrarily) fixed the directionality between $x$ and $y$; intuitively, one may think of $x$ as being ``on the pointy side'' of $\befeq$, while $y$ is ''on the open side''.
We next summarize some trivial observations, which will be used implicitly.

\begin{observation}
    We have the following:
    \begin{itemize}
        \item The relation $\befeq$ is reflexive and transitive, i.e., it is a preorder.
        
        \item $u \befeq v$ and $v \befeq u$ iff $u$ and $v$ are in the same $x,y$-unit, i.e. $\phi_{x,y} (u) = \phi_{x,y} (v)$.

        \item $u \befeq v$ iff in every $x,y$-mincut with $u$ on the side of $y$, $v$ is also on the side of $y$.
        
        \item $u \not\befeq v$ iff there exists some $x,y$-mincut with $v$ on the side of $x$ and $u$ on the side of $y$.
    \end{itemize}
\end{observation}


\paragraph{Directionality augmentations.}
The construction of the data structure is precisely as in~\Cref{sec:ds-construction-units}, only we augment each explicitly stored projection $\pi_\mu (v)$ with additional $O(1)$ information regarding ``directionality properties'' of $\mu$-mincuts that correspond to it.
This information comes from the following lemma, which is well-known in the ``carcass world'' (e.g.~\cite{DBLP:conf/soda/BaswanaP22,BhanjaPP26}).
For completeness, we give the proof in~\Cref{sec:missing-proofs}.

\begin{lemma}\label{lem:numbering-for-tree-edges}
    Let $f$ be a tree-edge in the skeleton $\cH_\mu$, let $\gamma$ be an endpoint of $f$, and let $S$ be the sub-skeleton rooted by $f$ in direction $\gamma$.
    There is a numbering $\sigma_{f,\gamma} : V \to [n]$ such that, for every two vertices $u,v$ with $f \in \pi_
    \mu (u) \cap \pi_\mu (v)$, the following hold:
    \begin{itemize}
        \item $\sigma_{f,\gamma} (u) < \sigma_{f,\gamma} (v)$ implies that there is a $\mu$-mincut $A$ with $A \cap \mu = S$, $u \in A$ and $v \notin A$.
        \item $\sigma_{f,\gamma} (u) = \sigma_{f,\gamma} (v)$ if and only if $\phi_\mu (u) 
        = \phi_\mu (v)$.
    \end{itemize}
\end{lemma}
\def\SIGMANUMBERS{
\begin{proof}[Proof of~\Cref{lem:numbering-for-tree-edges}]
    We omit the subscript $f,\gamma$ for readability.
    Let $H$ be a directed graph with vertex set $V$, such that the directed edge $(u,v)$ exists if and only if every $\mu$-mincut $A$ such that $A \cap \mu = S$ and $v \in A$ also has $u \in A$.
    Contract each SCC (strongly connected component) of $H$, and find a topological ordering of the resulting DAG: $\sigma(v)$ is the place of $v$'s SCC in this ordering. 
    Observe that $H$ is transitively closed, i.e., if it has edges $(u,v)$ and $(v,w)$, it also has an edge $(u,w)$: Indeed, if $A$ is a $\mu$-mincut with $A \cap \mu = S$ and $w \in A$, then by $(v,w) \in H$  we have $v \in A$, and now by $(u,v) \in H$ we get $u \in A$.

    Now, let $u,v \in V$ with $f \in \pi_\mu (u) \cap \pi_\mu (v)$, so the canonical cut defined by $f$ distinguishes both $u$ and $v$ by~\Cref{thm:proj-vertex}.
    By~\Cref{lem: Distinctness}, it holds that $\phi_\mu (u) = \phi_\mu (v)$ iff no $\mu$-mincut $A$ with $A \cap \mu = S$ (i.e., a $\mu$-mincut represented by $f$) separates $u$ from $v$.
    But this happens iff both $(u,v)$ and $(v,u)$ are edges in $H$ by its definition, which is equivalent to $u,v$ being in the same SCC as $H$ is transitively closed, and this happens iff $\sigma(u) = \sigma(v)$.
    This proves the second item.
    For the first item, $\sigma(u) < \sigma(v)$ means $(v,u)$ is not an edge in $H$ (as it would violate the topological order), so there is a $\mu$-mincut $A$ s.t.\ $A \cap \mu = S$, $u \in A$ and $v \notin A$.
\end{proof}
} 

Whenever we explicitly store a projection $\pi_\mu (v)$ which is a proper path, we add the following augmentation to the previous ones:
For each node $\gamma$ which is one of the two endpoints of $\pi_\mu (v)$, let $f \in \pi_\mu (v)$ be the tree-edge adjacent to $\gamma$, and store $\sigma_{f,\gamma} (v)$.

\paragraph{New invariant: maintaining a topological ordering.}
We closely follow the algorithm for reporting the units from~\Cref{sec:algorithm-units}, but add a third invariant:
\begin{itemize}
    \item[{\bf (I3)}] Immediately after $\mu_i$ is processed, we have a topological ordering of $\cU$.
\end{itemize}
We maintain the ordered collection $\cU$ using a standard linear-space order-maintenance data structure~\cite{DBLP:conf/esa/BenderCDFZ02} supporting precedence queries in $O(1)$ time and insertion at a specified position in $O(1)$ amortized time.

\subsection{Processing the LCA}

\paragraph{Ordering the $\alpha\beta$-components.}
Let $\mu := \mu_0 = \LCA_{\cT} (x,y)$ and let $\alpha,\beta$ be its children so that $x \in \alpha$ and $y \in \beta$.
We traverse the cactus path $P_{\alpha\beta}$ in the direction from $\alpha$ to $\beta$, and report the $\alpha\beta$-components in the order they are encountered.
Whenever this traversal enters a cycle $\cO$ with entry node $\gamma$ and exit node $\delta$, we first report the components on one of the two semi-cycles in the order from $\gamma$ to $\delta$, then report the components on the other semi-cycle in the same manner. Denote the resulting list by $T_1, T_2 \dots, T_k$.
By inspecting~\Cref{fig:inside-lca}, it is not hard to see that this is a valid topological order:

\begin{lemma}
    The list $T_1,T_2, \dots, T_k$ is a topological order of the $\alpha\beta$-components.
\end{lemma}
\begin{proof}
    Let $T_i$ and $T_j$ with $i < j$.
    If there is some tree-edge of $P_{\alpha\beta}$ between $T_i$ and $T_j$, then any $\mu$-mincut represented by it is an $x,y$-mincut keeping $T_i$ with $\alpha \ni x$ and $T_j$ with $\beta \ni y$, so $T_j \not\befeq T_i$.
    Otherwise, $T_i$ and $T_j$ are from the same cycle $\cO$. If they are from the same semi-cycle, then take an $x,y$-mincut represented by a canonical cut with one cycle-edge between $T_i$ and $T_j$ in their semi-cycle, and the other cycle-edge from the other semi-cycle. As before, this cut shows that $T_j \not\befeq T_i$.
    Finally, it could be that $T_i$ and $T_j$ are from two different semi-cycles of $\cO$.
    In this case, we consider the canonical cut consisting of the cycle-edge immediately \emph{after} $T_i$ in its semi-cycle, and the one immediately \emph{before} $T_j$ in the other semi-cycle, and again arrive at $T_j \not\befeq T_i$.
\end{proof}

\paragraph{Handling $R_0$: clean-up and strategy.}
We now come to handle the vertices in $R_0$.
As in~\Cref{alg:process-LCA}, we merge every $v \in R_0$ whose projection is contained in one $\alpha\beta$-component $T_i$ into $T_i$, resulting in an extended topological ordering $U_1, U_2, \dots, U_k$ where each $U_i$ contains $T_i$ and the $R_0$-vertices merged with it.
We are then left with the ``split'' $R_0$ vertices. 
Again, as in~\Cref{alg:process-LCA}, these are grouped into new $x,y$-units according to their $\phi_\mu (\cdot)$ values.
The goal is to extend the current topological ordering $U_1, U_2, \dots, U_k$ to also include these new units.
For each new unit consisting of split vertices from $R_0$, choose one arbitrary vertex; denote the set of chosen split vertices by $S_0$, so every two vertices in $S_0$ have different $\phi_\mu (\cdot)$ values.

Thus, our goal is to extend the topological ordering to include not only $U_1, \dots, U_k$, but also each vertex from $S_0$, with each $v \in S_0$ extended afterwards to a set including all $u \in R_0$ s.t.\ $\phi_\mu (u) = \phi_\mu (v)$.
The $S_0$ vertices will be placed ``in between'' the sets of the already known ordering $U_1, \dots, U_k$, as follows:
\begin{itemize}
    \item We open a new \emph{bucket} $B_i$ to the left of every $U_i$.
    \item We first place $S_0$ within the buckets, with each bucket still internally unordered, so that no violations are created between the buckets and the existing sets, nor between the different buckets.
    \item Finally, we create an internal valid topological order inside each bucket.
\end{itemize}

\paragraph{Placing $S_0$ in the buckets.}
Fix some $v \in S_0$.
As $v$ is a ``split'' vertex, its projection $\pi_\mu (v)$ must be a proper path that intersects the cactus path $P_{\alpha\beta}$.
We orient this intersection in the direction from $\alpha$ to $\beta$ (and this orientation is well-defined since a proper path cannot go through two different semi-cycles of the same cycle).
If the intersection ends inside some $\alpha\beta$-component $T_i$, we call $T_i$ the \emph{rightmost} component of $v$ and denote $T_r (v) = T_i$.
Else, the intersection ends in an entry node of some cycle $\cO$ on $P_{\alpha\beta}$.
In this case, we take $T_r (v)$ to be the first $\alpha\beta$-component from $\cO$ by order of appearance in $T_1, \dots T_k$.
(Note that $T_r (v)$ is never $T_1$.)
By the projection theorem~\Cref{thm:proj-vertex}, we have the following:
\begin{observation}\label{obs:rightmost-component-in-LCA}
    For $1 < j \leq k$, let $C_j$ be the canonical cut corresponding to the valid partition of $\mu$ into $T_1 \cup \cdots \cup T_{j-1}$ (containing $\alpha \ni x$) and $T_j \cup \cdots \cup T_k$ (containing $\beta \ni y$).
    Suppose $T_r (v) = T_i$.
    Then:
    \begin{itemize}
        \item $C_i$ distinguishes $v$.
        \item Whenever $j > i$, $C_j$ keeps all of $\pi_\mu (v)$ on the side with $\alpha$.
    \end{itemize}
\end{observation}

We place $v$ in the bucket $B_i$ such that $T_i = T_r (v)$.

\begin{lemma}
    The placement of vertices from $S_0$ in the buckets corresponding to their rightmost $\alpha\beta$-components does not violate the topological ordering.
\end{lemma}
\begin{proof}
    Let $v \in S_0$ with $T_r (v) = T_i$.
    As $C_i$ distinguishes $v$, we get two $\mu$-mincuts represented by $C_i$, one keeping $v$ with $T_1 \cup \cdots T_{i-1}$ on the side of $x$, and the other keeping it with $T_i \cup \cdots \cup T_k$ on the side of $y$.
    This shows that when $j < i$ then $v \not\befeq T_j$ (and hence $v \not\befeq U_j$), and when $j \geq i$ then $v \not\afteq T_j$ (and hence $v \not\afteq U_j$).
    So, it remains to check the comparability between $v$ and vertices $u$ placed in different buckets $B_j$ with $j \neq i$.
    If $j < i$, then we take a $\mu$-mincut $A$ corresponding to $C_i$ such that $x, u\in A$ and $v, y \notin A$, which exists by~\Cref{obs:rightmost-component-in-LCA}.
    If $j > i$, then take a $\mu$-mincut $B$ corresponding to $C_j$ such that $x,v \in B$ and $u,y \notin B$, again by~\Cref{obs:rightmost-component-in-LCA}.
    This shows that indeed $v \not\befeq u$ when $j < i$, and $v \not\afteq u$ when $j > i$.
\end{proof}

\paragraph{Internally ordering each bucket.}
Fix a bucket $B := B_i$, and denote $T = T_i$, so $T$ is the rightmost $\alpha\beta$-component of every $v \in B$.
If $\pi_\mu (v)$ intersects $T$, then denote by $\gamma_v$ the endpoint of $\pi_\mu (v)$ within $T$.
Else, $\pi_\mu (v)$ ends at the entry node of the cycle $\cO$ with $T$ being the first $\alpha\beta$-component on one of its semi-cycles.
In this case, the endpoint $\gamma_v$ is this entry node to $\cO$.
Denote by $f_v$ the tree-edge of $\pi_\mu (v)$ adjacent to $\gamma_v$.

Since $T$ is the rightmost $\alpha\beta$-component of every vertex in $B$, and these are ``split'' vertices, we have the following:
\begin{itemize}
    \item If $T$ is the first $\alpha\beta$-component on one of the semi-cycles of some cycle $\cO$, all projections of vertices in $B$ must share the tree-edge of $P_{\alpha\beta}$ preceding $\cO$.

    \item Else, all projections of vertices in $B$ must share the last edge from $P_{\alpha\beta}$ before $T$ (again, viewing $P_{\alpha\beta}$ as oriented from $\alpha$ to $\beta$).
\end{itemize}
Thus, the edge $f_v$ is uniquely determined from $\gamma_v$ (as the tree-edge adjacent to $\gamma_u$ on the unique proper path going from the common edge from $P_{\alpha\beta}$ to $\gamma_v$).
Namely, two vertices $u,v \in B$ have $\gamma_u = \gamma_v$ iff $f_u = f_v$.

The following lemma allows us to determine some valid ordering between any two vertices $u,v \in B$.

\begin{lemma}[Internal bucket order in the LCA]\label{lem:order-inside-bucket-LCA}
    Let $u,v \in B$.
    \begin{enumerate}[label=(\roman*)]
        \item If $f_u \notin \pi_\mu (v)$, then $u \not\befeq v$.
        \item Suppose $f = f_u = f_v$ and $\gamma = \gamma_u = \gamma_v$.
        If $\sigma_{f,\gamma} (u) < \sigma_{f,\gamma} (v)$ then $u \not\befeq v$.
    \end{enumerate}
\end{lemma}
\begin{proof}
    Item (i): 
    In this case, $f_u$ must lie inside $T$; otherwise, $f_u$ could only be a common tree-edge from $P_{\alpha\beta}$ preceding $T$, but then $\pi_\mu (v)$ does not contain this common edge, a contradiction.
    Let $S_u$ be the sub-skelton of $f_u$ in direction $\gamma_u$, so $S_u \subseteq T$.
    Note that $\pi_\mu (v)$ cannot intersect $S_u$: otherwise, in order to leave $S_u$ and go outside of $T$, $\pi_\mu (v)$ must go through $f_u$, but $f_u \notin \pi_\mu (v)$.
    Thus, by~\Cref{thm:proj-vertex}, there exists a $\mu$-mincut $A$ with $A \cap \mu = S_u \subseteq T$, $u \in A$ and $v,x,y \notin A$ (with $x,y \notin A$ because $\alpha$ and $\beta$ are not in $S_u$).

    Now, let $X$ be an $(x,y)$-mincut with $v \in X$, $T \subseteq \overline{X}$; such a cut exists by the first item of~\Cref{obs:rightmost-component-in-LCA}, since $T$ is the rightmost $\alpha\beta$-component of $v$.
    Then $A \setminus X$ is a $\mu$-cut (separating $S_u$ from $x$), and $X \setminus A$ is an $(x,y)$-cut.
    By posimodularity (\Cref{lem:sub-posi-general}), $X \setminus A$ is an $(x,y)$-mincut, with $v$ in and $u$ out, so $u \not\befeq v$.

    Item (ii): If $f$ is contained within $T$, then define $S_u$ as in the previous item.
    Now, as $\sigma_{f,\gamma}(u) < \sigma_{f,\gamma} (v)$, there exists a $\mu$-mincut $A$ with $A \cap \mu = S_u \subseteq T$, $u \in A$ and $v \notin A$.
    Continue as in Item (i) to obtain $u \not\befeq v$.
    
    Otherwise, $f$ is a tree-edge of $P_{\alpha\beta}$, with $\gamma$ being its endpoint closer to $\beta$.
    As $\sigma_{f,\gamma}(u) < \sigma_{f,\gamma} (v)$, there is a $\mu$-mincut $A$ represented by $f$ that keeps $v$ on the side of $x$ and $u$ on the side of $y$, so again $u \not\befeq v$.
\end{proof}

We now explain how we can get a valid ordering for the entire bucket $B$.
Let $\delta$ be the node in which $T$ intersects $P_{\alpha\beta}$.
Consider the tree obtained by starting with $T$, contracting all cycles within it, and rooting at $\delta$.
For every node $\gamma \in T$, let $d_T (\gamma)$ be its depth in this rooted tree (or, if $\gamma$ was a cycle node, the depth of the supernode to which it was contracted).
Note that almost every $\gamma_v$ for $v \in B$ is indeed a node in $T$; the only exception is that $\gamma_v$ might be the entry node of the cycle with $T$ being the first $\alpha\beta$-component in one of its semicycles; in this case we define $d_T (\gamma_v) = -1$.
Since the projections of vertices in $B$ are proper paths ending at tree-edges and that pass $\delta$ or end at the entry node of the cycle containing $\delta$ as a first node in one semi-cycle, we have that if $d_T (\gamma_v) \leq d_T (\gamma_u)$ and $\gamma_v \neq \gamma_u$, then $f_u \notin \pi_\mu (v)$.
Thus, to order the bucket $B$, we first group the vertices by their $\gamma_v$ endpoints (and hence also their $f_v$ last edges), report the groups in increasing order of depth, where the vertices within each group are reported in decreasing order of $\sigma_{f,\gamma}$ (where $\gamma$ and $f$ are the common endpoint and its adjacent tree-edge of every vertex in the group).
\Cref{lem:order-inside-bucket-LCA} ensures that this is indeed a valid topological ordering.

Once every bucket is internally ordered, we obtain a valid topological order on the partition of all of $\mu \cup R_0$ by the $x,y$-units, hence ensuring (I3) as required.

\paragraph{Running time.}
The main implementation details regarding the running time concern the way we create the internal order within each bucket. Let $B$ be a bucket associated with $T$, and let $U$ be the set in $\cU$ containing $T$ and all vertices from $R_0$ merged to $T$.
First, we mark all endpoints $\gamma_v$ of  $v \in B$ within $T$.
Then, we execute a BFS search in the tree obtained from $T$ by contracting cycles.
The order in which the marked nodes are found sorts them by $d_T (\cdot)$. (If there is an extra node $\gamma$ which is the entry cycle node, so $d_T (\gamma) = -1$, we place it before the rest).
The internal sorting of vertices with projection ending at the same tree-edge $f$ and marked endpoint $\gamma$ according to $\sigma_{f,\gamma} (\cdot)$ can be computed during preprocessing and attached to $\gamma$ (each stored projection is attached to the lists at its two endpoints, so the space remains bounded by $O(n)$).
Thus when a marked node $\gamma$ is reached by the BFS, we scan its sorted list and report the vertices from $B$ there in (reverse) order of appearance.
The marking plus BFS takes $O(|B| + |T|)$ time, so summed over all $B$, this is at most $O(|R_0| + |\mu_0|)$.
As for scanning the pre-sorted lists at each endpoint: this takes a total of $O(|P_0|)$ time where $P_0$ denotes the set of projections explicitly stored at $\mu_0$, since each such projection is visited at most twice in this scanning process.
Thus, the total running time for processing $\mu_0$ is $O(|\mu_0| + |R_0| + |P_0|)$.

\subsection{Processing a node above the LCA}

Let $\mu := \mu_i$ with $i \geq 1$ be a node on the query path above the LCA, and denote by $\alpha := \mu_{i-1}$ its child with $x,y \in \alpha$.
By the maintained invariants (I1-3), when we start processing $\mu$, the known collection $\cU$ correctly defines the partition of $\bigcup \cU = \mu_{i-1} \cup R_{\leq i-1} \supseteq R(\alpha,\mu)$ by $x,y$-units, and is ordered in a valid topological ordering $U_1, \dots, U_h$.

\paragraph{Extending the ordering to $\mu \cup R_{\leq i-1}$.}

As in~\Cref{alg:process-ancestor}, we first merge each $\alpha$-component $T$ with $v_T \in R(\alpha,\mu)$ into the set $U \in \cU$ containing $v_T$.
We then treat the Type-3 $\alpha$-components coming from the middle of a cycle $\cO$.
Again, if the boundary components of $\cO$ were merged to the same $U \in \cU$, then we merge all of the $\alpha$-components originating from $\cO$ to this $U$.
Otherwise, denote the cycle components originating from $\cO$ by $T_1, \dots, T_k$ in order of their appearance on $\cO$, such that the boundary component $T_1$ precedes $T_k$ in the known ordering $U_1, \dots, U_h$.
As before, we should insert $T_j$ for $j = 2,3,\dots, k-1$ as $k-2$ new sets in $\cU$, but now we have to place them in between the known sets while maintaining a valid topological ordering.
The following lemma shows that we can place $T_2, T_3, \dots, T_{k-1}$, \emph{in this order}, anywhere between $T_1$ and $T_k$.

\begin{lemma}[Comparability of cycle components] \label{lem:comp-cycle-components}
    Let $g = (\alpha, \beta)$ be a tree-edge adjacent to $\alpha$, with $\beta = \beta_0$ on cycle $\cO = (\beta_0, \beta_1, \dots, \beta_k, \beta_0)$.
    Denote the corresponding Type-3 $\alpha$-components, according to the cycle order in $\cO$, by $T_1,\dots,T_k$.
    Suppose that $T_1 \not\afteq T_k$.
    Then the following hold.
    \begin{enumerate}[label=(\roman*)]
        \item For every $1 \leq i < j \leq k$ it holds that $T_i \not\afteq T_j$.
        \item Let $v$ be a vertex which is not tied to any of $T_1, \dots, T_k$ and let $1 \leq i \leq k$.
        \begin{itemize}
            \item If $v \not\befeq T_1$ then $v \not\befeq T_i$.
            \item If $v \not\afteq T_k$ then $v \not\afteq T_i$.
        \end{itemize}
    \end{enumerate}
\end{lemma}
\begin{proof}
    Item (i):
    Let $A$ and $A'$ be $\mu$-mincuts with $A \cap \mu = T_1 \cup \cdots \cup T_i$ and $A' \cap \mu = T_{i+1} \cup \cdots \cup T_k$.
    Such $A$ and $A'$ exist, as they partition $\mu$  corresponding to canonical cuts of either two non-adjacent cycle-edges from $\cO$, or one tree-edge leaving $\cO$.
    Observe that $x,y \notin A \cup A'$.
    We use $A$ and $A'$ in steps as follows.
    
    Let $X$ be an $(x,y)$-mincut with $T_1 \subseteq X$ and $T_k \subseteq \overline{X}$.
    Then $A \cap X$ is a $\mu$-cut (separating $T_1$ from $T_k$), and $X' := X \cup A$ is an $(x,y)$-cut.
    By submodularity (\Cref{lem:sub-posi-general}), $X'$ is an $(x,y)$-mincut with $T_i \subseteq X'$ and $T_k \subseteq \overline{X'}$.
    Now, $A' \setminus X'$ is a $\mu$-cut (separating $T_k$ from $T_i$), and $X'' := X' \setminus A'$ is an $(x,y)$-cut.
    By posimodularity (\Cref{lem:sub-posi-general}), $X''$ is an $(x,y)$-mincut, with $T_i \subseteq X''$ and $T_j \subseteq \overline{X''}$, as needed.
    
    Item (ii):
    We show the first sub-item (the second is symmetric).
    The proof is essentially identical to the `otherwise' case in~\Cref{lem:cyc-dichotomy}.
    Let $H$ be the sub-skeleton rooted by $g$ in direction $\beta$.
    As $\pi_\mu (v)$ cannot be contained within the cycle $\cO$ (by~\Cref{thm:proj-vertex}), it must have an endpoint outside $H$.
    Thus, by~\Cref{thm:proj-vertex}, there is a $\mu$-mincut $A$ with $A \cap 
    \mu = H$ and $v \notin A$, and note that $x,y \notin A$ as $\alpha$ is not in $H$.
    Now, let $X$ be an $(x,y)$-mincut with $T_1 \subseteq X$ and $v \notin X$.
    Then $X \cap A$ is a $\mu$-cut (separating $T_1$ from $y$), and $X \cup A$ is an $(x,y)$-cut.
    By submodularity (\Cref{lem:sub-posi-general}), $X \cup A$ is an $(x,y)$-mincut, keeping $T_i$ with $x$ and $v$ with $y$.
\end{proof}

From this point on, we use $U_1, \dots, U_h$ to denote the resulting topological ordering on all of $\mu \cup R_{\leq i-1}$, and our task is thus to extend it to also include $R_i$.

\paragraph{Handling $R_i$: clean-up and strategy.}
We now come to handle the vertices in $R_i$.
As before, using the procedure $\Call{IsMerged}{}$, we determine the ``merged'' or ``split'' dichotomy for each $v \in R_i$, the first kind being merged to existing sets in $\cU$ containing their tied $\alpha$-components, and the latter partition by their $\phi_\mu (v)$ values to form new sets that are now inserted to $\cU$.
Our task is therefore to extend the $U_1, \dots, U_h$ by placing these new sets in between the sets in this ordering.

For each such new set of ``split'' $R_i$ vertices with the same $\phi_\mu (\cdot)$ value, choose one vertex, and collect all chosen vertices as $S_i$, so every two vertices in $S_i$ have different $\phi_\mu (\cdot)$ values.
The $S_i$ vertices will be placed ``in between'' the sets of the already known ordering $U_1, \dots, U_h$, as follows:
\begin{itemize}
    \item We open \emph{two new buckets} for every $U_j$, placed to its left and to its right.
    \item We first place $S_i$ within the buckets, with each bucket still internally unordered, so that no violations are created between the buckets and the existing sets, nor between the different buckets.
    So now, between a consecutive pair $U_j, U_{j+1}$ in the current ordering, there are two new buckets: the right bucket of $U_j$, and the left bucket of $U_{j+1}$, with the former appearing before the latter.

    \item The bucketing scheme for a given ``split'' representative $v \in S_i$ is the following:
    \begin{itemize}
        \item If $v$ has two tied $\alpha$-components $T_1$ and $T_2$, with $T_1$ appearing before $T_2$ in the current ordering, we place $v$ in the left bucket next to $T_2$.
        \item If $v$ has one tied $\alpha$-component $T$, and the vertex $z$ from the maximizer is found to the right (resp., left) of $T$, then we place $v$ in the bucket to the right (resp., left) of $T$.
    \end{itemize}
    
    \item Finally, we create an internal valid topological order inside each bucket.
\end{itemize}

\paragraph{Determining comparability with tied $\alpha$-components.}
Let $v \in S_i$, and $T$ be an $\alpha$-component tied to $v$.
Our first insight is that $v$ and $T$ are comparable:

\begin{lemma}[Tied component is comparable]\label{lem:tied-is-comparable}
    If an $\alpha$-component $T$ is tied to $v$, then $T$ and $v$ are comparable.
\end{lemma}
\begin{proof}
    If not, then there are $(x,y)$-mincut $X$ and $(y,x)$-mincut $Y$ such that $v \in X \cap Y$ and $T \subseteq \overline{X \cup Y}$.
    Since $v$ is tied to $T$, there is also a $\mu$-mincut $A$ with $A \cap \mu = T$ and $v \in A$ (\Cref{obs:tied-comp}).
    Then $X \setminus (Y \cup A)$ and $Y \setminus (X \cup A)$ are $x,y$-cuts, while $A \setminus (X \cup Y)$ is a $\mu$-cut (separating $T$ from, say, $x$).
    So, by~\Cref{lem: simple-four-comps}, $\emptyset = X \cap Y \cap A \ni v$, a contradiction.
\end{proof}

We now turn to determining the correct order between $v$ and $T$.

\begin{lemma}[Comparability with two tied components]\label{lem:two-comps-ordering}
    If $v$ is tied to two $\alpha$-components $T_1$ and $T_2$, then either $T_1, T_2$ and $v$ are all in the same $x,y$-unit, or
    $T_1 \bef v \bef T_2$, or $T_2 \bef v \bef T_1$.
\end{lemma}
\begin{proof}
    Suppose $T_1,T_2,v$ are not all in one $x,y$-unit, so by~\Cref{lem:two-tied-comps}, they are in three different units.
    As $v$ is comparable to both $T_1$ and $T_2$ (\Cref{lem:tied-is-comparable}), we should rule out the possibility that $v \bef T_1,T_2$ or that $v \aft T_1,T_2$.
    Seeking contradiction, suppose one of them holds, say $v \bef T_1,T_2$ (the other is symmetric, by swapping $x$ and $y$).
    Then, there are two $(x,y)$-mincuts $X_i$ with $v \in X_i$ and $T_i \subseteq \overline{X_i}$, for $i \in \{1,2\}$.
    By submodularity (\Cref{lem:sub-posi-general}), $X = X_1 \cap X_2$ is an $(x,y)$-mincut with $v \in X$ and $T_1,T_2 \subseteq \overline{X}$.
    As $v$ is tied to $T_1$ and $T_2$, by~\Cref{obs:tied-comp} there are two $\mu$-mincuts $A_j$ with $A_j \cap \mu = T_j$ and $v \in A_j$, for $j \in \{1,2\}$.
    Then $X \setminus (A_1 \cup A_2)$ is an $(x,y)$-cut, while $A_1 \setminus (A_2 \cup X)$ and $A_2 \setminus (A_1 \cup X)$ are $\mu$-cuts (both separating $T_1$ and $T_2$).
    So, by~\Cref{lem: simple-four-comps}, $\emptyset = X \cap A_1 \cap A_2 \ni v$, a contradiction.
\end{proof}

\begin{lemma}[Comparability with one tied component] \label{lem:comp-with-one-tied-component}
    Suppose vertex $v$ has exactly one tied $\alpha$-component $T$, and that $v$ and $T$ are in different $x,y$-units.
    Let $M$ be the $(w,z)$-mincut for $w,z \in \alpha$ which is the maximizer of $\cC_{T,v}$ from~\Cref{lem:maximizer}.
    If $z \not\befeq T$ (resp., $z \not\afteq T$), then $v \not\befeq T$ (resp., $v \not\afteq T$).
\end{lemma}
\begin{proof}  
    (The `respectively' part is proved by swapping $x,y$.)
    Let $X$ be an $(x,y)$-mincut with $T \subseteq X$ and $z \notin X$.
    If $v \notin X$ then we are done.
    Otherwise, as $v$ and $T$ are not in the same $x,y$-unit, $M \cap \{x,y\} \neq \emptyset$ by~\Cref{lem:using-maximizer}.  
    Now we proceed exactly as in direction $(\Leftarrow)$ of~\Cref{lem:using-maximizer} to deduce that $X \cap M$ must be an $(x,y)$-mincut, keeping $T$ with $x$ and $v$ with $y$. 
    (The only difference is that here, $x \in X$ and $y \notin X$ is enforced by using $z \not\befeq T$, while in the proof of~\Cref{lem:using-maximizer}($\Leftarrow)$ this was assumed without loss of generality).
\end{proof}

\paragraph{Placing $S_i$ into buckets.}
As explained in the strategy, if $v$ has two tied $\alpha$-components $T_1$ and $T_2$, with $T_1$ appearing before $T_2$ in the current ordering, we place $v$ in the left bucket created for the set containing $T_2$ in the current ordering.
If $v$ has one tied $\alpha$-component $T$, then we place it in the bucket next to the set containing $T$ which is in the direction towards the location of $z$ in the current order.

Thus, by the above lemmas, our bucketing strategy indeed places $v$ on the correct side of its tied $\alpha$-component(s).
Furthermore, the two buckets to the left and to the right of each set $U$ do not create any conflict, because if $v$ is in the left bucket and $u$ in the right one, then $v \prec U \prec u$ so in particular $u \not\befeq v$.

It remains to show that a bucket associated with some known $U_i$ does not create conflicts (i) with another $U_j$, nor with (ii) the buckets of $U_j$.
These correspond precisely to items (i) and (ii) of the following lemma.

\begin{lemma}\label{lem:buckets-correctness-lemma}
    Let $v$ be a vertex tied to $\alpha$-component $T$, and $u$ be a vertex which is \emph{not} tied to $T$.
    \begin{enumerate}[label=(\roman*)]
        \item If $u \not\befeq T$ (respectively, $u \not\afteq T$), then $u \not\befeq v$ (respectively, $u \not\afteq v$).
        \item If $u$ is tied to $\alpha$-component $T'$ and $T' \not\befeq T$ (respectively, $T' \not\afteq T$), then $u \not\befeq v$ (respectively, $u \not\afteq v$).
    \end{enumerate}
\end{lemma}
\begin{proof}
    (The `respectively' part is proved by swapping $x,y$.)
    Since $v$ is tied to $T$ and $u$ is not, there is a $\mu$-mincut $A$ with $A \cap \mu = T$, $v \in A$, $u,x,y \notin A$ (\Cref{obs:tied-comp}).
    We will use $A$ to prove both items.

    (i) Let $X$ be an $(x,y)$-mincut with $T \subseteq X$, $u \in \overline{X}$.
    Then $X \cap A$ is a $\mu$-cut (separating $T$ from $y$), and $X \cup A$ is an $(x,y)$-cut.
    By submodularity (\Cref{lem:sub-posi-general}), $X' := X \cup A$ is an $(x,y)$-mincut with $v \in X'$, $u \notin X'$, as needed.

    (ii)
    By Item (i) (with $u$  replaced by $T'$), there is an $(x,y)$-mincut $X'$ with $v \in X'$, $T' \subseteq \overline{X'}$.
    Since $u$ is tied to $T'$, there is a $\mu$-mincut $B$ with $B \cap \mu = T'$, $u \in B$, $x,y\notin B$.
    Note that $A \setminus B$ and $B \setminus A$ are $\mu$-cuts (both separate $T$ and $T'$), so by posimodularity (\Cref{lem:sub-posi-general}), $B' := B \setminus A$ is a $\mu$-mincut with $u \in B'$, $v,x,y \notin B'$.
    Now, $B' \setminus X'$ is a $\mu$-cut (separating $T'$ from $x$), and $X' \setminus B'$ is an $(x,y)$-cut.
    By posimodularity again, $X' \setminus B'$ is an $(x,y)$-mincut with $v$ inside and $u$ outside, as needed.
\end{proof}

For (i), note that $U_j$ must contain some vertex $u \in \mu \cup R_{\leq i-1}$, so $u$ is not tied to any $\alpha$-component $T$ contained in $U_i$: If $u \in \mu$ this is clear, and if $u \in R_{\leq i-1} \setminus \mu$ this is by~\Cref{claim:projection-is-alpha}.


For (ii), note that a vertex $u$ from a bucket of $U_j$ and a vertex $v$ from a bucket of $U_i$ cannot agree on their tied $\alpha$-components.
Indeed, if they agreed, then both would have a tied component in $U_i$ and in $U_j$, but this means we would place them in the same bucket (namely, the left bucket of rightmost set among $U_i,U_j$), a contradiction.

This concludes the validity argument for our bucketing strategy.

\paragraph{Subdividing the buckets by tied $\alpha$-components.}
We now focus on determining a valid internal ordering within each bucket.
Let $B$ be a bucket of some known set $U = U_i$ in the ordering.
Note that $U$ could contain several different $\alpha$-components.
However, a vertex $v \in B$ can be tied to only one of these $\alpha$-components; otherwise, it would have two tied $\alpha$-components merged to the same $x,y$-unit, so it would be merged with them into $U$, a contradiction.
We first observe that if $u,v \in B$ have different tied $\alpha$-components within $U$, then $u$ and $v$ are incomparable.

\begin{lemma}\label{lem:different-tied-components-in-one-bucket}
    Let $U$ be an $x,y$-unit which contains two different $\alpha$-components $T$ and $T'$.
    Let $v$ be tied to $T$, and $u$ be tied to $T'$.
    If $u,v \prec U$ (respectively, $u,v \aft U$), then $u,v$ are incomparable w.r.t.\ $x,y$.
\end{lemma}
\begin{proof}
    (The `respectively' part is proved by swapping $x,y$.)
    We will show that $v \not\befeq u$, and $u \not\befeq v$ is symmetric.
    Let $X$ be an $(x,y)$-mincut with $u,v \in X$, $U \subseteq \overline{X}$.
    By~\Cref{obs:tied-comp}, there is a $\mu$-mincut $A$ with $A \cap \mu = 
    T$, $v \in A$, $x,y \notin A$.
    Note that $u$ cannot be tied to $T$, because then we would have $u \in U$ by~\Cref{lem:two-tied-comps}.
    Thus, it must be that $u \notin A$ by~\Cref{thm:proj-vertex}.
    Now, $A \setminus X$ is a $\mu$-cut (separating $T$ from $y$), and $X \setminus A$ is an $(x,y)$-cut.
    By posimodularity (\Cref{lem:sub-posi-general}), $X \setminus A$ is an $(x,y)$-mincut, keeping $u$ inside and $v$ outside.
\end{proof}

Thus, we can divide $B$ into sub-buckets according to the corresponding $\alpha$-components within $U$, and order these sub-buckets arbitrarily without creating conflict.
We therefore have to focus on the internal order within each sub-bucket.
To simplify the notation, from here on we will use the letter $B$ for one arbitrary sub-bucket associated with an $\alpha$-component $T$ within $U$.

\paragraph{Internally ordering a sub-bucket.}
Because every $v \in B$ is tied to $T$ but not merged with it, the projection $\pi_\mu (v)$ must be a proper path which goes through the rooting tree-edge of $T$, then continues inside $T$.
Denote by $\gamma_v$ the endpoint within $T$, and by $f_v$ the tree-edge of $\pi_\mu (v)$ adjacent to it.
Note that $f_v$ is uniquely determined from $\gamma_v$ because there is a unique proper path from the rooting edge of $T$ to $\gamma_v$.
Namely, for two vertices $u,v \in B$, we have $\gamma_u = \gamma_v$ iff $f_u = f_v$.

The following lemma allows us to determine a valid order for every two vertices of the sub-bucket $B$, essentially in the same manner as we did when processing the LCA
(\Cref{lem:order-inside-bucket-LCA}).

\begin{lemma}[Comparability with shared tied component] \label{lem:ancestor-comparability-shared-tied-comp}
    Let $u$ and $v$ be two vertices tied to $\alpha$-component $T$ such that $u, v \bef T$ (resp., $u,v \aft T$).
    Let $\gamma_u$ denote an endpoint of $\pi_\mu (u)$ within $T$, and $f_u$ denote the tree-edge of $\pi_\mu (u)$ adjacent to $\gamma_u$.
    Denote $\gamma_v$ and $f_v$ analogously.
    \begin{enumerate}[label=(\roman*)]
        \item If $f_u \notin \pi_\mu (v)$, then $u \not\befeq v$ (resp., $u \not\afteq v$).
        Symmetrically, if $f_v \notin \pi_\mu (u)$, then $v \not\befeq u$ (resp., $v \not\afteq u$).
        \item Suppose $f = f_u = f_v$ and $\gamma = \gamma_u = \gamma_v$.
        If $\sigma_{f,\gamma} (u) < \sigma_{f,\gamma} (v)$ then $u \not\befeq v$ (resp., $u \not\afteq v$).
        Symmetrically, if $\sigma_{f,\gamma} (v) < \sigma_{f,\gamma} (u)$, then $v \not\befeq u$ (resp., $v \not\afteq u$).
    \end{enumerate}
\end{lemma}
\begin{proof}
    (The `symmetrically' parts are proved by swapping $u,v$, and the `respectively' by swapping $x,y$.)
    
    In both items, we will show a $\mu$-mincut $A$ with $A \cap 
    \mu \subseteq T$, $u \in A$ and $v \notin A$. 
    Note that $x,y \notin A$ (as $\alpha$ is not in $T$).
    This will yield the conclusion $u \not\befeq v$ as follows:
    Let $X$ be an $(x,y)$-mincut with $v \in X$, $T \subseteq \overline{X}$ (which exists since $v \bef T$).
    Then $A \setminus X$ is a $\mu$-cut (separating $A \cap \mu$ from $x$), and $X \setminus A$ is an $(x,y)$-cut.
    By posimodularity (\Cref{lem:sub-posi-general}), $X \setminus A$ is an $(x,y)$-mincut, keeping $v$ with $x$ and $u$ with $y$.

    Item (i):
    Because $u,v \bef T$, \Cref{lem:projection-inside} implies that $\pi_\mu (u), \pi_\mu (v)$ must also have endpoints outside $T$.
    Thus,
    the sub-skeleton $S_u$ rooted by $f_u$ in direction $\gamma_u$ is contained within $T$.
    Note that $\pi_\mu (v)$ cannot intersect $S_u$: otherwise, in order to leave $S_u$ and go outside of $T$, $\pi_\mu (v)$ must go through $f_u$, contradicting our assumption.
    Then, by~\Cref{thm:proj-vertex}, there exists a $\mu$-mincut $A$ with $A \cap \mu = S_u \subseteq T$, $u \in A$ and $v \notin A$.

    Item (ii):
    Define $S = S_u$ as in Item (i).
    Then by~\Cref{lem:numbering-for-tree-edges}, $\sigma_{f,\gamma} (u) < \sigma_{f,\gamma} (v)$ implies a $\mu$-mincut $A$ with $A \cap \mu = S \subseteq T$, $u \in A$ and $v \notin A$.
\end{proof}

We can now employ the same strategy from the LCA (which is now even a bit simpler, as every vertex in $B$ indeed has an endpoint of its projection within $T$).
Let $d_T (\gamma)$ denote the depth of node $\gamma$ in the tree obtained from $T$ by contracting each cycle within $T$, and rooting it (as one would expect) at at the endpoint of $T$’s rooting edge that lies in $T$.
Again, for $u, v \in B$, if $d_T (\gamma_v) \leq d_T (\gamma_u)$ and $\gamma_u \neq \gamma_v$, then $f_u \notin \pi_\mu (v)$.
Thus, to order the sub-bucket $B$, we act as follows.
Suppose the sub-bucket $B$ is placed to the left of $T$ (if it is to the right, swap ``increasing'' and ``decreasing'' in the following).
We first group the vertices in $B$ by their $\gamma_v$ endpoints (and hence also their $f_v$ last edges), then report the groups in increasing order of depth, where each group is reported in decreasing order of $\sigma_{f,\gamma}$ (where $\gamma$ and $f$ are the common endpoint and its adjacent tree-edge of every vertex in the group).
\Cref{lem:ancestor-comparability-shared-tied-comp} ensures that this is indeed a valid topological ordering of the sub-bucket $B$.

Once every sub-bucket is internally ordered, we obtain a valid topological order on the partition of all of $\mu \cup R_{\leq i}$ by the $x,y$-units, hence ensuring (I3) as required.

\paragraph{Running time.}
Similarly to the LCA case, the main implementation details concern the internal ordering within (sub-)buckets.
As the internal ordering procedures in LCA or above it are essentially identical, the same arguments as in the LCA show that apart from scanning the pre-sorted lists of projections associated with each possible ending tree-edge within $\cH_\mu$, ordering each sub-bucket takes $O(|B| + |T|)$ time.
Summing this over all sub-buckets counts each $O(|T|)$ term twice, since the $\alpha$-component $T$ has at most two sub-buckets associated with it, so this amounts to $O(|R_i| + |\mu_i \setminus \mu_{i-1}|)$ time.
Scanning the pre-sorted lists again takes $O(|P_i|)$ time with $P_i$ being the set of explicitly stored projections at $\mu_i$, as each is visited a constant number of times by these scans.
So, the running time for processing $\mu_i$ is $O(|\mu_i \setminus \mu_{i-1}| + |R_i| + |P_i|)$.


Finally, adding the cost of processing $\mu_0$ summing over all $1 \leq i \leq \ell$, the total running time calculation is as in~\Cref{eq:query-time-calculation}, but adding a term of $\sum_{i=0}^\ell |P_i|$.
ince the data structure stores $O(n)$ projection records in total, $\sum_{i=0}^{\ell}|P_i|=O(n)$.
\medskip
This concludes the proof of the first item of~\Cref{thm:main}; the second item follows immediately as noted in the introduction.

\bibliographystyle{alphaurl}
\bibliography{references}

\appendix

\section{Comparison with \cite{BhanjaKP26}}
\label{sec:comparison-BKP}

Our construction builds on the nearest mincut hierarchy
of Bhanja, Kenneth-Mordoch, and Petruschka~\cite{BhanjaKP26},
which occupies $O(n)$ space and reports the \emph{nearest} $(x,y)$-mincut
in $O(n)$ time for any ordered pair $x,y\in V$.
Thus, their structure recovers one distinguished $x,y$-unit:
the one containing $x$.
In our work, we recover the entire partition into $x,y$-units,
together with an ordering whose nonempty proper prefixes
are $x,y$-mincuts.
This allows us to report an $x,y$-mincut separating
any additional pair of vertices whenever one exists.
Although both constructions use a skeleton hierarchy
and proceed upward from $\LCA_{\cT}(x,y)$,
the stronger task requires new structural tools
already at this starting point.

\paragraph{The starting point: the LCA.}
Let $\mu=\LCA_{\cT}(x,y)$, and let $\alpha,\beta$
be its children containing $x,y$, respectively.
At this level, $x,y$-mincuts are precisely the
$\mu$-mincuts represented by canonical cuts of
$\cH_\mu$ separating $\alpha$ from $\beta$.
For the nearest-mincut query of~\cite{BhanjaKP26},
it suffices to consider the canonical cut closest
to $\alpha$ in the direction of $\beta$ in the
simplified tree skeleton.
The vertices whose projections lie entirely
on its $\alpha$-side are exactly those belonging
to $\Near(x,y)$.
Thus, the task is to identify the vertices confined
to one prescribed region of the skeleton.
Our query must instead recover the distinctions
induced by \emph{all} canonical cuts separating
$\alpha$ from $\beta$ in the full cactus skeleton.
These cuts partition $\mu$ into
$\alpha\beta$-components, each contained in a
different $x,y$-unit.

The handling of projections is also significantly different.
In~\cite{BhanjaKP26}, the classification is binary:
a vertex belongs to $\Near(x,y)$ if its projection
lies entirely on the $\alpha$-side of the canonical
cut closest to $\alpha$, and lies outside otherwise.
Our classification requires two separate steps.
First, we determine whether a vertex $v$ belongs
to the $x,y$-unit of some $\alpha\beta$-component.
This happens exactly when $\pi_\mu(v)$ is contained
in that component; otherwise, its unit does not
intersect $\mu$.
Second, for the vertices not merged with any
$\alpha\beta$-component, we must determine which
of them belong to the same $x,y$-unit.
Thus, being classified as outside the nearest
mincut concludes the earlier task, whereas our
query must further identify the units of these
vertices. 

\paragraph{Differences above the LCA.}
At a strict ancestor $\mu$ of $\LCA_{\cT}(x,y)$,
let $\alpha$ be the child containing both $x$ and $y$.
The differences between our construction and
that of~\cite{BhanjaKP26} extend to the decomposition
of the skeleton, the interpretation of projections,
and the information needed to classify representatives.

\begin{enumerate}
    \item \textbf{From skeleton subtrees to
    $\alpha$-components.}
    Crucially exploiting properties of nearest mincuts, BKP show that each sub-skeleton of $\cH_\mu\setminus\alpha$
    lies entirely inside or entirely outside
    the nearest $x,y$-mincut.
    However, an arbitrary $x,y$-mincut may partition
    such a sub-skeleton.
    Thus, vertices in one subskeleton that behave identically with
    respect to nearest-mincut membership need
    not belong to the same $x,y$-unit. Furthermore, the construction of~\cite{BhanjaKP26} uses
    simplified tree skeletons, which is sufficient for representing nearest mincuts. However,
    we need to work with the full cactus skeleton and
    introduce a finer decomposition of subskeletons into
    $\alpha$-components in a way that they are \emph{atomic}, i.e., each such contained in one
    $x,y$-unit for every $x,y\in\alpha$.

    This refinement also changes the space analysis.
    In~\cite{BhanjaKP26}, one representative per
    subtree suffices, and the number of subtrees
    is bounded by $\deg_{\cH_\mu}(\alpha)$.
    Our decomposition may have many more components
    because a neighboring cycle can contribute
    several of them.
    We therefore cannot afford to store a
    representative from every component.
    We establish new structural insights about these components, namely, the \emph{cyclic consistency} property, such that it restores the degree-based bound
    on the number of representatives and
    preserves $O(n)$ total space.

    \item \textbf{From projection membership
    to assignment among units.}
In~\cite{BhanjaKP26}, the subtrees intersected by a vertex's
projection serve as anchors for deciding its
nearest-mincut membership.
Their inside/outside classification allows the
known portion of the nearest mincut to be extended
upward through the hierarchy.
Here, several $\alpha$-components may be merged
into one $x,y$-unit or belong to different units.
We must characterize how a projected vertex
relates to these components in each case.
We introduce the concept of \emph{tied $\alpha$-components}
as anchors and provide characterizations
that determine whether the vertex joins their
unit or creates new.
These characterizations allow us to grow
individual units from the LCA upward
through the hierarchy.

    \item \textbf{From chain maximizers
    to maxima of crossing cut families.}
    One of the main tools developed
in~\cite{BhanjaKP26} is the chain maximizer.
For an appropriate subtree $T$ having one endpoint of the projection of a
vertex $v$, they consider the family
\[
    \mathcal{N}_{T,v}
    :=
    \{\Near(a,b) \mid a,b\in\alpha,\;
    T\subseteq\Near(a,b),\;
    v\notin\Near(a,b)\},
\]
where $\Near(a,b)$ denotes the nearest
$a,b$-mincut.
They prove that this family forms a chain
under containment and summarize it
by its largest member.
Our corresponding family replaces nearest
mincuts by arbitrary mincuts.
    In this family, the mincuts may
    cross; see~\Cref{lem:maximizer}.
    Thus, the chain property used in
    the earlier construction no longer holds.
    Nevertheless, we prove that this family
    is closed under unions.
    Whenever nonempty, it consequently
    has a unique maximum containing every
    member of the family.
    In~\cite{BhanjaKP26}, the witness vertices $w,z$
alone suffice, since the maximum is uniquely
determined as the nearest $(w,z)$-mincut.
Our maximum need not be the nearest $(w,z)$-mincut,
so the witness pair alone does not identify it.
We additionally store its canonical cut in
the skeleton at $\LCA_{\cT}(w,z)$.
    This constant-size augmentation lets
    us decide whether $v$ joins the unit
    containing $T$.

\end{enumerate}

The following two paragraphs are entirely new and have no counterpart in~\cite{BhanjaKP26}.

\paragraph{Distinguishing vertices
    that do not join existing units.}
    Once a vertex is classified using its projection as outside
    the nearest mincut, its classification
    in~\cite{BhanjaKP26} is complete.
    In our construction, determining that
    a vertex belongs to no unit intersecting
    $\mu$ completes only the first step.
    We must still decide whether it belongs
    to the same unit as another such vertex.

    At the LCA, classical carcass distinctness
    supplies the required criterion.
    Above the LCA, $x,y$ lie in the same
    terminal node $\alpha$, and their mincuts
    are not $\mu$-mincuts.
    We prove an extended distinctness lemma:
    for vertices $u,v$ whose $x,y$-units avoid
    $\mu$, provided at least one has a tied
    $\alpha$-component,
    \[
        \phi_{x,y}(u)=\phi_{x,y}(v)
        \quad\Longleftrightarrow\quad
        \phi_\mu(u)=\phi_\mu(v).
    \]
    Storing $\mu$-unit identifiers with the
    projections therefore allows us to group
    these vertices into their respective
    $x,y$-units. This extended distinctness can be viewed as a new carcass tool, which is of independent interest.


\paragraph{Reporting a topological order.}
Recovering all $x,y$-units determines exactly
which pairs of vertices can be separated by an
$x,y$-mincut, but it does not by itself provide
such a cut.
To report a separating mincut, we further need
an ordering of the units whose nonempty proper
prefixes are $x,y$-mincuts.
Thus, unlike in~\cite{BhanjaKP26}, where identifying
the nearest $x,y$-unit completes the query, here
recovering the units is only part of the task:
we must additionally understand how all these
units are ordered with respect to one another.

This introduces new challenges already at the
LCA.
We must order all the
$\alpha\beta$-components along the full cactus
skeleton.
Cycles make this substantially more delicate:
in particular, empty cycle-nodes do not themselves
correspond to $x,y$-units, yet they affect the
relative placement of the surrounding components
and of vertices whose projections interact with
the cycle.
Thus, even after the $x,y$-units at the LCA have
been identified, obtaining an ordering consistent
with all $x,y$-mincuts requires additional
structural arguments.


The difficulty only increases above the LCA.
We extend the ordering incrementally along the query path. At each ancestor, once the $x,y$-units inside the child have already been ordered, we show that the following steps must be executed in order. First, we order the tied $\alpha$-components consistently with the $x,y$-mincut relation that are arising from a cycle, that is Type-3. Second, for every newly created unit, we determine where it should be inserted relative to the already ordered $x,y$-units. The tied components identify the relevant existing unit, while the structural dichotomy, and in the one-tied-component case the maximizer information, determines on which side of that unit the new unit must appear. Finally, several new units may be assigned to the same position. At this point, the distinctness arguments used to identify these units no longer provide any information about their relative order. Determining this order requires finer structural information, obtained from their tied components, the positions of their projections, and additional ordering information stored in the structure.

\section{Missing Proofs}\label{sec:missing-proofs}
\SUBANDPOSI
\FOURCOMPS
\SIGMANUMBERS
\end{document}